\documentclass[12pt]{article}
\usepackage{epsfig, amssymb}
\usepackage{amsmath}
\usepackage{graphicx,epsfig}
\usepackage{pdflscape}
\usepackage{color} 
\usepackage{cite}
\usepackage[title]{appendix}
\newcommand\ddfrac[2]{\frac{\displaystyle #1}{\displaystyle #2}}

\begin{document}
\thispagestyle{empty}
\begin{flushright}
\end{flushright}

\bigskip

\begin{center}
\noindent{\Large \textbf
{Momentum-space Langevin dynamics of holographic Wilsonian RG flow: self-interacting massive scalar field with marginal deformation
}}\\ 
\vspace{2cm} \noindent{Ji-seong Chae${}^{a}$\footnote{e-mail:jiseongchae17@gmail.com} and 
Jae-Hyuk Oh${}^{a}$\footnote{e-mail:jaehyukoh@hanyang.ac.kr}}

\vspace{1cm}
  {\it
Department of Physics, Hanyang University, Seoul 04763, Korea${}^{a}$\\
 }
\end{center}

\vspace{0.3cm}
\begin{abstract}
\noindent
We explore the mathematical relationship between the holographic Wilsonian renormalization group(HWRG) and stochastic quantization(SQ) motivated by the similarity of the monotonicity in RG flow with Langevin dynamics of non-equilibrium thermodynamics. We look at scalar field theory in AdS space with its generic mass, self-interaction, and marginal boundary deformation in the momentum space. Identifying the stochastic time $t$ with radial coordinate $r$ in AdS, we establish maps between the fictitious time evolution of stochastic multi-point correlation function and the radial evolution of multi-trace deformation, which respectively, express the relaxation process of Langevin dynamics and holographic RG flow. We show that the multi-trace deformations in the HWRG are successfully captured by the Langevin dynamics of SQ. 
\end{abstract}
\newpage
\tableofcontents 

\section{Introduction}

The notion of holography naturally arises from Black hole(brane) thermodynamics, in a sense that the black hole(brane) entropy is proportional to its surface area even though the entropy is an extensive quantity that must be proportional to the system volume\cite{Stephens:1993an}. Therefore, one might guess that the holography implies that the boundary of a black hole(brane) might have the information of the black hole(brane) itself\cite{Susskind:1994vu}. One of the noticeable examples of this is AdS/CFT correspondence. The precise map of AdS/CFT correspondence is that the generating functional, $W(J)$ of a $d$-dimensional field theory defined on $d-$dimensional boundary of
AdS$_{d+1}$ spacetime
is the same with the appropriately renormalized on-shell action, $S_{os}(J)$ of $d+1$-dimensional dual gravitational theory defined in AdS$_{d+1}$, where the $J$ is the boundary value of the gravity field on the AdS boundary\cite{Witten:1998qj}. 

One of the interesting topics in such research fields is the holographic Wilsonian renormalization group(HWRG), which provides the energy scale dependency of a composite operator in the boundary field theory \cite{Witten:2001ua}.
In the context of holography, one identifies the energy scale of $d$-dimensional field theory to the radial direction, $r$ in AdS$_{d+1}$ spacetime, where the dual gravitational theory is defined \cite{Maldacena:1997re, Peet:1998wn}. As a consequence, the renormalization group flow of a certain deformation in the boundary field theory can be obtained from the radial flow on the cut-off hypersurface being normal to the radial coordinate \cite{Alvarez:1998wr,Akhmedov:1998vf}. This rigorous formulation efficiently enables us to describe the low-energy behavior of the boundary field theory system. Integrating out the bulk geometry along the radial direction corresponds to integrating out the high energy degrees of freedom in the field theory, which is called a holographic Wilsonian renormalization group ﬂow(HWRG)\cite{Faulkner:2010jy,Heemskerk:2010hk, Sin:2011yh}.

On the other hand, another interesting framework that might be related to HWRG is stochastic quantization(SQ).
Stochastic quantization is an alternative methodology to quantize Euclidean field theory by employing a first-order differential equation, called Langevin equation\cite{Damgaard:1987rr}. The Langevin equation describes a relaxation process from a non-equilibrium state to thermal equilibrium, where the evolution parameter is $t$, which is so-called stochastic time. The form of the Langevin equation is given by
\begin{equation}
\label{Langevin-eq-introduction}
\frac{\partial \Phi(x,t)}{\partial t}=-\frac{1}{2}\frac{\delta S_E}{\delta\Phi(x,t)}+\eta(x,t),
\end{equation} 
where $\Phi$ is a field living in $d$-dimensional Euclidean space, and $x_i$ is $d$-dimensional coordinate variable in Euclidean space. $S_E$ is a theory that we want to quantize and $\eta$ is called Gaussian white noise, which provides randomness to our system. The Gaussian white noise enjoys the following partition function: 
\begin{equation}
\label{parti-eta}
\mathcal Z=\int [\mathcal D \eta(x,t)]\exp\left\{ -\frac{1}{2}\int^t_{t_0}dtd^dx {\ }\eta^2(x,t) \right\}
\end{equation} 
This allows us to construct a stochastic partition function in terms of the field $\Phi$, which is given by
\begin{equation}
 \mathcal Z=\int [\mathcal D \Phi(x,t)]\exp\left\{-S_{FP}(\Phi,t)\right\},
\end{equation}
where to get the partition function from(\ref{parti-eta}), we change the variable $\eta$ to $\Phi$ by using the Langevin equation (\ref{Langevin-eq-introduction}). $S_{FP}(\Phi,t)$ is called Fokker-Plank action.
We notice that the role of the term of $S_E$ in the Langevin equation (\ref{Langevin-eq-introduction}) is potential, which gives a force of dissipation to the system. It forces the system to settle down to thermal equilibrium by either gaining or losing its energy to its surroundings. This is related to the famous fluctuation-dissipation theorem\cite{Callen:1951vq,Nyquist:1928zz}.

One can also formulate the probability distribution of the noise field $\eta$, which is given by
\begin{equation}
P[\phi(x,t);t]=\frac{1}{\mathcal Z}\int [\mathcal D \eta(x,t)]\exp\left\{ -\frac{1}{2}\int^t_{t_0}dtd^dx {\ }\eta^2(x,t) \right\}\prod_{y}\delta[\phi(y,t_0)-\phi(y)],
\end{equation}
The probability is conditional since it depends on the initial boundary condition given by the Dirac delta function in it. It turns out that this probability distribution function satisfies the equation of stochastic time evolution,
\begin{equation}
\frac{\partial P[\phi(x,t);t]}{\partial t}=\int d^{d+1}x \frac{\delta}{\delta \phi(x,t)}\left(
 \frac{\delta S_E}{\delta \phi(x,t)}+ \frac{\delta}{\delta \phi(x,t)}
 \right)P(\phi(x,t);t).
\end{equation} 

Several noticeable similarities between renormalization group flow and SQ are as follows. Firstly, Langevin equation is first-order differential equation in stochastic time and has a form of diffusion equation. The renormalization equation is also first-order equation in energy scale and converges to a few infra-red fixed points. Secondly, the stochastic process is a memory loss process, which is also known as Markovian process. Regardless of which state the system starts with, it goes into thermal equilibrium at a large stochastic time. The renormalization group flow has a similar property that regardless of which ultraviolet theory is, it gets into a few infra-red fixed points in low energy. Integrating out high energy degrees is also the process of losing initial information about the system

Finally, we would like to discuss entropy production\cite{Udo}. The entropy of the field of $\phi$ is defined as
\begin{equation}
 s(\tau) = -\log P(\phi(x,t),t),
\end{equation}
and its expectation value is 
\begin{equation}
  S(\tau) \equiv \langle s(t) \rangle = -\int d\phi \ p(\phi(x,t),t)\log p(\phi(x,t),t),
\end{equation}
Since there is a dissipative force term in Langevin equation, the field $\phi$ will lose its energy to its surroundings. The amount of energy loss per unit of stochastic time(power loss) is 
\begin{equation}
\frac{dW}{dt}=\frac{1}{2}\frac{\delta S_E}{\delta \phi}\dot \phi(x,t)  \rightarrow \dot s_{\rm surr}
\end{equation}
and it is the entropy production into the medium or surroundings. Once one defines the total entropy of the system as 
\begin{equation}
s_{tot}=s+s_{\rm surr},
\end{equation}
what the authors show in \cite{Udo} is that the expectation value of the $\dot s_{tot}$ is positive-semi definite. It implies a monotonic flow of the stochastic process which has a similarity with the monotonicity of RG flow\cite{Zamolodchikov:1986gt, Cardy:1988cwa, Komargodski:2011vj}.

In the stochastic quantization scheme\cite{Parisi:1980ys}, the correlation function of Euclidean $d$-dimensional field theory of $S_E$ corresponds to the equilibrium state of the correlation function of $d+1$-dimensional field theory described by the Fokker-Plank equation after a large fictitious time ($t\rightarrow \infty$). Similarly, for AdS/CFT description, the correlation function of a $d$-dimensional field theory corresponds to the renormalized-on-shell action of a $d+1$-dimensional gravitational theory. This observation leads to the following studies that explore the connection between stochastic quantization and holography. In the study of \cite{Lifschytz:2000bj}, authors relate the holographic directional coordinate $r$ to the fictitious time $t$ of stochastic quantization and argue that Schwinger-Dyson equations are conditions for equilibrium state in holographic dual gauge theory. Also, it is suggested the partition function of the holographic theory is related to the stochastic partition function \cite{Mansi:2009mz}.  
In particular, the authors in \cite{ Kim:2023gaa} associate the monotonicity of non-equilibrium thermodynamics with holographic RG flow by an emergent extradimension of Wilsonian RG transformation. 

These results motivate us to extend the analysis that the stochastic quantization is associated with AdS/CFT to the holographic Wilsonian renormalization group. Recent studies propose an interesting connection between HWRG and stochastic quantization. If one identifies the holographic directional coordinate $r$ to the fictitious time $t$, one can suggest that the Euclidean action $S_E$ and holographic boundary action $S_B$ have the following relation: $S_E=-2S_B$ \cite{Mansi:2009mz}. It allows the Fokker-Plank action to be identified as the bulk action in gravitational theory. In \cite{Oh:2012bx}, authors use the Hamiltonian description of the holographic Wilsonian renormalization group to yield the Schrodinger-type equation:
\begin{align}
    \partial_\epsilon \psi_H(\phi,r)=-\int_{r=\epsilon}d^dx \ \mathcal{H}_{RG}\left(-\frac{\delta}{\delta\phi},\phi\right)\psi_H(\phi,r)
\end{align}
where $\mathcal{H}_{RG}$ is the Hamiltonian of the bulk theory of the HWRG side which is obtained from Legendre transformation of the Lagrangian of the bulk gravity theory and the wave function is given as $\psi_H=e^{-S_B}$. Also, the Hamiltonian formalism of the stochastic quantization is given by
\begin{align}
    \partial_t\psi_S(\phi,t)=-\int_{r=\epsilon} \ \mathcal{H}_{FP}\left(-\frac{\delta}{\delta\phi},\phi\right)\psi_S(\phi,t)
\end{align}
where $\mathcal{H}_{FP}$ is the Fokker-Plank Hamiltonian which is obtained by Legendre transformation of the Fokker-Plank Lagrangian. The wave function is given by $\psi_S(\phi,t)=P(\phi,t)e^{\frac{S_E(\phi(t))}{2}}$.
Identification of the Hamiltonian and wave function of both sides of HWRG and stochastic quantization produces a relation $S_B=S_P-\frac{S_E}{2}$, where $P(\phi,t)=e^{-S_P}$. One can express $S_P$ in terms of stochastic correlation functions and it formulates radial flows of holographic multi-trace deformation which constitutes the boundary action $S_B$. 
By employing such a relation, the authors in \cite{Oh:2012bx,Jatkar:2013uga,Oh:2015xva,Oh:2013tsa, Moon:2017btx} construct a mathematical relationship between stochastic time evolution of stochastic 2-point correlation function and the radial flow of double trace operator in holography as
\begin{align}
\label{2-relation-intro}
\left.\frac{\delta^2S_B}{\prod^2_{i=1}\delta\Phi(k_i,r)}\right\rvert^{r=t}=\left\langle \Phi(k_1,t)\Phi(k_2,t)\right\rangle^{-1}_\textrm{S}-\frac{1}{2}\frac{\delta^2S_E}{\prod^2_{i=1}\delta\Phi(k_i,t)},
\end{align}
where the left-hand side of the relation is double trace deformation in holography whereas the first term on the right-hand side is the inverse of the stochastic 2-point correlation function in tree level.

The relation is also extended to higher-order stochastic function and multiple trace operators.
In \cite{Oh:2021bxx}, authors firstly suggested the relation of triple trace operator in HWRG and 3-point function in stochastic quantization, by employing conformally coupled scalar
with $\phi^3$ interaction in AdS$_5$.
The relation between quadruple trace operator and stochastic 4-point function by employing $\phi^4$ theory in AdS$_4$\cite{Lee:2023ynb}.
In \cite{Kim:2023bhp}, the authors extend the relation to a case in a more general boundary condition imposed on the holographic theory. Especially, the authors consider marginal multiple trace deformation on the conformal boundary and discuss the relation between multiple trace deformation in holography and multi-point function in stochastic quantization. However, they restrict themselves to zero boundary momentum case only.

In this study, we develop a computational method for the mathematical relationship between stochastic multi-point functions and multiple trace deformation in holography in non-zero momentum case.  
Our research is quite general in the sense that we consider scalar theory in AdS space with arbitrary mass, self-coupling, and boundary deformation as a holographic model. To be more precise, 
we consider marginal multiple trace deformation on the conformal boundary.

Together with the relation(\ref{2-relation-intro}), we obtain the relation between $n$-multiple trace deformation and stochastic $n$-point function, which is given by
\begin{align}
\label{n-relation-intro}
\left.\frac{\delta^nS_B}{\prod^n_{i=1}\delta\Phi(k_i,r)}\right\rvert^{r=t}=-\left\langle \prod^n_{i=1}\Phi(k_i,t)\right\rangle_\textrm{S}\prod^n_{j=1}\left\langle\Phi(k_j,t)\Phi(-k_j,t)\right\rangle^{-1}_\textrm{S}-\frac{1}{2}\frac{\delta^nS_E}{\prod^n_{i=1}\delta\Phi(k_i,t)}.
\end{align}
We also derive the relation of $(2n-2)$-multiple trace deformation and the stochatic $(2n-2)$-function being given by
\begin{align}
\label{2n-2-relation-intro}
\left.\frac{\delta^{2n-2}S_B}{\prod^{2n-2}_{i=1}\delta\Phi(k_i,r)}\right\rvert^{r=t}&=-\left\langle \prod^{2n-2}_{i=1}\Phi(k_i,t)\right\rangle_\textrm{S}\prod^{2n-2}_{j=1}\left\langle\Phi(k_j,t)\Phi(-k_j,t)\right\rangle^{-1}_\textrm{S}
\\ \nonumber
&+\frac{(2n-2)!n^2}{2(n!)^2}\prod^{2n-2}_{i=1}\left\langle\Phi(k_i,t)\Phi(-k_i,t)\right\rangle^{-1}_\textrm{S}\times\textrm{Perm}\left[\left\langle\left\{\prod^{n-1}_{j=1}\Phi(k_j,t)\right\}\Phi(q,t)\right\rangle_\textrm{S}\right.
\\ \nonumber
&\left.\times\left\langle\Phi(q,t)\Phi(-q,t)\right\rangle^{-1}_\textrm{S}\left\langle\left\{\prod^{2n-2}_{l=n}\Phi(k_l,t)\right\}\Phi(-q,t)\right\rangle_\textrm{S}\right]
\\ \nonumber
&-\frac{1}{2}\frac{\delta^{2n-2}S_E}{\prod^{2n-2}_{i=1}\delta\Phi(k_i,t)}.
\end{align}
We note that once one turns off the boundary directional momentum, the above relations are reduced to the relation of the zero-momentum case in \cite{Kim:2023bhp}.
We check the above relations for our holographic model and it turns out that
the relations (\ref{2-relation-intro}-\ref{2n-2-relation-intro}) successfully capture the HWRG flow by the process of SQ.

To check the relations(\ref{2-relation-intro}-\ref{2n-2-relation-intro}), we follow the prescription suggested in  \cite{Oh:2012bx,Jatkar:2013uga,Oh:2015xva,Oh:2013tsa, Moon:2017btx,Oh:2021bxx,Lee:2023ynb,Kim:2023bhp}, which is
$S_E = 2S_B=2(S_{\text{ct}} + S_{\text{def}})$, where $S_{\text{ct}}$ is the boundary counter term and $S_{\text{def}}$ is the boundary deformation in our holographic model. For our holographic model, we consider the self-interaction Lagrangian density up to $\mathcal{L}_{int} \sim \lambda_{2n-2} \phi^{(2n-2)}$, where $\lambda_{2n-2}$ is the self-interaction coupling constant. In this theory, the multiple trace coupling becomes marginal if $\nu=\frac{n-2}{2n}d$, where $\nu \equiv \sqrt{m^2+\frac{d^2}{4}}$ and $m$ is a mass of the scalar field, which is in the range that $-\frac{d^2}{4}\leq m^2 \leq -\frac{d^2}{4}+1$. In the range of the mass of the scalar field, the bulk theory allows alternative quantization, where one can impose the Neumann boundary condition as well as the Dirichlet boundary condition. We note that this model is the same as that in \cite{Kim:2023bhp}, but for simplicity, the authors in \cite{Kim:2023bhp} consider the zero-boundary-momentum case. Our work is more general in the way that the non-zero boundary momentum case is considered. 

In Sec.\ref{Holographic Wilsonian Renormalization Group}, we perform a calculation on holographic Wilsonian renormalization group flow with boundary multi-trace deformation. In Sec.\ref{Stochastic Quantization}, we discuss the non-equilibrium thermodynamic method to quantize the Euclidean field, called stochastic quantization. Especially, we construct Euclidean field theory in the stochastic frame by identifying the boundary action of our holographic model with Euclidean action with the relation $S_E=-2S_B$. We show that the identification of the holographic bulk action and the Fokker-Plank action reproduces the same Euclidean action in Sec.\ref{Evaluation of the Stochastic Kernel and Construction of the Euclidean Action}. 
In section \ref{The precise map of stochastic correlation functions and multi-trace deformations}, we establish the precise map of stochastic correlation functions and multi-trace deformations in our holographic model. 
We show that the relations are well established for our model.
Appendix \ref{2n-2-correlation-function} and \ref{2n-2-relation-appendix} contain technical methods of calculating stochastic correlation function in momentum space.
\section{A Brief Review of the Relation between HWRG and SQ}
In this section, we mostly follow the method of \cite{Oh:2012bx} to review the relationship between the Holographic Wilsonian renormalization group and stochastic quantization. We will show if we identify the Hamiltonian and wave functions of two different frames, we obtain a precise map between the multi-trace deformations in HWRG and the multi-point functions in SQ. 

\subsection{Holographic Wilsonian renormalization group}
We start with a brief review of the HWRG. Consider a holographic model defined in Euclidean AdS$_{d+1}$ given by
\begin{align}
\label{simple action}
    S=\int_{r>\epsilon} dr d^dx \sqrt{g} \mathcal{L}(\phi,\partial\phi)+S_B,
\end{align}
where $S_B$ is the boundary action defined on $r=\epsilon$ hypersurface and $\phi$ is the field in AdS$_{d+1}$. The background metric is given by
\begin{eqnarray}
\label{AdS metric}
    \quad ds^2=g_{\mu\nu}dx^\mu dx^\nu=\frac{1}{r^2}\left(dr^2+\sum_{i=1}^ddx^idx^j\delta_{ij}\right),
\end{eqnarray}
where $r$ is radial coordinate of Euclidean $AdS$ and $x^i$ is the boundary directional coordinates. The conformal boundary is defined at $r=0$ and the Poincaré horizon is at $r = \infty$. The coordinate indices $\mu,\nu$ run from 1 to $d+1$ (so $i,j$ run from $1$ to $d$) and we define the radial coordinate of AdS as $ x^{d+1} \equiv r$. 
One can define the canonical momentum of the field $\phi$ by requiring the condition that the variation of the action vanishes even in the $r=\epsilon$ boundary:
\begin{align}
    \Pi=\sqrt{g}\frac{\partial \mathcal{L}}{\partial (\partial_r\phi)}=\frac{\delta S_B}{\delta \phi(x)}.
\end{align}
Since the cut-off $\epsilon$ is arbitrary and action (\ref{simple action}) should not depend on it $\frac{dS}{d\epsilon}=0$, we obtain 
\begin{align}
\label{flow equation}
    \partial_\epsilon S_B=-\int_{r=\epsilon} d^dx \left(\frac{\delta S_B}{\delta \phi}\partial_r \phi-\mathcal{L}(\phi,\partial \phi)\right)=\int_{r=\epsilon} d^dx \mathcal{H}_{RG}(\frac{\delta S_B}{\delta \phi},\phi),
\end{align}
where we perform Legendre transformation for the second equality and obtain Hamiltonian density $\mathcal{H}_{RG}$. If we define the following wave function:
\begin{align}
    \psi_H=\exp(-S_B),
\end{align}
we obtain a Schrodinger-type equation as 
\begin{align}
\label{holographic Schrodinger-type equation}
    \partial_\epsilon \psi_H(\phi,r)=-\int_{r=\epsilon}d^dx \ \mathcal{H}_{RG}\left(-\frac{\delta}{\delta\phi},\phi\right)\psi_H(\phi,r).
\end{align}
 For the semi-classical limit $\frac{\delta^2 S_B}{\delta \phi^2} \ll \left(\frac{\delta S_B}{\delta \phi}\right)^2$, the equation (\ref{holographic Schrodinger-type equation}) approximatly becomes the equation (\ref{flow equation}).
\subsection{Stochastic quantization}
The Hamiltonian description of the stochastic quantization also provides the evolution of the system along the stochastic fictitious time $t$. The basic notion of stochastic quantization comes from the partition function of the Euclidean field theory is similar to the partition function of an equilibrium statistical system. We consider the correlation function of $d$-dimensinal Euclidean field as 
\begin{align}
    \langle \phi(x_1) \cdots \phi(x_n) \rangle=\frac{1}{\bar{\mathcal{Z}}}\int [\mathcal{D}\phi] e^{-\frac{S_E(\phi)}{\hbar}}\phi(x_1) \cdots \phi(x_n),
\end{align}
where $S_E$ is an Euclidean action and
\begin{align}
    \bar{\mathcal{Z}}=\int[\mathcal{D}\tilde{\phi}] e^{-\frac{S_E(\tilde{\phi})}{\hbar}}
\end{align}
With the condition $\hbar=k_BT$, the above correlation function is merely a statistical partition function in an equilibrium state with temperature $T$. 

However, the stochastic process describes a non-equilibrium state with a probabilistic distribution function $P(\phi,t)$ as
\begin{align}
\label{probaility distribution}
    \langle \phi(x_1) \cdots \phi(x_n) \rangle=\int [\mathcal{D}\phi] P(\phi,t) \phi(x_1) \cdots \phi(x_n).
\end{align}
SQ explains that the stochastic correlation of the scalar field $\Phi$ in $d+1$-dimensional space gives the correlation function of the $d$-dimensional Euclidean field at a large fictitious time ($t\rightarrow \infty$). The extra dimension of the fictitious time ``$t
$" emerges by promoting the field to be time-dependent as
\begin{align}
\phi(x) \rightarrow \phi(x,t).
\end{align}
 In the stochastic process, the scalar field interacts with an imaginary thermal reservoir causing the non-equilibrium stochastic process and settling down to an equilibrium distribution for a large fictitious time. This relaxation process can be expressed by the Langevin equation:
\begin{equation}
\frac{\partial \phi(x,t)}{\partial t}=-\frac{1}{2}\frac{\delta S_E}{\delta\phi(x,t)}+\eta(x,t),
\end{equation}
where the white noise field $\eta(x)$ has the Gaussian probability distribution and its correlation function is given by
\begin{equation}
\langle \eta(x,t) \eta(x',t') \rangle=\delta^d(x-x')\delta(t-t'),
\end{equation}
which comes from the Markov property of the stochastic process. 
The stochastic partition function is given by the noise field distribution:
\begin{equation}
\mathcal Z=\int [\mathcal D \eta(x,t)]\exp\left\{ -\frac{1}{2}\int^t_{t_0}d\tilde{t} \int d^dx {\ }\eta^2(x,\tilde{t}) \right\}.
\end{equation}
Once we employing the Langevin equation and changing variable $\eta$ to $\phi$, we get
\begin{align}
    \mathcal Z=\int [\mathcal D \phi(x,t)]\det\left(\frac{\delta\eta}{\delta\phi}\right)P(\phi,t_0)\exp\left\{ -\frac{1}{2}\int^t_{t_0}d\tilde{t} \int d^dx \left(\frac{\partial \phi(x,\tilde{t})}{\partial \tilde{t}}+\frac{1}{2}\frac{\delta S_E}{\delta\phi(x,\tilde{t})}\right)^2 \right\},
\end{align}
where the initial probability distribution is
\begin{align}
    P(\phi,t_0)=\prod_x\delta^{(d)}(\phi(x,t_0)-\phi_0(x)),
\end{align}
and the Jacobian factor comes from the Langevin equation given by
\begin{align}
    \det\left(\frac{\delta\eta}{\delta\phi}\right)=\exp\left[\frac{1}{4}\int^t_{t_0}d\tilde{t}\int d^dx \frac{\delta^2 S_E}{\delta\phi^2(x,\tilde{t})}\right].
\end{align}
Now, the complete partition function is given by
\begin{align}
\label{full partition function}
    \mathcal{Z}=\int\mathcal{D}\phi(x,t_0) P(\phi,t_0)e^{\frac{S_E(\phi(t_0))}{2}}\mathcal{D}\phi(x,t)e^{-\frac{S_E(\phi(t))}{2}}[\mathcal{D}\phi]\exp\left(-\int^t_{t_0}d\tilde{t}\int d^dx \mathcal{L}_{FP}(\phi(x,\tilde{t}))\right),
\end{align}
where
\begin{align}
    [\mathcal{D}\phi]=\prod_{t_0<\tilde{t}<t}\mathcal{D}\phi(x,\tilde{t}),
\end{align}
and the Fokker-Planck Lagrangian density is given by
\begin{align}
\label{Fokker-Planck Lagrangian}
    \mathcal{L}_{FP}=\frac{1}{2}\left(\frac{\partial\phi(x)}{\partial t}\right)^2+\frac{1}{8}\left(\frac{\delta S_E}{\delta \phi(x)}\right)^2-\frac{1}{4}\frac{\delta^2 S_E}{\delta \phi^2(x)}.
\end{align} 
By comparison of (\ref{full partition function}) with (\ref{probaility distribution}), we obtain the probability distribution as 
\begin{align}
    P(\phi,t)=\exp\left[-\frac{S_E(\phi(t))}{2}-\int^t_{t_0}d\tilde{t}\int d^dx \mathcal{L}_{FP}(\phi(x,\tilde{t}))\right].
\end{align}
It is well known that $P(\phi,t)$ satisfies the following functional equation:
\begin{align}
    \frac{\partial P(\phi,t)}{\partial t}=\frac{1}{2}\int d^dx \frac{\delta}{\delta\phi(x,t)}\left(\frac{\delta S_E}{\delta \phi(x,t)}+\frac{\delta}{\delta\phi(x,t)}\right)P(\phi,t),
\end{align}
and that is related to time evolution with the Fokker-Planck Hamiltonian. 
To see this, we define the wavefunction as
\begin{align}
    \phi_S(\phi,t)\equiv P(\phi,t)e^{\frac{S_E}{2}},
\end{align}
and then we can show that the wavefunction satisfies the following Schrodinger-type equation:
\begin{align}
\label{stochastic Schrodinger-type equation}
    \partial_t\psi_S(\phi,t)=-\int d^dx \mathcal{H}_{FP}(\frac{\delta}{\delta\phi},\phi)\psi_S(\phi,t)
\end{align}
where the Fokker-Planck Hamiltonian is 
\begin{align}
    \mathcal{H}_{FP} &\equiv \frac{1}{2}\left(-\frac{\delta}{\delta\phi(x)}+\frac{1}{2}\frac{\delta S_E}{\delta\phi(x)}\right)\left(\frac{\delta}{\delta\phi(x)}+\frac{1}{2}\frac{\delta S_E}{\delta\phi(x)}\right)
    \\ \nonumber &=-\frac{1}{2}\frac{\delta^2}{\delta\phi^2(x)}+\frac{1}{8}\left(\frac{\delta S_E}{\delta \phi(x)}\right)^2-\frac{1}{4}\frac{\delta^2 S_E}{\delta \phi^2(x)}.
\end{align}
One can also derive the Fokker-Planck Hamiltonian by Legendre transformation of the Fokker-Planck Lagrangian (\ref{Fokker-Planck Lagrangian}).
\subsection{One-to-one mapping between HWRG and SQ}
The Schrodinger type-equation of the HWRG (\ref{holographic Schrodinger-type equation}) and SQ(\ref{stochastic Schrodinger-type equation}) have similar Hamiltonian descriptions of the system. By identifying the Hamiltonian $\mathcal{H}_{RG}(r)=\mathcal{H}_{FP}(t)$, the two wave functions of the two different frames are identified as
\begin{align}
    \psi_H =e^{-S_B}=\psi_S=P(\phi,t)e^{\frac{S_E}{2}}.
\end{align}
From this identification, we obtain a relation 
\begin{align}
\label{fundamental relation}
    S_B=S_P-\frac{S_E}{2},
\end{align}
where we rewrite the probability distribution with $S_P$ as
\begin{align}
P(\phi,t)\equiv e^{-S_P}.
\end{align}
\subsubsection{Free theory relation}
Recalling the definition of the stochastic correlation function (\ref{probaility distribution}), the stochastic 2-point function in the momentum space is given by
\begin{align}
    \langle\phi (k_1,t_1)\phi(k_2,t_2)\rangle_S=\int [\mathcal{D}\phi] e^{-S_P(t)}\phi (k_1,t_1)\phi(k_2,t_2).
\end{align}
We assume that the $S_P$ has a form:
\begin{align}
    S_P(t)=\frac{1}{2}\int d^dk P_2(k,t)\phi(k,t)\phi(-k,t)
\end{align}
From this assumption, we obtain the stochastic 2-point correlation function as
\begin{align}
    \langle\phi (k_1,t_1)\phi(k_2,t_2)\rangle_S=\frac{1}{P_2(k,t)}\delta^{(d)}(k_1+k_2).
\end{align}
On the other hand, we define a kernel of the double-trace operator in holographic effective action given by
\begin{align}
    \langle \phi(k_1,r_1)\phi(k_2,r_2)\rangle_H^{-1}=\frac{\delta^2 S_B}{\delta\phi(k_1,r_1)\delta\phi(k_2,r_2)}.
\end{align}
According to the relation (\ref{fundamental relation}), by adding the kernel of Euclidean action $S_E$, we obtain the relation of the double-trace deformation and stochastic correlation function as
\begin{align}
\label{relation-2point-and-radial-flow}
    \left.\frac{\delta^2S_B}{\prod^2_{i=1}\delta\Phi(k_i,r)}\right\rvert^{r=t}=\left\langle \Phi(k_1,t)\Phi(k_2,t)\right\rangle^{-1}_\textrm{S}-\frac{1}{2}\frac{\delta^2S_E}{\prod^2_{i=1}\delta\Phi(k_i,t)}.
\end{align}
Note that we impose the identification $r=t$ and ignore all the momentum $\delta$-functions in the relation. To review the general case, we mostly follow \cite{Kim:2023bhp}.
\subsubsection{General case}
For a later discussion, we consider the general case of the relation between multi-trace deformations and stochastic multi-point functions. We define the most  general case of the probability distribution function as
\begin{align} 
P(\Phi(k,t);t) = \exp\left\{-\sum^{\infty}_{i=2}\left[\prod^i_{j=1}\int d^dk_j\, \Phi(k_j,t) \right] P_n(k_1,\cdots,k_i;t)\,\delta^{(5)}\left(\sum^i_{l=1}k_l\right) \right\}.
\end{align} 
For simplicity, we consider the existence of bi-linear, $n$-multiple, and $,(2n-2)$-multiple terms of the field $\Phi$ only in its Lagrangian density of both the stochastic theory and holographic model.
Then we can reduce the stochastic partition function as 
\begin{align}
\label{stochastic-partition-function-P}
Z &= \int\left[\mathcal{D}\Phi(k)\right] e^{-S_p}
\\ \nonumber
&= \int\left[\mathcal{D}\Phi(k)\right]\exp\left[-\int P_2(k_1,k_2)\delta^{(d)}(k_1+k_2)\prod^2_{s=1}\Phi(k_s)d^dk_s +\int J(k)\Phi(k)d^dk \right.
\\ \nonumber
&\hphantom{= \int\left[\mathcal{D}\Phi(k)\right]\exp[}- \int P_n(k_1,\cdots,k_n)\delta^{(d)}\left(\sum^n_{i=1}k_i\right)\prod^n_{j=1}\Phi(k_j)d^d\vec{k}_j
\\ \nonumber
&\left. \hphantom{= \int\left[\mathcal{D}\Phi(k)\right]\exp[} -\int P_{2n-2}(k_1,\cdots,k_{2n-2})\delta^{(d)}\left(\sum^{2n-2}_{l=1}k_l\right) \prod^{2n-2}_{m=1}\Phi(k_m)d^d\vec{k}_m \right].
\end{align}
An external source $J$ plays a role to get a generating functional of the partition function. Now we expand the exponents by assuming that the higher order interaction couplings $P_n$ and $P_{2n-2}$ are smaller than $P_2$ as $|P_2|\gg|P_n|\gg|P_{2n-2}|$ to evaluate the higher order interaction perturbatively. Then, the partition function becomes
\begin{align}
\label{Z-manipulated-partition-1}
Z&= \int\left[\mathcal{D}\Phi(k)\right]\exp\left[-\int P_2(k_1,k_2)\delta^{(d)}(k_1+k_2)\prod^2_{s=1}\Phi(k_s)d^dk_s +\int J(k)\Phi(k)d^dk \right]
\\ \nonumber
&\times \left\{ 1 - \int P_n(k_1,\cdots,k_n)\delta^{(d)}\left(\sum^n_{i=1}k_i\right)\prod^n_{j=1}\Phi(k_j)d^d\vec{k_j}\right.
\\ \nonumber
&\hphantom{\times [} +\frac{1}{2!}\int P_n(k_1,\cdots,k_n)\delta^{(d)}\left(\sum^n_{i=1}k_i\right)\prod^n_{j=1}\Phi(k_j)d^d\vec{k_j} \int P_n(k_{n+1},\cdots,k_{2n})\delta^{(d)}\left(\sum^{2n}_{l=n+1}k_l\right)\prod^{2n}_{m=n+1}\Phi(k_m)d^dk_m
\\ \nonumber
&\left.\hphantom{\times [} -\int P_{2n-2}(k_1,\cdots,k_{2n-2})\delta^{(d)}\left(\sum^{2n-2}_{l=1}k_l\right)\prod^{2n-2}_{m=1}\Phi(k_m)d^d\vec{k}_m+... \right\}.
\end{align}
By replacing every $\Phi$ with $\frac{\delta}{\delta J}$ in the curly bracket in (\ref{Z-manipulated-partition-1})and integrating out the field $\Phi$, we obtain the following form of the generating functional:
\begin{align}
Z&= \left\{ 1 - \int P_n(k_1,\cdots,k_n;t)\delta^{(d)}\left(\sum^n_{i=1}k_i\right)\prod^n_{j=1}\frac{\delta}{\delta J(k_j)}d^d\vec{k}_j \right.
\\ \nonumber
&\hphantom{=[} +\frac{1}{2!}\int P_n(k_1,\cdots,k_n;t) P_n(k_{n+1},\cdots,k_{2n};t) \delta^{(d)}\left(\sum^n_{i=1}k_i\right) \delta^{(d)}\left(\sum^{2n}_{j=n+1}k_j\right) \prod^{2n}_{l=1}\frac{\delta}{\delta J(k_l)}d^d\vec{k}_l
\\ \nonumber
&\left. \hphantom{=[} - \int P_{2n-2}(k_1,\cdots,k_{2n-2};t) \delta^{(d)}\left(\sum^{2n-2}_{l=1}k_l\right) \prod^{2n-2}_{m=1}\frac{\delta}{\delta J(k_m)}d^d\vec{k}_m \right\}
\\ \nonumber
&\times \exp \left[\frac{1}{4} \int d^dp_1 d^dp_2 \frac{\delta^{(d)}(p_1+p_2)}{P_2(p_1,p_2)} J(p_1)J(p_2) \right]
\end{align}
Now, we are ready to compute the stochastic multi-point correlation functions from the generating functional. The $n$-point correlation function is given by
\begin{align}
\label{generating functional n point}
\left\langle \prod^n_{i=1} \Phi(k_i,t) \right\rangle_\textrm{S}
&= \frac{\delta^n \log Z}{\prod^n_{i=1} \delta J(k_i)} = -n! P_n(k_1,\cdots,k_n;t) \prod^n_{i=1} \frac{1}{2 P_2(k_i,-k_i;t)}\,\delta^{(d)}\left(\sum^n_{j=1}k_j\right).
\end{align}
And $(2n-2)$-point correlation function is given by
\begin{align}
\label{generating functional 2n-2 point}
\left\langle \prod^{2n-2}_{i=1} \Phi(k_i,t) \right\rangle_\textrm{S}
&= \frac{\delta^{2n-2} \log Z}{\prod^{2n-2}_{i=1} \delta J(k_i)}
\\ \nonumber
&= -(2n-2)! P_{2n-2}(k_1,\cdots,k_{2n-2};t) \prod^{2n-2}_{i=1} \frac{1}{2 P_2(k_i,-k_i;t)}\,\delta^{(d)}\left(\sum^{2n-2}_{j=1}k_j\right)
\\ \nonumber
&\hphantom{=} +\frac{1}{2!}\cdot n^2 \frac{1}{2P_2(q,-q;t)} P_n(k_1,\cdots,k_{n-1},q;t)P_n(k_n,\cdots,k_{2n-2},-q;t)
\\ \nonumber
&\hphantom{=} \times (2n-2)!\prod^{2n-2}_{i=1}\frac{1}{2P_2(k_i,-k_i;t)} \delta^{(d)}\left(\sum^{2n-2}_{j=1} k_j\right)
\\ \nonumber
&= (2n-2)!\prod^{2n-2}_{i=1}\frac{1}{2P_2(k_i,-k_i;t)} \delta^{(d)}\left(\sum^{2n-2}_{j=1} k_j\right)
\\ \nonumber
&\hphantom{=} \times \textrm{Perm}\left[ \frac{n^2 P_n(k_1,\cdots,k_{n-1},q;t)P_n(k_n,\cdots,k_{2n-2},-q;t)}{4 P_2(q,-q;t)} - P_{2n-2}(k_1,\cdots,k_{2n-2};t) \right],
\end{align}
where Perm denotes all possible permutations of the momentum labels as
\begin{equation}
{\rm Perm}\{A(k_1,k_2,...k_m)\}=\frac{1}{m!}\left\{A(k_1,k_2,...k_m)+ {\rm all\ possible\ permutation\ of\ } k_1, k_2,...k_m {\rm\ in\ }A \right\}.
\end{equation}
The inverse relations of the correlation functions (\ref{generating functional n point}) and (\ref{generating functional 2n-2 point}) are written as
\begin{align}
\label{probability kernel n}
P_n(k_1,\cdots,k_n;t) = -\frac{1}{n!} \left\langle \prod^n_{i=1} \Phi(k_i,t) \right\rangle_\textrm{S} \prod^n_{j=1} \left\langle \Phi(k_j,t)\Phi(-k_j,t) \right\rangle_\textrm{S}^{-1},
\end{align}
and
\begin{align}
\label{probability kernel 2n-2}
&P_{2n-2}(k_1,\cdots,k_{2n-2};t)
\\ \nonumber
&= -\frac{1}{(2n-2)!} \left\langle \prod^{2n-2}_{i=1} \Phi(k_i,t) \right\rangle_\textrm{S} \prod^{2n-2}_{j=1} \left\langle \Phi(k_j,t)\Phi(-k_j,t) \right\rangle_\textrm{S}^{-1}
\\ \nonumber
&+ n^2\,\textrm{Perm}\left[ \frac{P_n(k_1,\cdots,k_{n-1},q;t)P_n(k_n,\cdots,k_{2n-2},-q;t)}{2 P_2(q,-q;t)} \right]
\\ \nonumber
&= -\frac{1}{(2n-2)!} \left\langle \prod^{2n-2}_{i=1} \Phi(k_i,t) \right\rangle_\textrm{S} \prod^{2n-2}_{j=1} \left\langle \Phi(k_j,t)\Phi(-k_j,t) \right\rangle_\textrm{S}^{-1}
\\ \nonumber
&+ \frac{n^2}{(n!)^2} \prod^{2n-2}_{j=1} \left\langle \Phi(k_j,t)\Phi(-k_j,t) \right\rangle_\textrm{S}^{-1} \textrm{Perm}\left[ \left\langle \prod^{n-1}_{l=1} \Phi(k_l,t)\Phi(q,t) \right\rangle_\textrm{S} \frac{\left\langle \Phi(q,t)\Phi(-q,t) \right\rangle_\textrm{S}^{-1}}{2} \left\langle \prod^{2n-2}_{m=n} \Phi(k_m,t)\Phi(-q,t) \right\rangle_\textrm{S} \right].
\end{align}
Inserting these results (\ref{probability kernel n}) and (\ref{probability kernel 2n-2}) into the relation (\ref{fundamental relation}), we get the following one-to-one mapping between the stochastic $n$-point, ($2n-2$)-point correlations and holographic $n$-multiple trace, ($2n-2$)-multiple trace operators respectively.

The relation between stochastic $n$-multiple-trace deformation and $n$-point function is given by
\begin{align}
\label{relation-npoint-ntrace}
\left.\frac{\delta^nS_B}{\prod^n_{i=1}\delta\Phi(k_i,r)}\right\rvert^{r=t}=-\left\langle \prod^n_{i=1}\Phi(k_i,t)\right\rangle_\textrm{S}\prod^n_{j=1}\left\langle\Phi(k_j,t)\Phi(-k_j,t)\right\rangle^{-1}_\textrm{S}-\frac{1}{2}\frac{\delta^nS_E}{\prod^n_{i=1}\delta\Phi(k_i,t)}.
\end{align}
The relation between and $(2n-2)$-multiple-trace operator and stochastic $(2n-2)$-point function is obtained as
\begin{align}
\label{relation-2n-2point-2n-2trace}
\left.\frac{\delta^{2n-2}S_B}{\prod^{2n-2}_{i=1}\delta\Phi(k_i,r)}\right\rvert^{r=t}&=-\left\langle \prod^{2n-2}_{i=1}\Phi(k_i,t)\right\rangle_\textrm{S}\prod^{2n-2}_{j=1}\left\langle\Phi(k_j,t)\Phi(-k_j,t)\right\rangle^{-1}_\textrm{S}
\\ \nonumber
&+\frac{(2n-2)!n^2}{2(n!)^2}\prod^{2n-2}_{i=1}\left\langle\Phi(k_i,t)\Phi(-k_i,t)\right\rangle^{-1}_\textrm{S}\times\textrm{Perm}\left[\left\langle\left\{\prod^{n-1}_{j=1}\Phi(k_j,t)\right\}\Phi(q,t)\right\rangle_\textrm{S}\right.
\\ \nonumber
&\left.\times\left\langle\Phi(q,t)\Phi(-q,t)\right\rangle^{-1}_\textrm{S}\left\langle\left\{\prod^{2n-2}_{l=n}\Phi(k_l,t)\right\}\Phi(-q,t)\right\rangle_\textrm{S}\right]
\\ \nonumber
&-\frac{1}{2}\frac{\delta^{2n-2}S_E}{\prod^{2n-2}_{i=1}\delta\Phi(k_i,t)}.
\end{align}
To confirm these relations (\ref{relation-2point-and-radial-flow}), (\ref{relation-npoint-ntrace}), and (\ref{relation-2n-2point-2n-2trace}), we compute the multi-trace operators in Sec. \ref{Holographic Wilsonian Renormalization Group} and stochastic correlation function with Euclidean action in Sec. \ref{Stochastic Quantization}, which constitute the left-hand and right-hand side of the relations respectively. Then we confirm the relation and find the matching conditions in Sec. \ref{The precise map of stochastic correlation functions and multi-trace deformations}.

\subsection{Summary of the Results}
\paragraph{Double trace deformation and stochastic 2-point function}
The correspondence between double trace deformation and stochastic 2-point function is given by
\begin{align}
    \left.\frac{\delta^2S_B}{\prod^2_{i=1}\delta\Phi(k_i,r)}\right\rvert^{r=t}&=\left\langle \Phi(k_1,t)\Phi(k_2,t)\right\rangle^{-1}_\textrm{S}-\frac{1}{2}\frac{\delta^2S_E}{\prod^2_{i=1}\delta\Phi(k_i,t)}.
\\ \nonumber \left.\partial_\epsilon \log \left[ \epsilon^{1/2}\left(I_\nu(\vert k \vert \epsilon)+c^{(1)}_{k}/c^{(2)}_{k}K_\nu(\vert k \vert \epsilon)\right)\right]\right\vert^{\epsilon=t}&= \partial_t \log \left[ t^{1/2}\left(I_\nu(\vert k \vert t)-\alpha_k(t_0) K_\nu(\vert k \vert t)\right)\right],
\end{align}
where the left-hand side result comes from the flow equation of the holographic model and the right-hand side result is derived in the stochastic frame. 
The relation is satisfied when 
\begin{align}
\label{constant relation}
c^{(1)}_{k}/c^{(2)}_{k}=-\alpha_k(t_0),
\end{align}
where $c^{(1)}_{k}$ and $c^{(2)}_{k}$ are the arbitrary coefficients of the solution of the bulk equation of motion, $ \mathcal{Q}_\nu(k, \epsilon)\equiv c^{(1)}_k \epsilon^{1/2} K_\nu(\vert k\vert \epsilon)+c^{(2)}_k \epsilon^{1/2}I_\nu (\vert k\vert \epsilon)$. $\alpha_k(t_0)$ is a constant depending on initial stochastic time, 
which is given by $\alpha_k(t_0)\equiv I_\nu(\vert k\vert t_0)/K_\nu(\vert k\vert t_0)$. This matching condition transfers the holographic quantities to the stochastic quantities and vice versa. 

\paragraph{n-multiple trace deformation and stochastic n-point function}
Similarly, we have found the correspondence between n-multiple trace deformation and stochastic n-point function as
\begin{align}
&\left.\frac{\delta^nS_B}{\prod^n_{i=1}\delta\Phi(k_i,r)}\right\rvert^{r=t}=-\left\langle \prod^n_{i=1}\Phi(k_i,t)\right\rangle_\textrm{S}\prod^n_{j=1}\left\langle\Phi(k_j,t)\Phi(-k_j,t)\right\rangle^{-1}_\textrm{S}-\frac{1}{2}\frac{\delta^nS_E}{\prod^n_{i=1}\delta\Phi(k_i,t)}
\\ \nonumber &\left.n!\cdot \ddfrac{\prod^{n}_{i=1}\bar{\sigma}_n^{1/n}\frac{1}{2^{-\nu+1}}\Gamma(\nu)\vert k_i\vert^{-\nu}  }{\prod^n_{j=1} \epsilon^{1/2}\left(K_\nu(\vert k_i\vert  \epsilon) -c^{(2)}_{k_i}/c^{(1)}_{k_i} I_\nu(\vert k_i\vert t)\right)}\right\vert^{\epsilon=t} =-\frac{n!}{2}\tau_n \prod^n_{i=1}\left[\frac{(-\alpha_{k_i}(t^\prime_0)/\alpha_{k_i}(t_0))+1}{t^{1/2}\left(K_\nu(\vert k_i\vert t) -\tilde{\alpha}_{k_i}(t_0) I_\nu(\vert k_i\vert t)\right)}\right],
\end{align}
which implies
\begin{equation}
\label{coupling constant relation}
\tau_n=\prod^{n}_{i=1}\left[(-2(\bar\sigma_n)^{1/n}\left\{\frac{1}{2}\Gamma(\nu)\right\}\left(\frac{\vert k_i\vert}{2}\right)^{-\nu}\right],
\end{equation}
while $\alpha_{k_i}(t_0^\prime)=0$ and $-\tilde{\alpha}_{k_i}(t_0)=-1/\alpha_{k_i}(t_0)=c^{(2)}_{k_i}/c^{(1)}_{k_i}$. 
$\bar{\sigma}_n$ is the coupling constant of the boundary deformation imposing the boundary condition of the marginal multiple trace deformation $\mathcal{D}^{(n)}(\epsilon\rightarrow 0)\sim \bar{\sigma}_n$ on the holographic side. $\tau_n$ is the coupling constant in the $\phi^n$-interaction term of the Euclidean action on the stochastic side. 
 In \eqref{coupling constant relation}, 
one can notice that the stochastic coupling constant $\tau_n$ from the stochastic calculation is proportional to marginal n-multiple trace coupling $\bar{\sigma}_n$.

\paragraph{(2n-2)-multiple trace deformation and stochastic (2n-2)-point function}
Finally, we have found the correspondence between (2n-2)-multiple trace deformation and stochastic (2n-2)-point function as
\begin{align}
\left.\frac{\delta^{2n-2}S_B}{\prod^{2n-2}_{i=1}\delta\Phi(k_i,r)}\right\rvert^{r=t}&=-\left\langle \prod^{2n-2}_{i=1}\Phi(k_i,t)\right\rangle_\textrm{S}\prod^{2n-2}_{j=1}\left\langle\Phi(k_j,t)\Phi(-k_j,t)\right\rangle^{-1}_\textrm{S}
\\ \nonumber
&+\frac{(2n-2)!n^2}{2(n!)^2}\prod^{2n-2}_{i=1}\left\langle\Phi(k_i,t)\Phi(-k_i,t)\right\rangle^{-1}_\textrm{S}\times\textrm{Perm}\left[\left\langle\left\{\prod^{n-1}_{j=1}\Phi(k_j,t)\right\}\Phi(q,t)\right\rangle_\textrm{S}\right.
\\ \nonumber
&\left.\times\left\langle\Phi(q,t)\Phi(-q,t)\right\rangle^{-1}_\textrm{S}\left\langle\left\{\prod^{2n-2}_{l=n}\Phi(k_l,t)\right\}\Phi(-q,t)\right\rangle_\textrm{S}\right]
\\ \nonumber
&-\frac{1}{2}\frac{\delta^{2n-2}S_E}{\prod^{2n-2}_{i=1}\delta\Phi(k_i,t)}.
\end{align}
The results of the left and right-hand sides are given by
\begin{align}
\label{section2 2n-2 relation}
     \text{L.H.S.}
    &=(2n-2)!\prod^{2n-2}_{i=1}\left\{\mathcal{Q}'_\nu(k_i, \epsilon)\right\}^{-1}\left[\int^\epsilon_0 d\epsilon' \left[  \frac{\bar{\lambda}_{2n-2}}{(2n-2)}{\epsilon'}^{(n-2)(d-1)-2}\prod^{2n-2}_{i=1}\mathcal{Q}'_\nu(k_i, \epsilon')\right]- \bar{\sigma}_{2n-2}\right.
    \\ \nonumber &\left.-\frac{1}{2}n^2\bar{\sigma}_n^2\left\{\frac{1}{2}\Gamma(\nu)\right\}^{2n} \left\{\prod^{2n-2}_{i=1}\left(\frac{\vert k_i\vert}{2}\right)^{-\nu}\right\}  \text{Perm}\left\{\frac{\epsilon^{1/2}I_\nu(\vert k\vert \epsilon)\left(\frac{\vert  k_1+k_2+\dots+k_{n-1}\vert}{2}\right)^{-2\nu}}{ \mathcal{Q}'_\nu(\vert k\vert k_j, \epsilon)}\right\}\right],
\end{align}
\begin{align}
\text{R.H.S.}&= (2n-2)!\prod^{2n-2}_{j=1}\tau_{k_j}\tilde{\mathcal{V}}'_{2n-2}(k_1,\dots,k_{2n-2};t) \Bigg[\frac{\bar{\lambda}_{2n-2}}{2n-2}\int^t_0 dt^\prime\left[{t^\prime}^{(n-2)(d-1)-2}\prod^{2n-2}_{i=1}\tilde{\mathcal{Q}}'_\nu(k_i, t^\prime) \right]-\frac{\tau_{2n-2}}{2}
\\ \nonumber &-\frac{n^2}{8}\text{Perm}\left[ \tau_{k_1+k_2+\cdots+k_{n-1}}^2\ddfrac{t^{1/2}\alpha_{k}(t_0) I_\nu(\vert k\vert t) }{\tilde{\mathcal{Q}}'_\nu(k,t)}\right]\Bigg],
\end{align}

where we define
\begin{align}
\tilde{\mathcal{Q}}'_\nu(k,t)&\equiv t^{1/2}\left[K_\nu(\vert k\vert t)-\tilde{\alpha}_k(t_0)I_\nu(\vert k\vert t)\right] ,
\\ \nonumber \mathcal{Q}'_\nu(k,\epsilon)&\equiv \epsilon^{1/2}\left[K_\nu(\vert k\vert \epsilon)-c^{(2)}_{k}/c^{(1)}_{k}I_\nu(\vert k\vert \epsilon)\right],
\end{align}
and 
\begin{align}
\tilde{\mathcal{V}}'_{2n-2}(k_1,\dots,k_{2n-2};t)\equiv \prod^{2n-2}_{i=1}\left[\tilde{\mathcal{Q}}'_\nu(k_i,t)\right]^{-1}.
\end{align}
We note that $\bar{\lambda}_{2n-2}$ is the coupling constant of the $\Phi^{2n-2}$-interaction term in the holographic model. This coupling, $\bar{\lambda}_{2n-2}$ also appears in the  $\Phi^{2n-2}$-interaction term in Euclidean action, $S_E$ in the stochastic frame.
$\bar{\sigma}_{2n-2}$ is the coupling constant that appears in the $(2n-2)$-multiple trace deformation and $\tau_{2n-2}$ is another coupling constant that appears in the $\Phi^{2n-2}$-interaction term of the Euclidean action. More precisely, each of the $(2n-2)$-trace deformation term in the holographic model and the coefficient of the $(2n-2)$-interaction term in Euclidean action has the following form:
\begin{align}
 \mathcal D^{(2n-2)}(k,\epsilon)&=\int d^dk \left(\bar{\sigma}_{2n-2}F_1(k,\epsilon)+ \bar{\lambda}_{2n-2}F_2(k,\epsilon)+\bar{\sigma}_n^2 F_3(k,\epsilon)\right),
\\ \bar {\mathcal G}_{2n-2}(k,\epsilon)&=\int d^dk \left(\tau_{2n-2}G_1(k,\epsilon)+ \bar{\lambda}_{2n-2}G_2(k,\epsilon)+\tau_n^2 G_3(k,\epsilon)\right),
\end{align}
where $F_i$ are solutions of the holographic RG flow equations up to each coupling constant, $O(\bar{\sigma}_{2n-2})$, $O(\bar{\sigma}_{n}^2)$ and $O(\bar{\lambda}_{2n-2})$. $G_i$ are terms that constitute the kernel of the $\Phi^{2n-2}$-interaction term and that is
$\bar {\mathcal G}_{2n-2}$ 
in the Euclidean action evaluated up to each coupling constant $O(\bar{\tau}_{2n-2})$, $O(\bar{\tau}_{n}^2)$ and $O(\bar{\lambda}_{2n-2})$.

From the previous relations (\ref{constant relation}), we have found $\tilde{\mathcal{Q}}'_\nu( k,t)=\mathcal{Q}'_\nu( k, \epsilon)\vert^{\epsilon=t}$.
To match the left and right-hand sides, we set $\alpha(t'_0)=\alpha(t''_0)=0$. 
The integration constants $\tau$ in the stochastic frame should have the following relation with the marginal coupling constant $\bar{\sigma}_n$ in holographic data: 
\begin{align}
    \prod^{2n-2}_{i=1}\tau_{k_i}&=\prod^{2n-2}_{i=1}\left[\left(-2\bar{\sigma}_n\right)^{1/n}\left\{\frac{1}{2}\Gamma(\nu)\right\}\left(\frac{\vert k_i\vert}{2}\right)^{-\nu}\right]
    \\ \nonumber
    \tau_{k_1+k_2+\cdots+k_{n-1}}^2&=\tau_{-k}^2=\tau_k^2=\left[\left(-2\bar{\sigma}_n\right)^{1/n}\left\{\frac{1}{2}\Gamma(\nu)\right\}\left(\frac{\vert k_1+k_2+\cdots+k_{n-1}}{2}\right)^{-\nu}\right]^2
    \\ \nonumber
    \tau_{2n-2}&=2\bar{\sigma}_{2n-2}.
\end{align}
\section{Holographic Wilsonian Renormalization Group of self-interacting scalar field theory with marginal deformation}
\label{Holographic Wilsonian Renormalization Group}
In this section, we illustrate the example of the holographic Wilsonian renormalization group of the scalar field theory with $\phi^{2n-2}$-self-interaction and marginal boundary deformation. We mostly follow \cite{Kim:2023bhp} to explain the procedure. The bulk action in Euclidean AdS$_{d+1}$ (\ref{AdS metric}) is given by  
\begin{eqnarray}
\label{original-action}
    S_{bulk}&=& \int_{r>\epsilon} dr  d^dx  \ \sqrt{g}\mathcal{L}(\phi, \partial \phi) + {S}_B
\\ &=&\int_{r>\epsilon} dr d^{d}x\sqrt{g}\left[\frac{1}{2}g^{\mu\nu}\partial_{\mu}\phi\partial_{\nu}\phi+\frac{1}{2}m^2\phi^2+\frac{\lambda_{2n-2}}{2n-2}\phi^{2n-2}\right]+{S}_B,
\end{eqnarray}
where $\epsilon$ is an arbitrary cut-off in the radial direction. $S_B$ is a boundary effective action at $r=\epsilon$ obtained by integrating out the degrees of freedom coming from $r < \epsilon$.
We consider a new field $\Phi$, to define the theory in effective flat space as,
\begin{equation}
    \phi=r^{\frac{d-1}{2} } \Phi.
\end{equation}
We also solve the problem in the boundary directional momentum space of the re-defined field by using the Fourier transform given by
\begin{equation}
  \Phi(x,r)=\frac{1}{(2\pi)^{\frac{d}{2}}}\int e^{-ik_ix_i}\Phi(\vec{k}, r)d^dk,
\end{equation}
where $k_ix_i=\sum_{i,j=1}^d k_ix_j\delta_{ij}$.
Then the bulk action becomes
\begin{align}
\label{re-bulk-action}
S_{\text{bulk} }&=\frac{1}{2}\int_{r>\epsilon} dr \ d^dk \left[ \partial_r \Phi(k,r) \partial_r\Phi(-k,r) +k^2\Phi(k,r)\Phi(-k,r)+\frac{1}{r^2}\left(m^2+\frac{d^2}{4}-\frac{1}{4}\right)\Phi(k,r)\Phi(-k,r)\right]
\\ \nonumber &+\int dr \frac{\lambda_{2n-2}}{2n-2} r^{(n-2)(d-1)-2}\frac{1}{(2\pi)^{d(n-2)}}\int \left(\prod^{2n-2}_{i=1} d^dk_i \Phi(k_i,r)\right)\delta^{(d)}\left(\sum^n_{j=1}\vec{k_j}\right)+S_B(\epsilon),
\end{align}
where the boundary action is also re-defined as 
\begin{equation}
S_B(\epsilon) \equiv {S'}_B(\epsilon) -\frac{d-1}{4\epsilon} \ \Phi^2(r=\epsilon).
\end{equation}

\subsection{Holographic Wilsonian renormalization group flow of multi-trace deformation}
We demand that the holographic model action does not depend on $\epsilon$, since the cut-off $\epsilon$ is an arbitrary choice. In the momentum space, this condition gives rise to the flow equation of the boundary action $S_B$:
\begin{align}
\label{Hamilton-Jacobi equation-SB'}
\partial_\epsilon {S}_B&=-\frac{1}{2}\int_{r=\epsilon}d^dk_i \left[\left(\frac{\delta {S'}_B}{\delta \Phi(k_i,r)}\right)\left(\frac{\delta {S'}_B}{\delta \Phi(-k_i,r)}\right)-\left(\vert k \vert^2+\frac{1}{r^2}\left(\nu^2-\frac{1}{4}\right)\Phi(k_i,r) \Phi(-k_i,r)\right)\right]
\\ \nonumber&+ \frac{\lambda_{2n-2}}{2n-2}r^{(n-2)(d-1)-2}\frac{1}{(2\pi)^{d(n-2)}}\int \left[\prod^{2n-2}_{i=1}d^dk_i\Phi(k_i,r)\right]\delta^{(d)}\left(\sum^{2n-2}_{j=1}\vec{k_j}\right).
\end{align}
where
\begin{equation}
\nu\equiv\sqrt{m^2+\frac{d^2}{4}}.
\end{equation}
Let us solve the equation by assuming that the trial solution is designed to be a weak field expansion in $\Phi$. The ansatz of the boundary action is given by
\begin{eqnarray}
\label{SB'-ansatz}
{S}_B&=&\Lambda_\Phi(\epsilon)+\int d^dk \mathcal{J}(k,\epsilon)\Phi(k,\epsilon)+\int d^dk \mathcal{D}^{(2)}(k,\epsilon)\Phi(k,\epsilon)\Phi(-k,\epsilon)
\\ \nonumber &+& \sum^\infty_{m=3}\int\left[ \prod^m_{i=1}d^dk_i\Phi({k_i},\epsilon)\right]\mathcal{D}^{(m)}_{k_1\dots k_m}(\epsilon)\delta^{(d)}\left(\sum^m_{j=1}k_j\right),
\end{eqnarray}
where $\Lambda(\epsilon)$ is a boundary cosmological constant. $J(\epsilon,k)$ and $D^{(n)}(\epsilon,k)$ are couplings related to single and multi-trace operators which are unknown functions of the radial cut-off $\epsilon$. We solve the unknown functions $\Lambda(\epsilon), J(\epsilon,k)$ and $D^{(n)}(\epsilon,k)$, by substituting the ansatz(\ref{SB'-ansatz}) into the Hamiltonian-Jacobi equation(\ref{Hamilton-Jacobi equation-SB'}). Comparison of the coefficients of multiple products of field $\Phi$ gives a set of equations. 

As discussed in \cite{Aharony:2015afa, Kim:2023bhp}, we focus on a case that $\mathcal D^{(2)}$ and $\mathcal D^{(n)}$ are turned on in the boundary action. Once we request 
$\delta S^{on-shell}_{\rm bulk} =0$
as $\epsilon \rightarrow 0$, we obtain an interesting boundary condition as 
\begin{align}
    \left(-2\nu A^{(2)}_k+n\bar{\sigma}_n \prod^{n-1}_{i=1} A^{(1)}_{k_i}\right)\delta A^{(1)}_k=0,
\end{align}
where $S^{on-shell}_{\rm bulk}$ is the on-shell action of the bulk action \eqref{re-bulk-action} and the solution of the equation of motion is given by
\begin{align}
    \Phi(k,r \rightarrow0)=A^{(1)}_k r^{\frac{1}{2}-\nu}+A^{(2)}_k r^{\frac{1}{2}-\nu}.
\end{align}
We note that such a boundary condition is obtained when we put a boundary action, $S_B$ in position space as
\begin{equation}
S_B=S_{\rm c.t}+S_{\rm def}=\frac{1}{2}\sqrt{\gamma(\epsilon)}\left(\frac{d}{2}-\nu\right)\phi^2(\epsilon)+\bar \sigma_n\sqrt{\gamma(\epsilon)}\phi^n(\epsilon),
\end{equation}
where $\gamma$ is the determinant of the induced metric, $\gamma_{ij}=\delta_{ij}/\epsilon^2$ and $\bar\sigma_n$ is the $n$-multiple trace coupling. We also impose a condition that $\nu=d\left(\frac{1}{2}-\frac{1}{n}\right)$, and it forces the boundary term become marginal.
Now, one can choose the Dirichlet boundary condition as
\begin{align}
  \delta A^{(1)}_k=0,
\end{align}
or another boundary condition as 
\begin{align}
     A^{(2)}_k=\frac{1}{2\nu}n\bar{\sigma}_n \prod^{n-1}_{i=1} A^{(1)}_{k_i},
\end{align}
which gives the marginal $n$-multiple trace deformation to the boundary field theory. In this paper, we will focus on the second boundary condition.

\paragraph{Single and Double trace deformation}
 We start with $m=1,2$ case and when $m\geq 2$. The equations are given by
 \begin{equation}
     \partial_\epsilon \mathcal{J}(k,\epsilon)=-2\mathcal{J}(k,\epsilon) \mathcal{D}^{(2)}(k,\epsilon)
 \end{equation}
\begin{equation}
    \partial_\epsilon \mathcal{D}^{(2)}(k,\epsilon)=-\frac{1}{2}\left[ 4\mathcal{D}^{(2)}(k,\epsilon)\mathcal{D}^{(2)}(-k,\epsilon)-\left(\vert k \vert^2+\frac{1}{\epsilon^2}\left(\nu^2-\frac{1}{4}\right)\right)\right],
\end{equation}
The most general form of the solution is given by
\begin{equation}
    \mathcal{J}(k,\epsilon)=-\frac{c^{(0)}(k)}{\mathcal{Q}_\nu(k, \epsilon)}
\end{equation}
\begin{equation}
    \mathcal{D}^{(2)}(k,\epsilon)=\frac{1}{2}\partial_\epsilon \log\left[\mathcal{Q}_\nu(k, \epsilon)\right],
\end{equation}
where
\begin{equation}
\label{bessel-bulk-solution}
   \mathcal{Q}_\nu(k, \epsilon)\equiv c^{(1)}_k \epsilon^{1/2} K_\nu(\vert k\vert \epsilon)+c^{(2)}_k \epsilon^{1/2}I_\nu (\vert k\vert \epsilon),
\end{equation}
where $c^{(0)}_k$, $c^{(1)}_k$ and $c^{(2)}_k$ are arbitrary constant with momentum label. We note that in the following, we concentrate on a case with $\mathcal J=0$. The reason is that once one considers connected-tree level correlation functions in the stochastic frame, they do not include zero- and one-point functions. One may consider another case with non-zero vev of the scalar field, $\langle\phi\rangle$, but we will save it for later discussion.

We note that $K_\nu$ and $I_\nu$ are the modified Bessel function. 
\paragraph{n-multiple trace deformation}
The equation for $\mathcal{D}^{(n)}(k,\epsilon)$ is given by
 \begin{eqnarray}
 \label{D-n-equation}
 \partial_\epsilon \mathcal{D}^{(n)}_{(k_1,\dots,k_n)}(\epsilon)\delta^{(d)}\left(\sum^n_{j=1}k_j\right)
 &=& -\frac{1}{2}\Bigg[4\left(\sum^n_{i=1}\mathcal{D}^{(2)}(-k_i,\epsilon)\right)\mathcal{D}^{(n)}_{(k_1,\dots,k_{n-1},k_n)}(\epsilon)\delta^{(d)}\left(\sum^n_{s=1}k_s\right)
 \\ \nonumber &+& (1-\delta_{3,n})\sum^{n-1}_{l=3}l(n-l+2)\text{Perm}\left\{\mathcal{D}^{(l)}_{(k_1,\dots,k_{l-1},k) }(\epsilon) \mathcal{D}^{(n-l+2)}_{(k_l,\dots,k_n,-k)}(\epsilon)\right\}
 \\ \nonumber &\times& \delta\left(\sum^m_{s=1}k_s\right)\Bigg].
 \end{eqnarray}
To simplify our situation, 
we restrict ourselves in the case that $\mathcal{D}^{(m)}=0$ for $2<m<n$, 
where the second term in the equation(\ref{D-n-equation}) vanishes, only leaving the contribution of double-trace deformation to $n$-multiple trace deformation. 
This choice is also consistent with a fact that we deform our AdS boundary with a marginal $n$-multiple trace operator only. The deformation becomes marginal once the condition, $\nu=d\left(\frac{1}{2}-\frac{1}{n}\right)$ is satisfied.
As a consequence, the multi-trace deformation is given by 
\begin{equation}
\label{sigma definition}
\mathcal{D}^{(n)}_{(k_1,...,k_n)}(\epsilon)=\ddfrac{\ C_n(k_1,...,k_n)}{\prod^n_{i=1}\mathcal{Q}_\nu(k_i, \epsilon)},
\end{equation}
where $C_n(k_1,...,k_n)$ is an arbitrary boundary momentum, $k_i$ dependent function and specified by considering the boundary condition on the conformal boundary. We impose boundary deformation which gives marginal n-multiple trace deformation at the boundary when $\nu=d\left(\frac{1}{2}-\frac{1}{n}\right)$: 
\begin{equation}
    \mathcal{D}^{(n)}_{(k_1,...,k_n)}(\epsilon\rightarrow 0)  \equiv\sqrt{\gamma}\epsilon^{n\left(\frac{d-1}{2}\right)}\bar{\sigma}_n(k_1,...,k_n)=\epsilon^{-d+n\left(\frac{d-1}{2}\right)}\bar{\sigma}_n(k_1,...,k_n),
\end{equation}
where $\bar{\sigma}_n$ is the marginal multi-trace coupling constant.
The asymptotics of the second kind of Bessel functions, $K_\nu$ and $I_\nu$ are given as
\begin{equation}
K_\nu(z\rightarrow 0) = \frac{1}{2}\Gamma(\nu)\left(\frac{1}{2}z\right)^{-\nu},{\rm\ and \ \ } I_\nu(z\rightarrow 0) = \frac{1}{\Gamma(\nu+1)}\left(\frac{1}{2}z\right)^{\nu}.
\end{equation}
This boundary condition on the conformal boundary fixes the function $C_n$ as
\begin{equation}
    C_n(k_1,...,k_n)=\bar{\sigma}_n\left\{\frac{1}{2}\Gamma(\nu)\right\}^n \left\{\prod^n_{i=1}c^{(1)}_{k_i}\left(\frac{\vert k_i\vert}{2}\right)^{-\nu}\right\}.
\end{equation}
Then, the final form of the solution of $\mathcal{D}^{(n)}$ is given by
\begin{equation}
\label{D-n-deformation}
    \mathcal{D}^{(n)}_{(k_1,...,k_n)}(\epsilon)= \ddfrac{\bar{\sigma}_n\left\{\frac{1}{2}\Gamma(\nu)\right\}^n \left\{\prod^n_{i=1}c^{(1)}_{k_i}\left(\frac{\vert k_i\vert }{2}\right)^{-\nu}\right\} }{\prod^n_{j=1}\mathcal{Q}_\nu(k_j, \epsilon)}= \ddfrac{\prod^{n}_{i=1}\bar{\sigma}_n^{1/n}\frac{1}{2^{-\nu+1}}\Gamma(\nu)\vert k_i\vert^{-\nu}  }{\prod^n_{j=1}\mathcal{Q}'_\nu(k_j, \epsilon)},
\end{equation}
where 
\begin{align}
\label{Q-prime-definition}
\mathcal{Q}'_\nu(k_j, \epsilon)\equiv\epsilon^{1/2}K_\nu(\vert k_j\vert \epsilon)+\frac{c^{(2)}_{k_i}}{c^{(1)}_{k_i}}\epsilon^{1/2}I_\nu(\vert k_j\vert \epsilon).
\end{align}
\paragraph{(2n-2)-multiple trace deformation: $m=2n-2$ case}
The equation for $\mathcal{D}^{(2n-2)}(k,\epsilon)$ is given by
 \begin{eqnarray}
 \label{D-2n-2-equation}
 \partial_\epsilon \mathcal{D}^{(2n-2)}_{(k_1,\dots,k_n)}(\epsilon)\delta^{(d)}\left(\sum^{2n-2}_{j=1}k_j\right)
 &=& -\frac{1}{2}\Bigg[4\left(\sum^{2n-2}_{i=1}\mathcal{D}^{(2)}(-k_i,\epsilon)\right)\mathcal{D}^{(2n-2)}_{(k_1,\dots,k_{2n-3},k_{2n-2})}(\epsilon)\delta^{(d)}\left(\sum^n_{s=1}k_s\right)
 \\ \nonumber &+& (1-\delta_{3,2n-2})\sum^{2n-3}_{l=3}l(2n-l)\text{Perm}\left\{\mathcal{D}^{(l)}_{(k_1,\dots,k_{l-1},k) }(\epsilon) \mathcal{D}^{(2n-l)}_{(k_l,\dots,k_{2n-2},-k)}(\epsilon)\right\}
 \\ \nonumber &\times& \delta\left(\sum^m_{s=1}k_s\right)\Bigg]
 \end{eqnarray}
 Since we assume $D^{(l)}=0$ for $l \leq n-1$,
\begin{eqnarray}
\partial_\epsilon \mathcal{D}^{(2n-2)}_{(k_1,...,k_{2n-2})}(\epsilon)&=&-\frac{1}{2}\Big[4\sum^{2n-2}_{i=1} \mathcal{D}^{(2)}(k_i,\epsilon)\mathcal{D}^{(2n-2)}_{(k_1,...,k_{2n-2})}(\epsilon)
\\ \nonumber &+& n^2\text{Perm}\left\{\mathcal{D}^{(n)}_{(k_1,...,k_{n-1},k)}(\epsilon)\mathcal{D}^{(n)}_{(k_n,...,k_{2n-2},-k)}(\epsilon)\right\}\Big]
\\ \nonumber &+& \frac{\lambda_{2n-2}}{(2n-2)(2\pi)^{d(n-2)}}\epsilon^{(n-2)(d-1)-2},
\end{eqnarray}
where $\text{Perm}\{A_{k_1,...,k_m}\}\equiv \frac{1}{m!}(A_{k_1,...,k_m}+{\rm all\ possible\ momentum\ permutations})$.
First, we consider the homogeneous solution which does not include the effect of coupling $\lambda_{2n-2}$:
\begin{equation}
     \mathcal{D}^{(2n-2)}_{(k_1,...,k_{2n-2})}(\epsilon)=\frac{ C_{2n-2}(k_1,...,k_{2n-2})}{\prod^{2n-2}_{i=1}\mathcal{Q}_\nu(k_i, \epsilon)}
\end{equation}
Then, along with the coupling $\lambda_{2n-2}$, the inhomogeneous solution has the following form as
\begin{equation}
    \mathcal{D}^{(2n-2)}_{(k_1,...,k_{2n-2})}(\epsilon) \equiv G(\epsilon)\prod^{2n-2}_{i=1}\left\{\mathcal{Q}_\nu(k_i, \epsilon)\right\}^{-1},
\end{equation}
which allows $G(\epsilon)$ to satisfy the following equation:
\begin{eqnarray}
\partial_\epsilon G(\epsilon) \prod^{2n-2}_{i=1}\left\{\mathcal{Q}_\nu(k_i, \epsilon)\right\}^{-1}&=& \frac{\lambda_{2n-2}}{(2n-2)(2\pi)^{d(n-2)}}\epsilon^{(n-2)(d-1)-2}
\\ \nonumber &-&\frac{1}{2}n^2\bar{\sigma}_n^2\frac{\left\{\frac{1}{2}\Gamma(\nu)\right\}^{2n-2} \left\{\prod^{2n-2}_{i=1}\frac{c^{(1)}_{k_i}}{2}\vert k_i\vert \right\}}{\prod^{2n-2}_{j=1}\mathcal{Q}_\nu(k_j, \epsilon)}
\\ \nonumber &\times& \text{Perm}\left\{\frac{\left(\left\{\frac{1}{2}\Gamma(\nu)\right\} \frac{c^{(1)}_{k_1+k_2+\dots+k_{n-1}}}{2}\vert  k_1+k_2+\dots+k_{n-1}\vert\right)^2}{\left\{ 
\mathcal{Q}_\nu(\sum^{n-1}_{j=1} k_j, \epsilon)\right\}^2}\right\}.
\end{eqnarray}
The solution of $G(\epsilon)$ is given by
\begin{align}
    G(\epsilon)&=\int d\epsilon' \left[ \frac{\lambda_{2n-2}}{(2n-2)(2\pi)^{d(n-2)}}{\epsilon'}^{(n-2)(d-1)-2}\prod^{2n-2}_{i=1}\mathcal{Q}_\nu(k_i, \epsilon')\right.
    \\ \nonumber &-\frac{1}{2}n^2\bar{\sigma}_n^2\left\{\frac{1}{2}\Gamma(\nu)\right\}^{2n-2} \left\{\prod^{2n-2}_{i=1}\frac{c^{(1)}_{k_i}}{2}\vert k_i\vert \right\} \left.\int^\epsilon_0 d\epsilon' \ 
     \text{Perm}\left\{\frac{\left(\left\{\frac{1}{2}\Gamma(\nu)\right\} \frac{c^{(1)}_{k_1+k_2+\dots+k_{n-1}}}{2}\vert  k_1+k_2+\dots+k_{n-1}\vert\right)^2}{\left\{\mathcal{Q}_\nu(\sum^{n-1}_{j=1} k_j, \epsilon')\right\}^2}\right\}\right].
\end{align}
Thus, ($2n-2$)-multiple trace deformation is given by
\begin{align}
\label{holographic-2n-2-trace-deformation}
     \mathcal{D}^{(2n-2)}_{(k_1,...,k_{2n-2})}(\epsilon)
    &=\prod^{2n-2}_{i=1}\left\{\mathcal{Q}'_\nu(k_i, \epsilon)\right\}^{-1}\left[\int^\epsilon_0 d\epsilon' \left[  \frac{\bar{\lambda}_{2n-2}}{(2n-2)}{\epsilon'}^{(n-2)(d-1)-2}\prod^{2n-2}_{i=1}\mathcal{Q}'_\nu(k_i, \epsilon')\right]- \bar{\sigma}_{2n-2}\right.
    \\ \nonumber &\left.-\frac{1}{2}n^2\bar{\sigma}_n^2\left\{\frac{1}{2}\Gamma(\nu)\right\}^{2n} \left\{\prod^{2n-2}_{i=1}\left(\frac{\vert k_i\vert}{2}\right)^{-\nu}\right\}  \text{Perm}\left\{\frac{\epsilon^{1/2}I_\nu(\sum^{n-1}_{j=1} k_j,\epsilon)\left(\frac{\vert  k_1+k_2+\dots+k_{n-1}\vert}{2}\right)^{-2\nu}}{ \mathcal{Q}'_\nu(\sum^{n-1}_{j=1} k_j, \epsilon)}\right\}\right],
\end{align}
where $\bar{\lambda}_{2n-2}=\frac{\lambda_{2n-2}}{(2\pi)^{d(n-2)}}$ and $\bar{\sigma}_{2n-2}=C_{2n-2} \left\{\frac{1}{2}\Gamma(\nu)\right\}^{2n-2} \left\{\prod^{2n-2}_{i=1}c^{(1)}_{k_i}\left(\frac{\vert k_i\vert}{2}\right)^{-\nu}\right\}$.

According to these results, we construct the holographic boundary action up to first-order in $\lambda_{2n-2}$,$\bar{\sigma}_{2n-2}$, and second-order in $\bar{\sigma}_n$ as
\begin{align}
\label{holographic-boundary-action}
\nonumber
    S_B=&\int d^dk \ \frac{1}{2}\partial_\epsilon \log\left[\mathcal{Q}_\nu(k, \epsilon)\right]\Phi(k, \epsilon)\Phi(-k, \epsilon)+ \int\left[ \prod^n_{i=1}d^dk_i\Phi({k_i},\epsilon)\right]\frac{\prod^{n}_{i=1}\left[(-2\bar{\sigma})^{1/n}\left\{\frac{1}{2}\Gamma(\nu)\right\}\left(\frac{\vert k_i\vert}{2}\right)^{-\nu}\right]  }{\prod^n_{j=1}\mathcal{Q}'_\nu(k_j, \epsilon)}
\\ \nonumber \times&\delta^{(d)}\left(\sum^n_{j=1}k_j\right)+ \int\left[ \prod^{2n-2}_{i=1}d^dk_i\Phi({k_i},\epsilon)\right]\prod^{2n-2}_{i=1}\left\{\mathcal{Q}'_\nu(k_i, \epsilon)\right\}^{-1}\left[\int^\epsilon_0 d\epsilon' \left\{  \frac{\bar{\lambda}_{2n-2}}{(2n-2)}{\epsilon'}^{(n-2)(d-1)-2}\prod^{2n-2}_{i=1}\mathcal{Q}'_\nu(k_i, \epsilon')\right\}\right.
    \\ \nonumber &\left.- \bar{\sigma}_{2n-2}-\frac{1}{2}n^2\bar{\sigma}_n^2\left\{\frac{1}{2}\Gamma(\nu)\right\}^{2n-2} \left\{\prod^{2n-2}_{i=1}\left(\frac{\vert k_i\vert}{2}\right)^{-\nu}\right\} \text{Perm}\left\{\frac{\epsilon^{1/2}I_\nu(\sum^{n-1}_{j=1} k_j,\epsilon)\left(\frac{\vert  k_1+k_2+\dots+k_{n-1}\vert}{2}\right)^{-2\nu}}{ \mathcal{Q}'_\nu(\sum^{n-1}_{j=1} k_j, \epsilon)}\right\}\right] 
    \\  &\times \delta^{(d)}\left(\sum^{2n-2}_{j=1}k_j\right).
\end{align}
\paragraph{The relation between the boundary action, $S_B$ and the Euclidean action, $S_E$} Keeping in mind that the relation between $S_E$ and $S_B$ suggested in \cite{Kim:2023bhp} is $S_E=-2S_B$, one can construct the Euclidean action $S_E$ by using the boundary action (\ref{holographic-boundary-action}). To perform this identification, we set $c^{(2)}_{k_i}=0$ in $\mathcal{Q}_\nu$ and $\mathcal{Q}'_\nu$ which are given in (\ref{bessel-bulk-solution}) and (\ref{Q-prime-definition}) respectively. 
The argument for this is, to construct on-shell action, we manipulate a regular solution in the bulk region but the second kind modified Bessel function, $I_\nu(\vert k\vert r)$, is divergent near $r=\infty$. Thus, we drop this term and keep $K_\nu(\vert k\vert r)$ solution only. Then the relation $S_E=-2S_B$ is consistent with identification $S_E=-2I_{os}$ in \cite{Oh:2012bx}, where $I_{os}$ is on-shell action since we use regular solution for constructing on-shell action for the holographic model. 
\section{Stochastic Quantization Frame}
\label{Stochastic Quantization}
In this section, we formulate HWRG by the Langevin dynamics of stochastic quantization. We consider the following form of the Euclidean action:
\begin{equation}
\label{Euclidean-action-position-space}
S_E=\int d^dx \sum_{n=2}^\infty \mathcal{G}_n(\nabla^2,t)\Phi^n(x),
\end{equation}
which contains stochastic time dependent kernel $g_n(\nabla^2,t)$ with Laplacian in $d-$dimensional Euclidean space $\nabla^2\equiv \delta^{\mu\nu} \frac{\partial^2}{\partial^\mu \partial^\nu}$. Now, we express the Euclidean action $S_E$ in the momentum space which is given by
\begin{equation}
\label{Euclidean-action-momentum-s}
S_E=\sum_{n=2}^\infty \int \frac{1}{(2\pi)^{n\frac{d}{2}-1}}\left(\prod_{i=1}^n d^d k_i \Phi(k_i,t)\right) \mathcal{G}_n(k_1, ... ,k_n ; t) \delta^{(d)}\left(\sum_{j=1}^n k_j\right).
\end{equation}
By employing the Langevin equation in momentum space:
\begin{equation}
\label{langevin-equation-momentum}
\frac{\partial \Phi(k,t)}{\partial t}=-\frac{1}{2} \frac{\delta S_E}{\delta \Phi(-k,t)}+\eta(k,t),
\end{equation}
we obtain the following equation as
\begin{equation}
\frac{\partial \Phi(k,t)}{\partial t}=-\frac{1}{2} \sum_{n=2}^\infty n\int \left(\prod_{i=1}^{n-1}d^d p_i \Phi(p_i,t)\right)\bar{\mathcal{G}}_n (p_1,...,p_{n-1},-k,t) \delta^{(d)} \left(\sum_{j=1}^{n-1} p_j-k\right) + \eta(k,t),
\end{equation}
where we newly defined the stochastic kernels $\bar {\mathcal{G}}_n$ as
\begin{equation}
\bar{\mathcal{G}}_n(p_i,t) \equiv \frac{1}{(2\pi)^{\frac{nd}{2}-1}} \mathcal{G}_n(p_i,t).
\end{equation}
The stochastic partition function in momentum space is given by
\begin{equation}
\label{partition-function-momentum}
\mathcal Z= \int[D\eta(k,t)]\exp \left[ -\frac{1}{2}\int^t_{t_0} dt d^d k{\ }\eta(k,t)\eta(-k,t) \right],
\end{equation}
where $t_0$ is the initial stochastic time. Once we replace the noise field $\eta$ by $\Phi$ by substitution of the Langevin equation (\ref{langevin-equation-momentum}) into this stochastic partition function (\ref{partition-function-momentum}), we obtain the partition function with the Fokker-Plank formalism:
\begin{equation}
\mathcal Z= \int[\mathcal D\Phi(k,t)]\exp\{-S_{\rm FP}(\Phi,t) \},
\end{equation}
where $S_{\rm FP}$ is called Fokker-Planck action.

\subsection{Evaluation of the stochastic kernel and construction of the Euclidean action}
\label{Evaluation of the Stochastic Kernel and Construction of the Euclidean Action}
We start with the Euclidean action given as (\ref{langevin-equation-momentum}). From the Langevin equation, we change the variable of the stochastic partition function with field $\eta \rightarrow \Phi$. Then we get the Fokker-Planck action:
\begin{equation}
S_{FP} \equiv \int dt {\ }\mathcal{L}_{FP}(\Phi,t),
\end{equation}
where the Fokker-Plank Lagrangian density is given by
\begin{eqnarray}
\label{Fokker-Planck-scalar-action1}
\nonumber
\mathcal{L}_{\rm FP}&=&-\frac{1}{2} \left[\int d^d k\ \left(\frac{\partial \Phi(k,t)}{\partial t}\right)\left(\frac{\partial \Phi(-k,t)}{\partial t}\right)\right.+\sum_{n=2}^\infty \left(\int \prod_{l=1}^{N} d^d k_l \Phi(k_l,t)\right)\delta^{(d)}\left(\sum_{s=1}^N k_s\right) 
\\ \nonumber
&\times&\left\{\frac{1}{4}\sum_{n=2}^N n(N+2-n) \bar{\mathcal{G}}_n\left(k_1,...,k_n,-\sum_{j=1}^{n-1}k_j ; t\right)
\bar{\mathcal{G}}_{N+2-n}\left(k_n,...k_N,-\sum_{l=n}^N k_l ;t\right) -\frac{\partial \bar{\mathcal{G}}_N (k_1,...,k_n;t)}{\partial t} 
  \right\} 
\\ 
&+&\left.\sum_{m=2}^\infty \frac{\partial}{\partial t} \left\{ \int \left(\prod_{l=1}^m d^d k_l \Phi(k_l;t)\right) \bar{ \mathcal{G}}_m(k_1,...,k_m;t)\delta^{(d)} \left(\sum_{s=1}^m k_s\right)\right\}\right].
\end{eqnarray}
By identifying $t=r$, one can compare the Fokker-Plank action to the bulk action (\ref{re-bulk-action}) and we get the series of equations of the stochastic kernel which constitutes the Euclidean action:

\begin{equation}
\label{G2-equation-g2}
    \bar{\mathcal{G}}_2(k,-k,t)^2-\frac{\partial\bar{\mathcal{G}_2}(k,-k,t)}{\partial t}=\left( \frac{m^2+\frac{d^2-1}{4}}{t^2}+p^2 \right)
\end{equation}
\begin{equation}
\label{gn-equation-gn}
    \frac{1}{2}\bar{\mathcal{G}}_n(k_1,...,k_n,t)\left(\sum^n_{j=1}\bar{\mathcal{G}}_2(k_j,-k_j,t)\right)-\frac{1}{2}\frac{\partial}{\partial t}\bar{\mathcal{G}}_n(k_1,..,k_n,t)=0
\end{equation}
\begin{align}
\label{g2n-2-equation}
    &\frac{1}{2}\bar{\mathcal{G}}_{2n-2}(k_1,...,k_{2n-2},t)\left(\sum_{l=1}^{2n-2}\bar{\mathcal{G}}_2(k_l, -k_l)\right)-\frac{1}{2}\frac{\partial \bar{\mathcal{G}}_{2n-2}(k_1,...,k_{2n-2},t) }{\partial t}  
    \\ \nonumber &+ \frac{n^2}{8} \text{Perm}\left\{\bar{\mathcal{G}}_n(k_1,..,k_{n-1}, -\sum^{n-1}_{j=1}k_j,t)\bar{\mathcal{G}}_n(k_n,..,k_{2n-2},-\sum^{2n-2}_{i=n}k_i,t)\right\}=\frac{\bar{\lambda}_{2n-2}}{2n-2}t^{\left(n-2\right)\left(d-1\right)-2}.
\end{align}
The solutions of the equation(\ref{G2-equation-g2}) and (\ref{gn-equation-gn}) are given by
\begin{eqnarray}
\label{tau definition}
    \bar{\mathcal{G}}_2(k,-k, t)=-\partial_t \log{\left[  t^{\frac{1}{2}} K_{\nu}(\lvert k \rvert t)+c_0 t^{\frac{1}{2}} I_{\nu}(\lvert k \rvert t) \right]}. 
    \\
        \bar{\mathcal{G}}_n(k_1,...,k_n,t) = \tau_n t^{-\frac{n}{2}} \prod^n_{j=1} \left\{   K_{\nu}(\lvert k_j \rvert t)+c_0 I_{\nu}(\lvert k_j \rvert t) \right\}^{-1},
\end{eqnarray}
where $\tau_n$ is an integral constant. The Euclidean action can be derived from the on-shell action by the relation $S_E \equiv -2S_B$, where we need to choose the regular solution which should not be divergent in AdS interior. This condition vanishes the irregular part of the solution, $c_0=0$.

The homogeneous solution of the equation (\ref{g2n-2-equation}) is given by
\begin{equation}
    \bar{\mathcal{G}}_{2n-2}(k_1,...,k_{2n-2},t) =  \ t^{-n+1}\prod^{2n-2}_{j=1}  \left\{  K_{\nu}(\lvert k_j \rvert t)\right\}^{-1}.
\end{equation}

By employing this solution, one can get the solution in $\mathcal{O}(\bar \lambda^1_{2n-2})$. We apply the following type of trial solution:

\begin{equation}
    \bar{\mathcal{G}}_{2n-2}(k_1,...,k_{2n-2},t)=\left[ \prod^{2n-2}_{j=1} \left(t^{\frac{1}{2}} K_{\nu}(\lvert k_j \rvert t) \right)\right]^{-1}R_{2n-2} ( \bar{\lambda}_{2n-2}, t),
\end{equation}
and the solution is given by
\begin{eqnarray}
    R_{2n-2} ( \bar{\lambda}_{2n-2}, t)&=&\int^t_{t_0} dt^\prime \  \Bigg[ \frac{n^2}{4}{t^\prime}^{-1}\prod^{2n-2}_{j=1}\tau_{k_j} \text{Perm}\left\{  \tau_{k_1+...+k_{n-1}}^2\left( K_\nu\left(\Bigg\vert-\sum^{n-1}_{j=1}k_j \Bigg\vert {t^\prime} \right)\right)^{-2}\right\} 
    \\ \nonumber &-& \frac{\bar{\lambda}_{2n-2}}{n-1}{t^\prime}^{(n-2)d-1} \prod^{2n-2}_{j=1}K_\nu(\vert k_j\vert {t^\prime})\Bigg].
\end{eqnarray}
By assuming that $\tau_n(k)=\tau_n(-k)$\footnote{
One way to get this relation, $\tau_n(k)=\tau_n(-k)$ is that we request the coupling constant be a $O(d)$ invariant. Parity, $k_i \rightarrow -k_i$ is a subgroup of $O(d)$-rotation, then we get the relation. Another way to look at this is as follows.
We assume that the scalar field $\phi(x)$ is a real function of the Euclidean coordinates $x_i$ in position space. This restricts the field in the momentum space as $\phi^*_k=\phi_{-k}$ once we consider its Fourier transform, where $^*$ denotes complex conjugate.
Also, we request that the interaction term in position space $L_n=\int \tau_n(t)\phi^n(x) d^dx$ should transform into $L_n=\int d^dk \frac{\tau_n(k_1,...,k_n)}{f(|k_i|t)}\phi_{k_1}\cdots \phi_{k_n} \delta^{(d)}(\sum^{n}_{i=1} k_n)$ in momentum space,where $f$ is a function of $O(d)$ invariants and stochastic time, $t$.
We note that the coupling $\tau_n(t)$ in position space is not just a function of $t$ but it could be a function of differential operators. This ensures that the coupling becomes a function of momenta in momentum space.
From the fact $L^*_n=-L_n$, we get $\tau_n^*(k_i)=\tau_n(-k_i)$, because Euclidean time satisfies $x_d^*=-x_d$, which is obtained by Wick rotation of Lorentzian time. If we assume that the momentum space coupling constant $\tau_n(k_i)$ is also real, we obtain $\tau_n(k_i)=\tau_n(-k_i)$.  }, the permutations in the curly bracket are given by
\begin{eqnarray}
    \text{Perm}\left\{\bar{\mathcal{G}}_n(k_1,..,k_{n-1}, -\sum^{n-1}_{j=1}k_j,t)\bar{\mathcal{G}}_n(k_n,..,k_{2n-2},-\sum^{2n-2}_{i=n}k_i,t)\right\}= \prod^{2n-2}_{j=1}\tau_{k_j}\left[ t^{\frac{1}{2}} K_{\nu}(\lvert k_j \rvert t)\right]^{-1} 
    \\ \nonumber \times \text{Perm}\left\{  \tau_{k_1+...+k_{n-1}}\tau_{k_n+...+k_{2n-2}}\frac{1}{t}\left( K_\nu\left(\Bigg\lvert -\sum^{n-1}_{j=1}k_j\Bigg\rvert t\right)\right)^{-1}\left(K_\nu\left(\Bigg\lvert-\sum^{2n-2}_{j=n}k_j\Bigg\rvert t\right)\right)^{-1}\right\}.
\end{eqnarray}
The solution of $\bar{\mathcal{G}}_{2n-2}(k_1,...,k_{2n-2},t)$ is given by
\begin{align}
    &\bar{\mathcal{G}}_{2n-2}(k_1,...,k_{2n-2},t)
    \\ \nonumber
    &=\prod^{2n-2}_{j=1}\left[ t^{\frac{1}{2}}K_\nu(\vert k_j\vert t)\right]^{-1}\int^t_{0} dt^\prime \  \Bigg[ \frac{n^2}{4}{t^\prime}^{-1}\prod^{2n-2}_{j=1}\tau_{k_j} \text{Perm}\left\{  \tau_{k_1+...+k_{n-1}}^2\left( K_\nu\left(\Bigg\vert-\sum^{n-1}_{j=1}k_j\Bigg\vert {t^\prime}\right)\right)^{-2}\right\} 
    \\ \nonumber &- \frac{\bar{\lambda}_{2n-2}}{n-1}{t^\prime}^{(n-2)d-1} \prod^{2n-2}_{j=1}K_\nu(\vert k_j \vert {t^\prime})\Bigg].
\end{align}
If we perform integration of the first term in the integration part, the simplest solution of $\bar{\mathcal{G}}_{2n-2}(k_1,...,k_{2n-2},t)$ is given by
\begin{align}
\label{tau 2n-2}
\bar{\mathcal{G}}_{2n-2}(k_1,...,k_{2n-2},t)&= \prod^{2n-2}_{j=1}\left[ {t}^{\frac{1}{2}}K_\nu(\vert k_j\vert {t})\right]^{-1} \Bigg[\frac{n^2}{4}\prod^{2n-2}_{j=1}\tau_{k_j} \text{Perm} \left\{ \tau^2_{k_1+...+k_{n-1}} \frac{I_\nu\left(\vert-\sum^{n-1}_{j=1}k_j\vert {t}\right)}{K_\nu\left(\vert-\sum^{n-1}_{j=1}k_j\vert {t}\right)}\right\}
    \\ \nonumber &- \frac{\bar{\lambda}_{2n-2}}{n-1} \int^t_{0} \left[{t^\prime}^{(n-2)d-1} \prod^{2n-2}_{j=1}K_\nu(\vert k_j \vert {t^\prime})\right] d{t^\prime}+\tau_{2n-2}\Bigg].
\end{align}
Now we construct the Euclidean action as 
\begin{align}
    S_E&=\int d^dk \ \Phi(k,t)\Phi(-k,t) \left[-\partial_t \log \left(t^{\frac{1}{2}}K_\nu(\vert k \vert t) \right)\right] 
   \\ \nonumber &+\int\left[\prod^{n}_{i=1} d^dk_i \Phi({k_i},t) \right] \left[\tau_n \prod^n_{j=1} \left( t^{\frac{1}{2}} K_\nu(\vert k_j \vert t) \right)^{-1}\right]\delta^{(d)} \left( \sum^n_{j=1} k_j\right)
    \\ \nonumber &+ \int\left[\prod^{2n-2}_{i=1} d^dk_i \Phi(k_i,t) \right] \Bigg[ \prod^{2n-2}_{j=1}\left[ {t}^{\frac{1}{2}}K_\nu(\vert k_j\vert {t})\right]^{-1} \Bigg[\frac{n^2}{4}\prod^{2n-2}_{j=1}\tau_{k_j} \text{Perm} \left\{ \tau_{k_1+...+k_{n-1}}^2 \frac{I_\nu\left(\vert-\sum^{n-1}_{j=1}k_j\vert {t}\right)}{K_\nu\left(\vert-\sum^{n-1}_{j=1}k_j\vert {t}\right)}\right\}
    \\ \nonumber &- \frac{\bar{\lambda}_{2n-2}}{n-1} \int^t_{0} \left[{t^\prime}^{(n-2)d-1} \prod^{2n-2}_{j=1}K_\nu(\vert k_j \vert {t^\prime})\right] d{t^\prime}+\tau_{2n-2}\Bigg]\Bigg]\delta^{(d)} \left( \sum^{2n-2}_{j=1} k_j\right).
\end{align}
Note that the Euclidean action is obtained by identifying the holographic bulk action with the Fokker-Planck action. One can also derive the Euclidean action using the relation $S_E=-2S_B$ with $c^{(2)}=0$. This relation fixes the integration constant $\tau$ by comparing it with the holographic boundary action as
\begin{align}
    \tau_n&=\prod^{n}_{i=1}\left[(-2\bar{\sigma})^{1/n}\left\{\frac{1}{2}\Gamma(\nu)\right\}\left(\frac{\vert k_i\vert}{2}\right)^{-\nu}\right]
    \\ \nonumber \prod^{2n-2}_{i=1}\tau_{k_i}&=\prod^{2n-2}_{i=1}\left[\left(-2\bar{\sigma}_n\right)^{1/n}\left\{\frac{1}{2}\Gamma(\nu)\right\}\left(\frac{\vert k_i\vert}{2}\right)^{-\nu}\right]
    \\ \nonumber
    \tau_{k_1+k_2+\cdots+k_{n-1}}^2&=\tau_{-k}^2=\tau_k^2=\left[\left(-2\bar{\sigma}_n\right)^{1/n}\left\{\frac{1}{2}\Gamma(\nu)\right\}\left(\frac{\vert k_1+k_2+\cdots+k_{n-1}}{2}\right)^{-\nu}\right]^2
    \\ \nonumber \tau_{2n-2}&=2\bar{\sigma}_{2n-2}.
\end{align}
However, we keep using the integration constant in our calculation and show that the same results can be obtained by the relation of the multi-trace deformation and stochastic correlation function in Sec.\ref{The precise map of stochastic correlation functions and multi-trace deformations}.
\subsection{Stochastic correlation functions}
\label{Stochastic correlation functions}
Before we calculate the correlation functions, we define the following quantity to assign a Feynman-like rule as
\begin{align}
    \text{propagator:} \ &\mathcal{K}_\nu(k,t;t') \equiv \frac{t^{1/2}K_\nu(\vert k\vert t)}{{t'}^{1/2}K_\nu(\vert k\vert t')}
    \\ \nonumber &\mathcal{K}_\nu(k,t) \equiv t^{1/2}K_\nu(\vert k\vert t)
    \\ \nonumber
    \text{propagator with noise:} \ &\mathcal{S}_n(k_1,\dots,k_n;t) \equiv \prod^n_{i=1} \int^t_{t_0} \frac{t^{1/2}K_\nu(\vert k\vert t)}{{t'_i}^{1/2}K_\nu(\vert k\vert t'_i)}\eta_{k_i}(t'_i)dt'_i
    \\ \nonumber
  \text{n-point vertex:} \  &-\frac{n}{2}\mathcal{\bar{V}}^n(k_1,\dots, k_n;t) \equiv -\frac{n}{2} \prod^n_{i=1} \left\{ t^{1/2} K_\nu(\vert k_i\vert t)\right\}^{-1}=  -\frac{n}{2}\prod^n_{i=1} \mathcal{K}^{-1}_\nu(k_i,t).
\end{align}
\begin{figure}[t!]
\centering
\includegraphics[width=180mm]{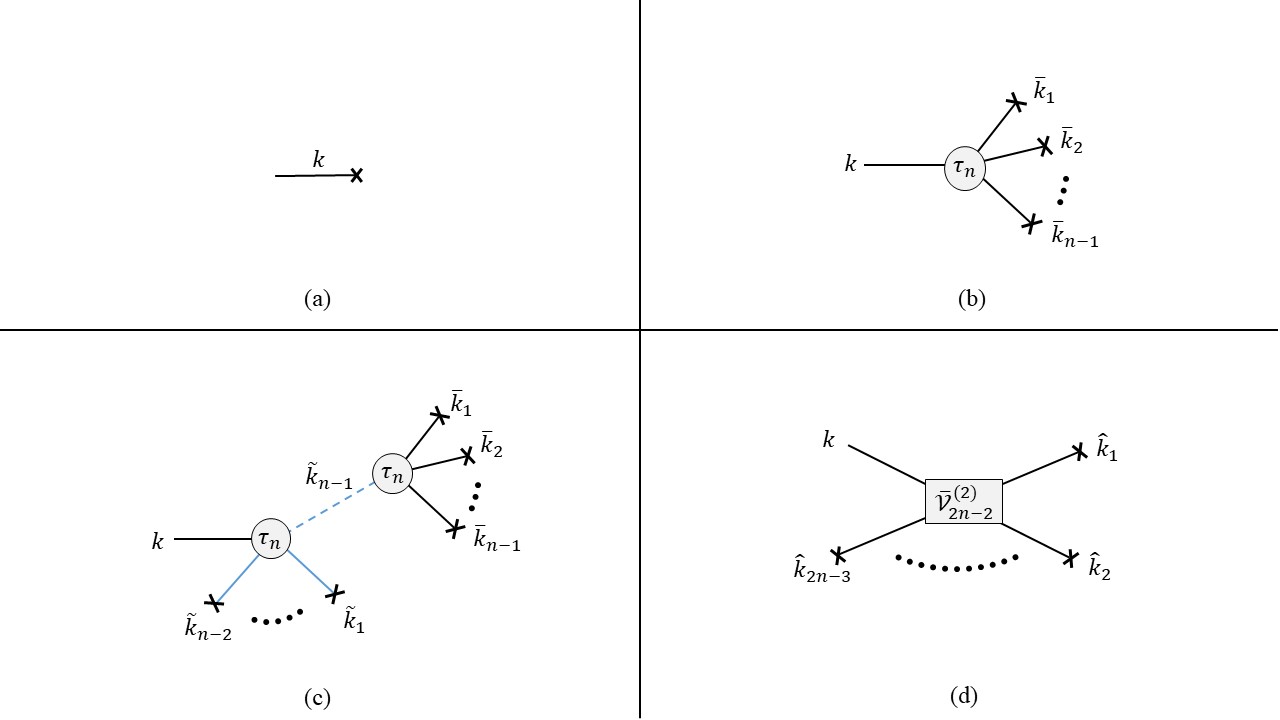}
\caption{Diagrams for the Feynman-like rule. (a) is the diagram of the propagator with noise. After the time integration, it represents $\Phi^{(2)}(k,t)$. (b) is composed of one propagator, $n$-point vertex, and $(n-1)$- propagator with noise. After the time integration, it represents the solution of the first order of $\tau$, $\Phi^{(n)}(k,t)$. (c) is the first type of the $2n-2$ solution, $\phi^{(2n-2,1)}(k,t)$. Note that it is composed of connected two different $n$-point vertex with different times. Also, the colored line denotes that we need to consider the permutation of it, where the dotted line is the propagator that connects each vertex. (d) is the second type of the $2n-2$ solution, $\phi^{(2n-2,2)}(k,t)$, integrating with time.
}
\label{scffig1}
\end{figure}
In upcoming section, we solve the equations with the weak expansion of the coupling constant $\tau_n$ and $\bar{\lambda}_{2n-2}$ order by order.
\subsubsection{Stochastic 2-point function}
Firstly, we employ the Langevin equation up to the zeroth order of $\tau_n$ from the Euclidean action: 
\begin{equation}
    \frac{\partial \Phi^{(2)} (k,t)}{\partial t}=\partial_t \log\left[\mathcal{K}_\nu(k,t)\right]\Phi^{(2)}(k,t)+\eta_k(t),
\end{equation}
where the solution of the equation is given by
\begin{equation}
    \Phi^{(2)}(k,t)=\int^t_{t_0} \mathcal{K}_\nu(k,t;t') \eta_k(t') dt'.
\end{equation}
From the result, one can compute a non-equal time stochastic 2-point function:
\begin{eqnarray}
\label{non-equal-2-point-function}
    \langle \Phi^{(2)}(k,t)\Phi^{(2)}(k',t')\rangle &=& \int^t_{t_0}\int^{t'}_{t_0}\mathcal{K}_\nu(k,t;\tilde{t})\mathcal{K}_\nu(k',t';\tilde{t}')\langle \eta_k(\tilde{t}) \eta_{k'}(\tilde{t'})\rangle d\tilde{t} d\tilde{t'}
\\ \nonumber &=& \mathcal{K}_\nu(k,t)\mathcal{K}_\nu(k,t')\frac{\tilde{\mathcal{Q}}_\nu(k,t)}{\mathcal{K}(k,t)}\delta^{(d)}(k+k')=\mathcal{K}_\nu(k,t)\tilde{\mathcal{Q}}_\nu(k,t)\delta^{(d)}(k+k'),
\end{eqnarray}
where we define a new quantity:
\begin{align}
\label{alpha definition}
    \tilde{\mathcal{Q}}_\nu(k,t)\equiv t^{1/2}\left[I_\nu(\vert k\vert t)-\alpha_k(t_0)K_\nu(\vert k\vert t) \right], \quad \alpha_k(t_0)\equiv \frac{I_\nu(\vert k\vert t_0)}{K_\nu(\vert k\vert t_0)},
\end{align}
and we use Markov property of noise correlator $\langle \eta_p(t)\eta_{k'}(t') \rangle=\delta^{(d)}(k+k')\delta(t-t')$. Also, note that the integration is given by
\begin{align}
    \int^t_{t_0} \mathcal{K}_\nu^{-2} (k,t')dt'=\frac{I_\nu(\vert k\vert t)}{K_\nu(\vert k\vert t)}-\alpha_k(t_0)=\frac{\tilde{\mathcal{Q}}_\nu(k,t)}{\mathcal{K}_\nu(k,t)}.
\end{align}
Them, the equal-time correlator is given by
\begin{align}
\label{2-point-function-final}
 \langle \Phi^{(2)}(k,t)\Phi^{(2)}(k',t)\rangle=\mathcal{K}_\nu(k,t)^2\frac{\tilde{\mathcal{Q}}_\nu(k,t)}{\mathcal{K}(k,t)}\delta^{(d)}(k+k')=\mathcal{K}_\nu(k,t)\tilde{\mathcal{Q}}_\nu(k,t)\delta^{(d)}(k+k').
\end{align}
\subsubsection{Stochastic $n$-point function}
To consider the sub-leading term in the Euclidean action, we solve the equation by perturbation of coupling $\tau_n$ given by
\begin{equation}
    \Phi^{(n)}(k,t)={\tau}_n\bar{\Phi}^{(n)}(k,t).
\end{equation}
Then, the Langevin equation in the first order of $\tau_n$ is given by
\begin{eqnarray}
    \frac{\partial \bar{\Phi}^{(n)}(k,t)}{\partial t}=\partial_t \ln\left(\mathcal{K}_\nu(k,t)\right) \bar{\Phi}^{(n)}(k,t)-\frac{n}{2}\int\left(\prod^{n-1}_{i=1} d^dk_i \Phi^{(2)}({k_i},t)\right)\bar{\mathcal{V}}_n(k_1,\dots,k_n;t) \delta^{(d)}\left(\sum^{n-1}_{j=1} k_j+k\right).
\end{eqnarray}
The solution of the equation is given by
\begin{align}
\label{phi-n-solution}
        \bar{\Phi}^{(n)}(k,t)&= -\frac{n}{2} \int^t_{t^\prime_0} \Bigg[\mathcal{K}_\nu(k,t;t'') \tau_n \bar{\mathcal{V}}_n(k_1,\dots,k_n;t)   \int \prod^{n-1}_{i=1}d^d \bar{k}_i \ \mathcal{S}_{n-1}(\bar{k}_1,\dots,\bar{k}_{n-1};t'') \Bigg] dt'' \delta^{(d)} \left(\sum^{n-1}_{j=1} \bar{k}_j +k\right).
\end{align}
The stochastic $n$-point function is given by
\begin{equation}
    \langle \Phi_{k_1}(t) \Phi_{k_2}(t) \cdots \Phi_{k_n}(t)\rangle = n \cdot \text{Perm}\left\{ \langle \overbrace{\Phi^{(2)}_{k_1} \cdots \Phi^{(2)}_{k_{n-1}}(t)}^{n-1} \Phi^{(n)}_{k_n}(t)\rangle \right\}.
\end{equation} 
From the solution of the Langevin equation, we get 
\begin{align}
    &\langle \overbrace{\Phi^{(2)}_{k_1} \cdots \Phi^{(2)}_{k_{n-1}}(t)}^{n-1} \Phi^{(n)}_{k_n}(t)\rangle
    \\ \nonumber
    &=\Bigg\langle -\frac{n}{2} \int^t_{t^\prime_0} \Bigg[\mathcal{K}_\nu(k_n,t;\bar{t}_n)\tau_n \bar{\mathcal{V}}_n(k_1,\dots,k_n;\bar{t}_n)  \int \prod^{n-1}_{i=1}d^d \bar{k}_i \ \mathcal{S}_{n-1}(\bar{k}_1,\dots,\bar{k}_{n-1};\bar{t}_n) \Bigg] d\bar{t}_n 
    \\ \nonumber &\times\ \mathcal{S}_{n-1}(k_1,\dots,k_{n-1};t) \Bigg\rangle \delta^{(d)} \left(\sum^{n-1}_{j=1} \bar{k}_j +k_n\right) .
\end{align}
This result also can be derived diagrammatically in Figure 2(b). We have one propagator, $n$-point vertex, and $n-1$ number of propagators with noise. Note that we have time integration $\int^t_{t'_0} dt''$ to the whole ingredients connected to a $n$-point vertex from the equation (\ref{phi-n-solution}).
The only correlated terms inside the bracket $\langle \cdot \rangle$ are the noise fields inside $\mathcal{S}_n$ with different times and momentum labels. If we collect the noise fields, we get
\begin{align}
&=  -\frac{n}{2} \int^t_{t^\prime_0} \Bigg[\mathcal{K}_\nu(k_n,t;\bar{t}_n)\tau_n \bar{\mathcal{V}}_n(k_1,\dots,k_n;\bar{t}_n)   \int \prod^{n-1}_{i=1}d^d \bar{k}_i \left\{ \int^{\bar{t}_n}_{t_0} \mathcal{K}_\nu(\bar{k}_i,\bar{t}_n;\bar{t}_i) d\bar{t}_i\right\} \Bigg] d\bar{t}_n 
\\ \nonumber &\times \delta^{(d)} \left(\sum^{n-1}_{j=1} \bar{k}_j +k_n\right)  \prod^{n-1}_{j=1}\left\{ \int^t_{t_0} \mathcal{K}_\nu(k_j,t;t_j') dt^\prime_j\right\} \left\langle \prod^{n-1}_{i=1} \prod^{n-1}_{j=1} \eta_{k_i}(\bar{t}_i) \eta_{k_j}(t^\prime_j)\right\rangle_{\text{connected,tree}}.
\end{align}
Note that the extraction of noise fields inside $\mathcal{S}_n$ gives a product of propagator $\mathcal{K}$. 
We consider the connected-tree level diagram as 
\begin{align}
    &\langle \overbrace{\Phi^{(2)}_{k_1} \cdots \Phi^{(2)}_{k_{n-1}}(t)}^{n-1} \Phi^{(n)}_{k_n}(t)\rangle
    \\ \nonumber &=-\frac{n}{2} \int^t_{t^\prime_0} \Bigg[\mathcal{K}_\nu(k_n,t;\bar{t}_n)\tau_n \bar{\mathcal{V}}_n(k_1,\dots,k_n;\bar{t}_n)   \int \prod^{n-1}_{i=1}d^d \bar{k}_i \left\{ \int^{\bar{t}_n}_{t_0} \mathcal{K}_\nu(\bar{k}_i,\bar{t}_n;\bar{t}_i) d\bar{t}_i\right\} \Bigg] d\bar{t}_n 
\\ \nonumber &\times \delta^{(d)} \left(\sum^{n-1}_{j=1} \bar{k}_j +k_n\right)  \prod^{n-1}_{j=1}\left\{ \int^t_{t_0} \mathcal{K}_\nu(k_j,t;t_j') dt^\prime_j\right\}(n-1)!\prod^{n-1}_{i=1} \delta^{(d)} (\bar{k}_i-k_i) \delta(\bar{t}_i-t^\prime_i)
    \\ \nonumber &=  -\frac{n!}{2} \prod^n_{i=1} \left[\mathcal{K}_\nu(k_i,t)\right]\int^t_{t^\prime_0} \Bigg[\mathcal{K}^{-2}_\nu(k_n, \bar{t}_n) \tau_n \prod^{n-1}_{i=1}\frac{\tilde{\mathcal{Q}}_\nu(k_i,\bar{t}_n)}{\mathcal{K}_\nu(k_i,\bar{t}_n)} \Bigg] d\bar{t}_n \delta^{(d)} \left(\sum^{n-1}_{j=1} {k}_j +k_n\right) .
\end{align}
We bring back the permutation to complete the $n$-point function given by
\begin{align}
\label{n-point-function-final}
    &n\cdot \text{Perm}\left\{\langle \overbrace{\Phi^{(2)}_{k_1} \cdots \Phi^{(2)}_{k_{n-1}}(t)}^{n-1} \Phi^{(n)}_{k_n}(t)\rangle\right\} 
    \\ \nonumber &=  -n \cdot \frac{n!}{2} \prod^n_{i=1} \left[\mathcal{K}_\nu(k_i,t)\right]  \text{Perm} \left\{\int^t_{t^\prime_0} \Bigg[\frac{1}{\mathcal{K}^2_\nu(k_n, \bar{t}_n)} \tau_n \prod^{n-1}_{i=1}\frac{\tilde{\mathcal{Q}}_\nu(k_i,\bar{t}_n)}{\mathcal{K}_\nu(k_i,\bar{t}_n)}\Bigg] d\bar{t}_n \right\} \delta^{(d)} \left(\sum^{n-1}_{j=1} {k}_j +k_n\right) 
\\ \nonumber &= -\frac{n!}{2} \cdot \tau_n \prod^n_{i=1} \left[\mathcal{K}_\nu(k_i, t)\right] \Bigg\{\prod^n_{i=1} \frac{\tilde{\mathcal{Q}}_\nu(k_i,t)}{\mathcal{K}_\nu(k_i,t)} -\prod^n_{i=1} \left[\alpha(t^\prime_0)-\alpha(t_0)\right]\Bigg\}\delta^{(d)} \left(\sum^{n-1}_{l=1} {k}_l +k_n\right).
\end{align}
In the last equality, we use the relation:
\begin{align}
    &\int  \text{Perm} \Bigg[\mathcal{K}^{-2}_\nu(k_n, \bar{t}_n)  \prod^{n-1}_{i=1}\left\{  \frac{\tilde{\mathcal{Q}}_\nu(k_i,\bar{t}_n)}{\mathcal{K}_\nu(k_i,\bar{t}_n)} \right\} \Bigg]d\bar{t}_n
    \\ \nonumber
    &=\int \frac{1}{n}\Bigg[\sum^n_{i=1}\mathcal{K}^{-2}_\nu( k_n , \bar{t}_n)  \prod^{n}_{j=1, i\neq j}\left\{  \frac{\tilde{\mathcal{Q}}_\nu(k_i,\bar{t}_n)}{\mathcal{K}_\nu(k_i,\bar{t}_n)} \right\} \Bigg]d\bar{t}_n= \prod^{n}_{i=1}\left\{  \frac{\tilde{\mathcal{Q}}_\nu(k_i,t)}{\mathcal{K}_\nu(k_i,t)} \right\}.
\end{align}
We note that the integrand is invariant under the permutation of the momentum label since $k_n=-\sum^{n-1}_{j=1}k_j$. Originally, we should have a permutation of the integrand to be solvable but invariance under permutation allows us to satisfy the following relation without permutation:
\begin{align}
   \mathcal{K}^{-2}_\nu\left(-\sum^{n-1}_{i=1}k_i, \bar{t}_n\right)  \prod^{n-1}_{i=1}\left\{  \frac{\tilde{\mathcal{Q}}_\nu(k_i,\bar{t}_n)}{\mathcal{K}_\nu(k_i,\bar{t}_n)} \right\} = \frac{1}{n}\Bigg[\sum^n_{i=1}\mathcal{K}^{-2}_\nu( k_n , \bar{t}_n)  \prod^{n}_{j=1, i\neq j}\left\{  \frac{\tilde{\mathcal{Q}}_\nu(k_i,\bar{t}_n)}{\mathcal{K}_\nu(k_i,\bar{t}_n)} \right\} \Bigg].
\end{align}
\subsubsection{Stochastic ($2n-2$)-point function}
In this case, we need to find the solution up to the order of $\tau_n^2$ and $\bar{\lambda}_{2n-2}$. The Langevin equation is given by 
\begin{align}
    &\frac{\partial \Phi^{(2n-2)}_k(t)}{\partial t}-\partial_t \log\left[ \mathcal{K}_\nu(k,t) \right] \Phi^{(2n-2)}_k(t)
    \\ \nonumber &= \frac{n^2(n-1)}{4}\text{Perm}\Bigg\{\int \prod^{n-2}_{i=1} d^d\tilde{k}_i \ \mathcal{S}_{n-2}(\tilde{k}_1,\dots,\tilde{k}_{n-2};t)  \int d^d\tilde{k}_{n-1}\int^{t^\prime}_{t_0} \Bigg[ \mathcal{K}_\nu(\tilde{k}_{n-1},t,t'') \tau_n\bar{\mathcal{V}}_n(k_1,\dots,k_n;t'') 
     \\ \nonumber &\times \int \prod^{n-1}_{i=1}d^d \bar{k}_i \ \mathcal{S}_{n-1}(\bar{k}_1, \dots,\bar{k}_{n-1};t'')  \Bigg]dt'' \delta^{(d)}\left(\sum^{n-1}_{j=1}\bar{k}_j-\tilde{k}_{n-1}\right)\Bigg\} \tau_n \bar{\mathcal{V}}_n (k_n,\dots,k_{2n-2},k;t)\delta^{(d)}\left(\sum^{n-1}_{j=1} k_j-k\right)
     \\ \nonumber &+(n-1) \int \prod^{2n-3}_{i=1} d^d \hat{k}_i \left[ \mathcal{S}_{2n-3}(\hat{k}_1,\dots,\hat{k}_{2n-3};t)  \bar{\mathcal{V}}^{(2)}_{2n-2}(k_1,\dots,k_{2n-2};t)\right]\delta^{(d)}\left(\sum^{2n-3}_{l=1}k_l-k\right),
\end{align}
where we define the second type of $(2n-2)$ vertex as
\begin{align}
     \bar{\mathcal{V}}^{(2)}_{2n-2}(k_1,\dots,k_{2n-2};t)&\equiv \prod^{2n-2}_{j=1}\tau_{k_j} \ \bar{\mathcal{V}}_{2n-2}(k_1,\dots,k_{2n-2};t)
     \\ \nonumber &\times\left[-\frac{n^2}{4}\tau_{k_1+k_2+\cdots+k_{n-1}}^2 \frac{I_\nu\left(\vert -\sum^{n-1}_{j=1} k_j\vert t\right)}{K_\nu\left(\vert -\sum^{n-1}_{j=1}k_j\vert t\right)}+\frac{\bar{\lambda}_{2n-2}}{n-1} \bar{K}_\nu^{(2n-2)}(k,t)-\tau_{2n-2}\right],
\end{align}
and
\begin{equation}
    \bar{K}_\nu^{(2n-2)}(k,t) \equiv \int dt^\prime \left[ {t^\prime}^{n(2\nu-1)} \prod^{2n-2}_{i=1}K_\nu(\vert k_i\vert t^\prime)\right].
\end{equation}
The solution of the differential equation is straightforward. However, for simplicity, we split the solution of the field $\Phi^{(2n-2)}$ into 2 different types whether the solution is related to the second type of $(2n-2)$-vertex $\bar{\mathcal{V}}^{(2)}_{2n-2}$ or not:
\begin{equation}
    \Phi^{(2n-2)}_k(t)=\sum^2_{i=1}\Phi^{(2n-2,i)}_k(t).
\end{equation}
The first type of $(2n-2)$ solution which is not related to $\bar{\mathcal{V}}^{(2)}_{2n-2}$ is given by
\begin{align}
    &\Phi^{(2n-2,1)}_k(t)= \frac{n^2(n-1)}{4}\int^t_{{t}_0''}\Bigg[\mathcal{K}_\nu(k,t;t')\text{Perm}\Bigg\{\int\prod^{n-2}_{i=1} d^d\tilde{k}_i \ 
\mathcal{S}_{n-2}(\tilde{k}_1, \dots, \tilde{k}_{n-2};t^\prime) 
\\ \nonumber &\times \int d^d\tilde{k}_{n-1}\int^{t^\prime}_{t_0}\Bigg[\mathcal{K}_\nu(\tilde{k}_{n-1},t';t'')
 \tau_n \bar{\mathcal{V}}^n(k_1,\dots,k_n;t'')\int \prod^{n-1}_{i=1}d^d \bar{k}_i \ \mathcal{S}_{n-1}(\bar{k}_1,\dots,\bar{k}_{n-1};t'') \Bigg]dt'' 
 \\ \nonumber &\times \delta^{(d)}\left(\sum^{n-1}_{j=1}\bar{k}_j+\tilde{k}_{n-1}\right)\Bigg\} \tau_n \bar{\mathcal{V}}^n(k_n,\dots,k_{2n-2},k;t')\delta^{(d)}\left(\sum^{n-1}_{j=1} k_j+k\right)\Bigg] dt^\prime,
 \end{align}
 and the second type of the solution which is related to $\bar{\mathcal{V}}^{(2)}_{2n-2}$ is given by
 \begin{align}
      \Phi^{(2n-2,2)}_k(t)&=(n-1)\int^t_{t_0''} \Bigg[\mathcal{K}_\nu(k,t;t')\int \prod^{2n-3}_{i=1} d^d \hat{k}_i \ S_{2n-3}(\hat{k}_1,\dots,\hat{k}_{2n-3};t^\prime) \bar{\mathcal{V}}^{(2)}_{2n-2}(k_1,\dots,k_{2n-2};t')
      \\ \nonumber &\times \delta^{(d)}\left(\sum^{2n-3}_{l=1}k_l+k\right)\Bigg]dt^\prime.
\end{align}
To simplify the calculation, we divide the stochastic $2n-2$-point function into three different types given by
\begin{equation}
    \langle \overbrace{\Phi_{k}(t) \cdots \Phi_{k}(t)}^{2n-2} \rangle^{[1]} = \frac{(2n-2)!}{2!(2n-4)!} \cdot \text{Perm}\left\{ \langle \overbrace{\Phi^{(2)}_{k_1} \cdots \Phi^{(2)}_{k_{2n-4}}(t)}^{2n-4} \Phi^{(n)}_{k_{2n-3}}(t)\Phi^{(n)}_{k_{2n-2}}(t)\rangle \right\}
\end{equation}
\begin{equation}
    \langle \overbrace{\Phi_{k}(t) \cdots \Phi_{k}(t)}^{2n-2} \rangle^{[2]} = (2n-2) \cdot \text{Perm}\left\{ \langle \overbrace{\Phi^{(2)}_{k_1} \cdots \Phi^{(2)}_{k_{2n-3}}(t)}^{2n-3} \Phi^{(2n-2,1)}_{k_{2n-2}}(t)\rangle \right\}
\end{equation}
\begin{equation}
     \langle \overbrace{\Phi_{k}(t) \cdots \Phi_{k}(t)}^{2n-2} \rangle^{[3]} = (2n-2) \cdot \text{Perm}\left\{ \langle \overbrace{\Phi^{(2)}_{k_1} \cdots \Phi^{(2)}_{k_{2n-3}}(t)}^{2n-3} \Phi^{(2n-2,2)}_{k_{2n-2}}(t)\rangle \right\}.
\end{equation}
\begin{figure}[b!]
\centering
\includegraphics[width=180mm]{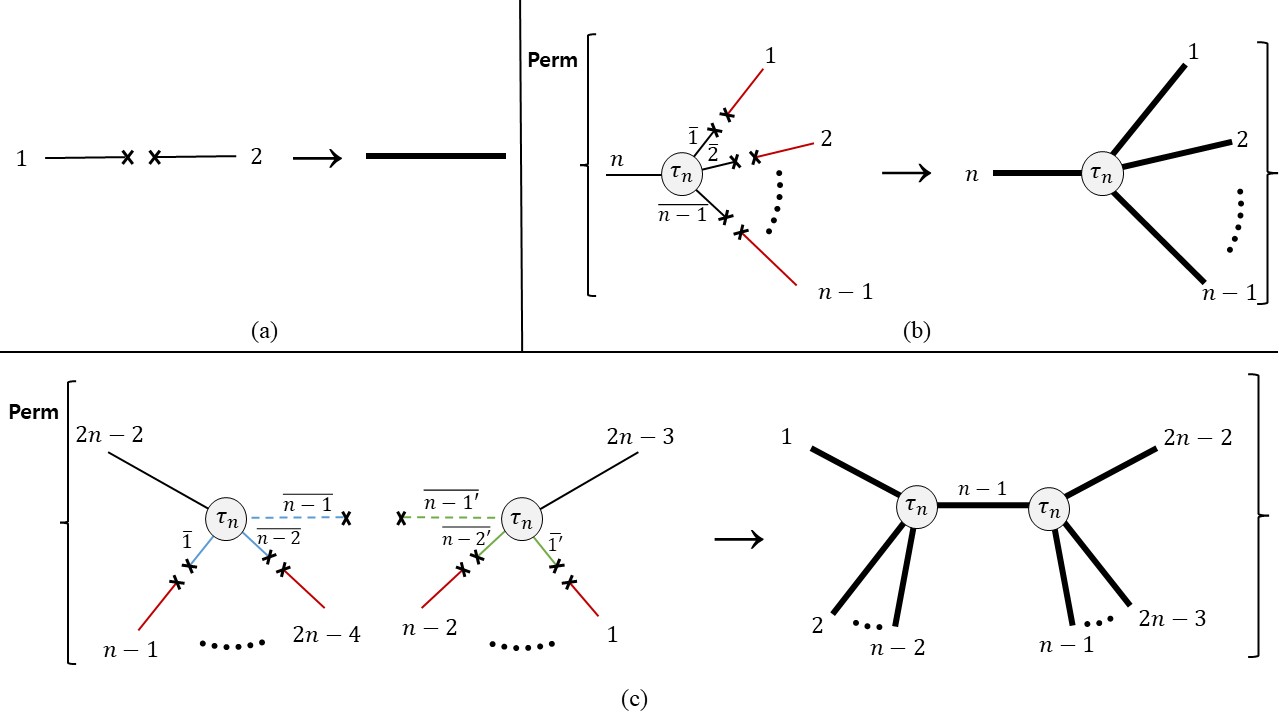}
\caption{Diagram of the correlation functions. (a) is a $2$-point correlation function. The noise fields are contracted to each other and become a thick line. (b) is a $n$-point correlation function. and (c) is the first type of $2n-2$-point correlation function. Note that the diagram represents the connection of two different $n$-point functions with different times, which means that we need to consider the order of time and this gives a step function. 
}
\label{scffig2}
\end{figure}
The detailed calculation of the $(2n-2)$ correlation functions is explained in Appendix \ref{2n-2-correlation-function}. In this section, we start with the result of the calculation of the correlation functions. The first type of $2n-2$-point function is given by
\begin{align}
    &\langle \overbrace{\Phi_{k}(t) \cdots \Phi_{k}(t)}^{2n-2} \rangle^{[1]}=\frac{(2n-2)!}{2!(2n-4)!}\text{Perm}\Bigg[\Bigg\langle S_{2n-4}(k_1,\dots,k_{2n-4};t)
\\ \nonumber
&\times -\frac{n}{2}\int^t_{t^\prime_0}\left[ \mathcal{K}_\nu(k_{2n-3},t;\bar{t}_n) \tau_n\bar{\mathcal{V}}_n(k_{n-1},\dots,k_{2n-3},k_{n-1};\bar{t}_n)\int\prod^{n-1}_{i=1}d^d\bar{k}_i \ S_{n-1}(\bar{k}_1,\dots,\bar{k}_{n-1};\bar{t}_n)     \right]d\bar{t}_n
\\ \nonumber
&\times -\frac{n}{2}\int^t_{t^\prime_0}\left[\mathcal{K}_\nu(k_{2n-2},t;\bar{t}'_n) \tau_n \bar{\mathcal{V}}_n(k_{1},\dots,k_{n-1},k_{2n-2};\bar{t}'_n)\int \prod^{n-1}_{l=1} d^d\bar{k}^\prime_l \ S_{n-1}(\bar{k}^\prime_1,\dots,\bar{k}^\prime_{n-1}; \bar{t}^\prime_{n}) \right]d\bar{t}^\prime_n
\\ \nonumber
&\times \delta^{(d)}\left(\sum^{n-1}_{m=1} \bar{k}_m +k_{2n-3}\right) \delta^{(d)}\left(\sum^{n-1}_{m^\prime=1}\bar{k}^\prime_{m^\prime} +k_{2n-2}\right)\Bigg\rangle\Bigg].
\end{align}
It can be interpreted as a diagrammatic expression of the first type of $(2n-2)$ correlation function in Figure 2(c). Note that the disconnected vertices at the earlier stage have different time integration $\int^t_{t'_0} d\bar{t}_n$ and $\int^t_{t'_0} d\bar{t}'_n$. After the calculation, the first type of $(2n-2)$-correlation function is given by 
\begin{align}
    & \langle \overbrace{\Phi_{k}(t) \cdots \Phi_{k}(t)}^{2n-2} \rangle^{[1]} 
    \\ \nonumber &=\frac{(2n-2)!(n-1)^2n}{4} \text{Perm}\Bigg[\prod^{2n-2}_{i=1}\tau^2_n\left\{\mathcal{K}_\nu(k_i,t)\right\} 
    \\ \nonumber
    &\times\Bigg[ \frac{1}{2n-1}\prod^{2n-2}_{i=1}\frac{\tilde{\mathcal{Q}}_\nu(k_i,t)}{\mathcal{K}_\nu(k_i,t)}\frac{\tilde{\mathcal{Q}}_\nu(k,t)}{\mathcal{K}_\nu(k,t)} - \frac{1}{2n-1}\prod^{2n-2}_{i=1}\left(\alpha_{k_i}(t^\prime_0)-\alpha_{k_i}(t_0)\right)\left(\alpha_{k}(t^\prime_0)-\alpha_{k}(t_0)\right) 
    \\ \nonumber
    &-\frac{1}{n-1}\prod^{2n-2}_{i=n}\frac{\tilde{\mathcal{Q}}_\nu(k_i,t)}{\mathcal{K}_\nu(k_i,t)}\prod^{n}_{i=1}\left(\alpha_{k_i}(t^\prime_0)-\alpha_{k_i}(t_0)\right) + \frac{1}{n-1}\prod^{2n-2}_{i=1}\left(\alpha_{k_i}(t^\prime_0)-\alpha_{k_i}(t_0)\right) \left(\alpha_{k}(t^\prime_0)-\alpha_{k}(t_0)\right) \Bigg].
\end{align}
The second type of $2n-2$ correlation function is given by
\begin{align}
    &\langle \overbrace{\Phi_{k}(t) \cdots \Phi_{k}(t)}^{2n-2} \rangle^{[2]}=(2n-2)\text{Perm}\Bigg[ \Bigg\langle S_{2n-3}(k_1,\dots,k_{2n-3};t) \frac{n^2(n-1)}{4}
    \\ \nonumber &\int^t_{t_0''} \Bigg[\mathcal{K}_\nu(k_{2n-2},t;t')\text{Perm}\Bigg\{\int\prod^{n-2}_{l=1}d^d\tilde{k}_l \ S_{n-2}(\tilde{k}_1,\dots,\tilde{k}_{n-2};t^\prime)
    \\ \nonumber
    &\times d^d\tilde{k}_{n-1}\left[\int^{t^\prime}_{t^\prime_0}\mathcal{K}_\nu(\tilde{k}_{n-1},t';t'')\tau_n\bar{\mathcal{V}}_n(k_1,\dots,k_n;t'')\int \prod^{n-1}_{m=1} d^d\bar{k}_m \ S_{n-1}(\bar{k}_1,\dots,\bar{k}_{n-1};t'')\right] dt'' \Bigg\}
    \\ \nonumber
    &\times \tau_n\bar{\mathcal{V}}_n(k_n,\dots,k_{2n-2},k_n;t')\Bigg]dt^\prime \delta^{(d)}\left(\sum^{n-1}_{p=1}\bar{k}_p-\tilde{k}_{n-1}\right)\delta^{(d)}\left(\sum^{n-1}_{q=1}k_q-k_{2n-2}\right)\Bigg].
\end{align}
\begin{figure}[t!]
\centering
\includegraphics[width=180mm]{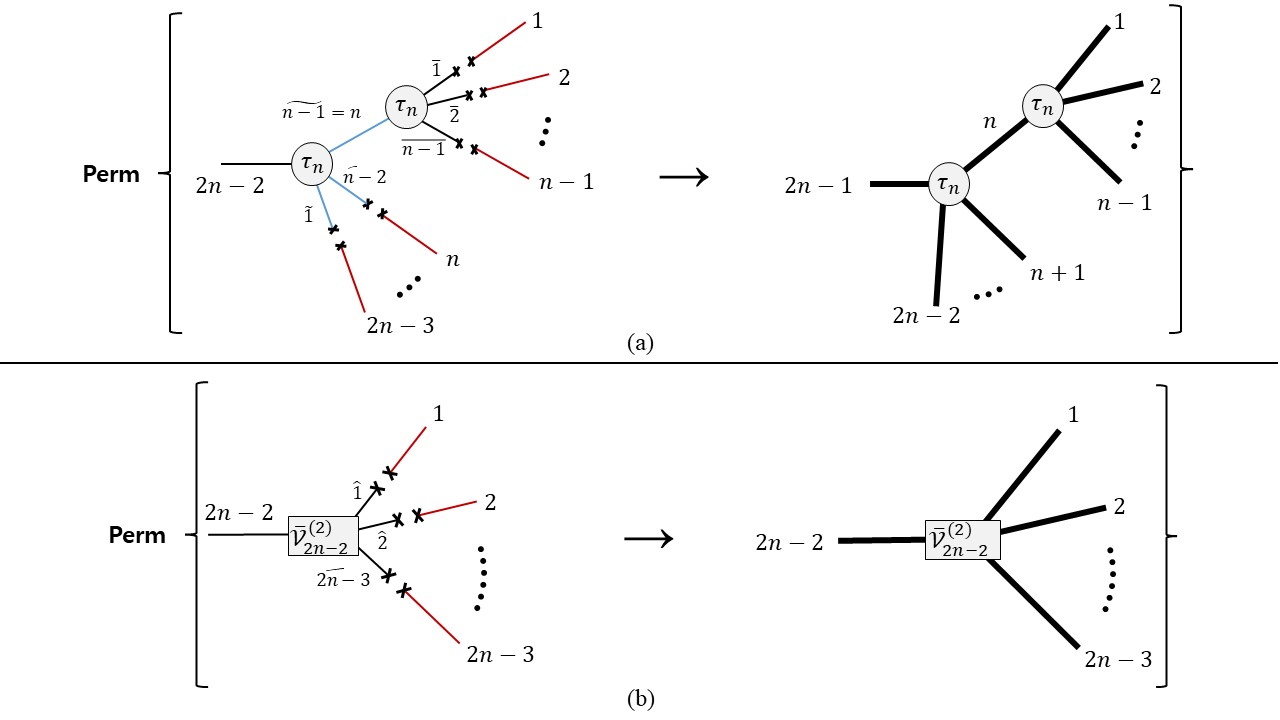}
\caption{Diagram of the correlation functions. (a) is the second type of $2n-2$-point correlation function. (b) is the third type of $2n-2$-point correlation function.
}
\label{scffig3}
\end{figure}
The diagrammatic expression of the second type of $(2n-2)$-point function is given in Figure 3(a). This looks similar to the first type diagram but it's slightly different from the contraction stage. Unlike the first type diagram has disconnected vertices at the early stage, the second type diagram has already connected vertices which means we don't have to assign a different time integration to the vertex contribution. They are topologically the same at the later stage but the difference in connection of vertices at the earlier stage gives different symmetry factors by the permutation. The result of the second type of $(2n-2)$ correlation function is given by
\begin{align}
&\langle \overbrace{\Phi_{k}(t) \cdots \Phi_{k}(t)}^{2n-2} \rangle^{[2]}
\\ \nonumber
        &= (2n-2)! \frac{n(n-1)}{4} \text{Perm}\Bigg\{\tau^2_n\prod^{2n-2}_{i=1}\left[\mathcal{K}_\nu(k_i,t)\right]\Bigg[\Bigg[\frac{1}{2n-1}\prod^{2n-2}_{i=1}\frac{\tilde{\mathcal{Q}}_\nu(k_i,t)}{\mathcal{K}_\nu(k_i,t)} \frac{\tilde{\mathcal{Q}}_\nu(k,t)}{\mathcal{K}_\nu(k,t)}
        \\ \nonumber
        &-\frac{1}{2n-1}\prod^{2n-2}_{i=1}\left[\alpha_{k_i}(t''_0)-\alpha_{k_i}(t_0)\right]\left[\alpha_{k}(t''_0)-\alpha_{k}(t_0)\right]
    -\frac{1}{n-1}\prod^{2n-2}_{i=n}\frac{\tilde{\mathcal{Q}}_\nu(k_i,t)}{\mathcal{K}_\nu(k_i,t)}\prod^{n}_{j=1}\left[\alpha_{k_j}(t^\prime_0)-\alpha_{k_j}(t_0)\right]
   \\ \nonumber
   &+\frac{1}{n-1}\prod^{2n-2}_{i=n}\left(\alpha_{k_i}(t''_0)-\alpha_{k_i}(t_0)\right)\prod^{n}_{j=1}\left[\alpha_{k_j}(t^\prime_0)-\alpha_{k_j}(t_0)\right]\Bigg]\Bigg\}.
\end{align}
We can notice that the factor $(n-1)$ is different from the first type diagram which is $(n-1)^2$ and this comes from the permutation of choosing the propagator to connect vertices. 
\\ The integral expression of the third type of correlation function is given by
\begin{align}
& \langle \overbrace{\Phi_{k}(t) \cdots \Phi_{k}(t)}^{2n-2} \rangle^{[3]}
\\ \nonumber 
&=(2n-2)(n-1)\text{Perm}\Bigg[\Bigg\langle S_{2n-3}(k_1,\dots,k_{2n-3};t)
\\ \nonumber &\int^t_{t_0''}dt^\prime\Bigg[\mathcal{K}_\nu(k_{2n-2},t;t')\int\prod^{2n-3}_{i=1}d^d\hat{k} \ S_{2n-3}(\hat{k}_1,\dots,\hat{k}_{2n-3};t^\prime) \bar{\mathcal{V}}^{(2)}_{2n-2}(k_1,\dots,k_{2n-2};t')\delta^{(d)}\left(\sum^{2n-3}_{j=1}k_j-k_{2n-2}\right)\Bigg\rangle\Bigg].
\end{align}
The result of the calculation of the third type of correlation function is given by
\begin{align}
& \langle \overbrace{\Phi_{k}(t) \cdots \Phi_{k}(t)}^{2n-2} \rangle^{[3]}=(2n-2
)!(n-1) \text{Perm}\Bigg[ \prod^{2n-2}_{i=1}\tau_{k_i}\left[\mathcal{K}_\nu( k_i, t)\right] \tau_{k_1+k_2+\cdots+k_{n-1}}^2\Bigg[
\\ \nonumber
&\times \frac{1}{2n-1}\Bigg[-\frac{n^2}{4}\left\{\prod^{2n-2}_{i=1}\frac{\tilde{\mathcal{Q}}_\nu(k_i,t)}{\mathcal{K}_\nu(k_i,t)}\frac{\tilde{\mathcal{Q}}_\nu(k,t)}{\mathcal{K}_\nu(k,t)}\right.
-\left.\prod^{2n-2}_{i=1}[\alpha_{k_i}(t''_0)-\alpha_{k_i}(t_0)][\alpha_{k}(t''_0)-\alpha_k(t_0)]\right\}\Bigg]
\\ \nonumber
&+\frac{1}{2n-2}\left[-\frac{n^2}{4}\alpha_k(t_0)\left\{\prod^{2n-2}_{i=1}\frac{\tilde{\mathcal{Q}}_\nu(k_i,t)}{\mathcal{K}_\nu(k_i,t)}-\prod^{2n-2}_{i=1}\left[\alpha_{k_i}(t''_0)-\alpha_{k_i}(t_0)\right]\right\} \right]\Bigg]
    \\ \nonumber 
    &-\frac{1}{2n-2}\Bigg[\prod^{2n-2}_{i=1}\left\{\tilde{\mathcal{Q}}_\nu(k_i,t)\right\}\frac{\bar{\lambda}_{2n-2}}{n-1}\bar{K}^{(2n-2)}_\nu(k,t)+\prod^{2n-2}_{i=1}\left\{\mathcal{K}_\nu(k_i,t)\right\}\frac{\bar{\lambda}_{2n-2}}{n-1}\int^t_0 {t^\prime}^{(n-2)d-2}\prod^{2n-2}_{i=1}Q_\nu(\vert k_i\vert t^\prime) dt^\prime
    \\ \nonumber
    &-\tau_{2n-2}\left[\prod^{2n-2}_{i=1}\tilde{\mathcal{Q}}_\nu(k_i,t)-\prod^{2n-2}_{i=1}\mathcal{K}_\nu(k_i,t)\left(\alpha_{k_i}(t''_0)-\alpha_{k_i}(t_0)\right)\right]\Bigg].
\end{align}
Note that the integration in the fourth line can not be solved analytically. However, we still can verify the relation of $(2n-2)$-stochastic correlation function and $(2n-2)$-trace deformation, maintaining this integral expression and using the integration by part. One can find the detailed calculation in Appendix \ref{2n-2-relation-appendix}.
\section{Reconstruction of Multi-trace Deformations by the Stochastic Frame}
\label{The precise map of stochastic correlation functions and multi-trace deformations}
In this section, we compute the stochastic correlation function to verify the relation between multiple trace deformations and the stochastic correlation function with the boundary directional momentum. 
\subsection{Double trace deformation and stochastic 2-point function}
\label{double-trace-relation}
We start with the double trace deformation to warm up. From the correlation function and Euclidean action, 
\begin{eqnarray}
    \left\langle \Phi(k_1,t)\Phi(k_2,t)\right\rangle^{-1}_\textrm{S}&=&\left[\mathcal{K}_\nu(k,t)\tilde{\mathcal{Q}}_\nu(k,t)\right]^{-1}
    \\ \nonumber -\frac{1}{2}\frac{\delta^2S_E}{\prod^2_{i=1}\delta\Phi(k_i,t)}&=&\partial_t \log\left(\mathcal{K}_\nu(k,t)\right)
\end{eqnarray}
Using the relation, 
\begin{equation}
    \partial_t\left(\frac{I_\alpha(x)}{K_\alpha(x)}\right)= \frac{1}{x\left[K_\alpha(x)\right]^2},
\end{equation}
we can easily obtain the following result.
\begin{align}
\label{double-trace-stochastic-relation}
    \left\langle \Phi(k_1,t)\Phi(k_2,t)\right\rangle^{-1}_\textrm{S}-\frac{1}{2}\frac{\delta^2S_E}{\prod^2_{i=1}\delta\Phi(k_i,t)}&=\partial_t \log \left[\tilde{\mathcal{Q}}_\nu(k,t)\right].
\end{align}
It implies the following correspondence between double trace deformation and stochastic 2-point function as
\begin{align}
    \left.\frac{\delta^2S_B}{\prod^2_{i=1}\delta\Phi(k_i,r)}\right\rvert^{r=t}&=\left\langle \Phi(k_1,t)\Phi(k_2,t)\right\rangle^{-1}_\textrm{S}-\frac{1}{2}\frac{\delta^2S_E}{\prod^2_{i=1}\delta\Phi(k_i,t)}.
\\ \nonumber \partial_t \log \left[ t^{1/2}\left(I_\nu(\vert k \vert t)+c^{(1)}_{k}/c^{(2)}_{k}K_\nu(\vert k \vert t)\right)\right]&= \partial_t \log \left[ t^{1/2}\left(I_\nu(\vert k \vert t)-\alpha_k(t_0) K_\nu(\vert k \vert t)\right)\right],
\end{align}
which is consistent with double trace deformation of HWRG once it satisfies $\mathcal{Q}_\nu(k,\epsilon)\vert_{\epsilon=t}=\tilde{\mathcal{Q}}_\nu(k,t)$, in other words, $c^{(1)}_{k}/c^{(2)}_{k}=-\alpha_k(t_0)$.
\subsection{$n$-multiple trace deformation and stochastic $n$-point function}
We can construct n-multiple trace deformation with the stochastic correlation function by the following relation:
\begin{align}
\left.\frac{\delta^nS_B}{\prod^n_{i=1}\delta\Phi(k_i,r)}\right\rvert^{r=t}=-\left\langle \prod^n_{i=1}\Phi(k_i,t)\right\rangle_\textrm{S}\prod^n_{j=1}\left\langle\Phi(k_j,t)\Phi(-k_j,t)\right\rangle^{-1}_\textrm{S}-\frac{1}{2}\frac{\delta^nS_E}{\prod^n_{i=1}\delta\Phi(k_i,t)}.
\end{align}
We already calculated 2-point(\ref{2-point-function-final}) and n-point correlation function(\ref{n-point-function-final}). The last term is related to nothing but the n-th stochastic kernel in the Euclidean action. If we insert these results into the relation, we get
\begin{align}
\label{n-point-relation-final}
=&  \Bigg[\frac{n!}{2}\tau_n \prod^n_{i=1} \left[\mathcal{K}_\nu( k_i, t)\right] \left\{\prod^n_{i=1} \frac{\tilde{\mathcal{Q}}_\nu(k_i,t)}{\mathcal{K}_\nu(k_i,t)} -\prod^n_{i=1}\left[\alpha_i(t^\prime_0)-\alpha_i(t_0) \right]\right\}
\\ \nonumber\times& \prod^n_{j=1}\left[\mathcal{K}_\nu(k_j,t)\tilde{\mathcal{Q}}_\nu(k_j,t)\right]^{-1}-\frac{n!}{2}\tau_n\bar{\mathcal{V}}_n(k_1,\dots,k_n;t)\Bigg]\delta^{(d)} \left(\sum^{n-1}_{l=1} {k}_l +k_n\right)
\\ \nonumber =&
-\frac{n!}{2}\tau_n\prod^n_{i=1}\left[\alpha_i(t^\prime_0)-\alpha_i(t_0) \right]\tilde{\mathcal{V}}_n(k_1,\dots,k_n;t)
\\ \nonumber =& -\frac{n!}{2}\tau_n \prod^n_{i=1}\left[\frac{(-\alpha_{k_i}(t^\prime_0)/\alpha_{k_i}(t_0))+1}{t^{1/2}\left(K_\nu(\vert k_i\vert t) -\tilde{\alpha}_{k_i}(t_0) I_\nu(\vert k_i\vert t)\right)}\right]\delta^{(d)} \left(\sum^{n-1}_{j=1} {k}_j +k_n\right),
\end{align}
where
\begin{align}
\tilde{\mathcal{V}}_n(k_1,\dots,k_n;t)\equiv\prod_{i=1}^n\left[\tilde{\mathcal{Q}}_\nu(k_i,t)\right]^{-1}.
\end{align}
We remember that the holographic data of n-multiple trace deformation is given by (\ref{D-n-deformation}).
The stochastic construction (\ref{n-point-relation-final}) coincides with the holographic n-multiple deformation if it satisfies
\begin{equation}
    \tau_n=\prod^{n}_{i=1}\left[(-2\bar{\sigma})^{1/n}\left\{\frac{1}{2}\Gamma(\nu)\right\}\left(\frac{\vert k_i\vert}{2}\right)^{-\nu}\right],
\end{equation}
while $\alpha_{k_i}(t_0^\prime)=0$ and $-\tilde{\alpha}_{k_i}(t_0)=-1/\alpha_{k_i}(t_0)=c^{(2)}_{k_i}/c^{(1)}_{k_i}$. It also matches with the result of the section \ref{double-trace-relation}. We find the integration constant $\tau_n$ from the stochastic side is proportional to marginal n-multiple trace coupling $\bar{\sigma}_n$.
\subsection{$(2n-2)$-multiple trace deformation and stochastic $(2n-2)$-point function}
\begin{align}
\left.\frac{\delta^{2n-2}S_B}{\prod^{2n-2}_{i=1}\delta\Phi(k_i,r)}\right\rvert^{r=t}&=-\left\langle \prod^{2n-2}_{i=1}\Phi(k_i,t)\right\rangle_\textrm{S}\prod^{2n-2}_{j=1}\left\langle\Phi(k_j,t)\Phi(-k_j,t)\right\rangle^{-1}_\textrm{S}
\\ \nonumber
&+\frac{(2n-2)!n^2}{2(n!)^2}\prod^{2n-2}_{i=1}\left\langle\Phi(k_i,t)\Phi(-k_i,t)\right\rangle^{-1}_\textrm{S}\times\textrm{Perm}\left[\left\langle\left\{\prod^{n-1}_{j=1}\Phi(k_j,t)\right\}\Phi(q,t)\right\rangle_\textrm{S}\right.
\\ \nonumber
&\left.\times\left\langle\Phi(q,t)\Phi(-q,t)\right\rangle^{-1}_\textrm{S}\left\langle\left\{\prod^{2n-2}_{l=n}\Phi(k_l,t)\right\}\Phi(-q,t)\right\rangle_\textrm{S}\right]
\\ \nonumber
&-\frac{1}{2}\frac{\delta^{2n-2}S_E}{\prod^{2n-2}_{i=1}\delta\Phi(k_i,t)}.
\end{align}
We split the first term of the right-hand side as
\begin{align}
     \left\langle \prod^{2n-2}_{i=1} \Phi(k_i,t) \right\rangle=\sum^{3}_{i=1}\left\langle \prod^{2n-2}_{i=1} \Phi(k_i,t) \right\rangle^{[i]},
\end{align}
where the $i$ denotes the first, second, and third type of the $2n-2$ correlation function. First, we calculate the following quantity:
\begin{align}
\label{2n-2-relation-first}
    &\Bigg[\frac{n^2}{2(n!)^2}\textrm{Perm}\Bigg[\left\langle\left\{\prod^{n-1}_{j=1}\Phi(k_j,t)\right\}\Phi(q,t)\right\rangle_\textrm{S}
\left\langle\Phi(q,t)\Phi(-q,t)\right\rangle^{-1}_\textrm{S}\left\langle\left\{\prod^{2n-2}_{l=n}\Phi(k_l,t)\right\}\Phi(-q,t)\right\rangle_\textrm{S}\Bigg]
\\ \nonumber 
&-\frac{1}{(2n-2)!}\sum_{i=1}^2\langle \overbrace{\Phi_{k}(t) \cdots \Phi_{k}(t)}^{2n-2} \rangle^{[i]}\Bigg]\prod^{2n-2}_{i=1}\left\langle\Phi(k_i,t)\Phi(-k_i,t)\right\rangle^{-1}
 \\ \nonumber &=\frac{n^2}{8} \text{Perm}\Bigg[\tau^2_n \tilde{\mathcal{V}}_{2n-2}(k_1,\dots,k_{2n-2};t) \Bigg[ \frac{1}{2n-1}\prod^{2n-2}_{i=1}\frac{\tilde{\mathcal{Q}}_\nu(k_i,t)}{\mathcal{K}_\nu(k_i,t)}\frac{\tilde{\mathcal{Q}}_\nu(k,t)}{\mathcal{K}_\nu(k,t)}
    -\frac{1}{2n-1}\prod^{2n-2}_{i=1}\left(-\alpha_{k_i}(t_0)\right)\left(-\alpha_{k}(t_0)\right) 
    \\ \nonumber 
    &+\prod^{2n-2}_{i=n}\left(-\alpha_{k_i}(t_0)\right)\left(-\alpha_{k}(t_0)\right)^2 \left(\frac{\tilde{\mathcal{Q}}_\nu(k,t)}{\mathcal{K}_\nu(k,t)}\right) ^{-1}-\prod^{2n-2}_{i=1}\left(-\alpha_{k_i}(t_0)\right) \left(-\alpha_{k}(t_0)\right) \Bigg]\Bigg].
\end{align}
Next, we calculate the left terms in the relation:
\begin{align}
\label{2n-2-relation-second}
   & -\frac{1}{(2n-2)!}\langle \overbrace{\Phi_{k}(t) \cdots \Phi_{k}(t)}^{2n-2} \rangle^{[3]}\prod^{2n-2}_{i=1}\left\langle\Phi(k_i,t)\Phi(-k_i,t)\right\rangle^{-1} -\frac{1}{2}\frac{1}{(2n-2)!}\frac{\delta^{2n-2}S_E}{\prod^{2n-2}_{i=1}\delta\Phi(k_i,t)}
   \\ \nonumber
 &=\frac{n^2}{8} \prod^{2n-2}_{j=1}\tau_{k_j}\text{Perm}\Bigg[ \tau^2_{k_1+k_2+\cdots+k_{n-1}} \tilde{\mathcal{V}}_{2n-2}(k_1,\dots,k_{2n-2};t)
\\ \nonumber
&\times \Bigg[-\frac{1}{2n-1}\prod^{2n-2}_{i=1}\frac{\tilde{\mathcal{Q}}_\nu(k_i,t)}{\mathcal{K}_\nu(k_i,t)}\frac{\tilde{\mathcal{Q}}_\nu(k,t)}{\mathcal{K}_\nu(k,t)}
+\frac{1}{2n-1}\prod^{2n-2}_{i=1}[\alpha_{k_i}(t''_0)-\alpha_{k_i}(t_0)][\alpha_{k_i}(t''_0)-\alpha_k(t_0)]
\Bigg]\Bigg]
    \\ \nonumber 
    &+\frac{\bar{\lambda}_{2n-2}}{2n-2}\tilde{\mathcal{V}}_{2n-2}(k_1,\dots,k_{2n-2};t)\left[\int^t_0 {t^\prime}^{(n-2)(d-1)-2}\prod^{2n-2}_{i=1}\mathcal{Q}_\nu( k_i, t^\prime) dt^\prime\right]
    \\ \nonumber
    &-\frac{\tau_{2n-2}}{2}\left[\prod^{2n-2}_{i=1}\left(\alpha_{k_i}(t''_0)-\alpha_{k_i}(t_0)\right)\tilde{\mathcal{V}}_{2n-2}(k_1,\dots,k_{2n-2};t)\right].
\end{align}
To construct $(2n-2)$-multiple deformation by the stochastic data, we should add up (\ref{2n-2-relation-first}) and (\ref{2n-2-relation-second}). Finally, we can compare $(2n-2)$-multiple trace deformation and stochastic $(2n-2)$-correlation function by 
\begin{align}
    &-\left\langle \prod^{2n-2}_{i=1}\frac{1}{(2n-2)!}\Phi(k_i,t)\right\rangle_\textrm{S}\prod^{2n-2}_{j=1}\left\langle\Phi(k_j,t)\Phi(-k_j,t)\right\rangle^{-1}_\textrm{S}
\\ \nonumber
&+\frac{n^2}{2(n!)^2}\prod^{2n-2}_{i=1}\left\langle\Phi(k_i,t)\Phi(-k_i,t)\right\rangle^{-1}_\textrm{S}\times\textrm{Perm}\left[\left\langle\left\{\prod^{n-1}_{j=1}\Phi(k_j,t)\right\}\Phi(q,t)\right\rangle_\textrm{S}\right.
\\ \nonumber
&\left.\times\left\langle\Phi(q,t)\Phi(-q,t)\right\rangle^{-1}_\textrm{S}\left\langle\left\{\prod^{2n-2}_{l=n}\Phi(k_l,t)\right\}\Phi(-q,t)\right\rangle_\textrm{S}\right]
-\frac{1}{2}\frac{1}{(2n-2)!}\frac{\delta^{2n-2}S_E}{\prod^{2n-2}_{i=1}\delta\Phi(k_i,t)}
\\ \nonumber
    &= \prod^{2n-2}_{j=1}\tau_{k_j}\tilde{\mathcal{V}}'_{2n-2}(k_1,\dots,k_{2n-2};t) \Bigg[-\frac{n^2}{8}\text{Perm}\left[ \tau_{k_1+k_2+\cdots+k_{n-1}}^2\ddfrac{\alpha_{k}(t_0) I_\nu(\vert k\vert t) }{I_\nu(\vert k\vert t)-\alpha_k(t_0)K_\nu(\vert k\vert t)}\right]
    \\ \nonumber 
    &+\frac{\bar{\lambda}_{2n-2}}{2n-2}\left[\int^t_0 {t^\prime}^{(n-2)(d-1)-2}\prod^{2n-2}_{i=1}\tilde{\mathcal{Q}}'_\nu(k_i, t^\prime) dt^\prime\right]-\frac{\tau_{2n-2}}{2}\Bigg],
\end{align}
where we define
\begin{align}
\tilde{\mathcal{Q}}'_\nu(k,t)&\equiv t^{1/2}\left[K_\nu(\vert k\vert t)-\tilde{\alpha}_k(t_0)I_\nu(\vert k\vert t)\right] \text{,}\quad \tilde{\alpha}_k(t_0)\equiv 1/\alpha_k(t_0),
\end{align}
and 
\begin{align}
\tilde{\mathcal{V}}'_{2n-2}(k_1,\dots,k_{2n-2};t)\equiv \prod^{2n-2}_{i=1}\left[\tilde{\mathcal{Q}}'_\nu(k_i,t)\right]^{-1}.
\end{align}
From the $2$ and $n$-point relations, we know $\tilde{\mathcal{Q}}'_\nu( k,t)=\mathcal{Q}'_\nu( k, \epsilon)\vert^{\epsilon=t}$.
To match with holographic data (\ref{holographic-2n-2-trace-deformation}), we set $\alpha(t'_0)=\alpha(t''_0)=0$. 
The coupling constant in the stochastic frame should have the following relation with the marginal coupling constant in holographic data: 
\begin{align}
    \prod^{2n-2}_{i=1}\tau_{k_i}&=\prod^{2n-2}_{i=1}\left[\left(-2\bar{\sigma}_n\right)^{1/n}\left\{\frac{1}{2}\Gamma(\nu)\right\}\left(\frac{\vert k_i\vert}{2}\right)^{-\nu}\right]
    \\ \nonumber
    \tau_{k_1+k_2+\cdots+k_{n-1}}^2&=\tau_{-k}^2=\tau_k^2=\left[\left(-2\bar{\sigma}_n\right)^{1/n}\left\{\frac{1}{2}\Gamma(\nu)\right\}\left(\frac{\vert k_1+k_2+\cdots+k_{n-1}}{2}\right)^{-\nu}\right]^2
    \\ \nonumber
    \tau_{2n-2}&=2\bar{\sigma}_{2n-2}.
\end{align}
 We explain the detailed calculation in appendix \ref{2n-2-relation-appendix}. This result is consistent with the n-multiple trace case, implying that the Langevin dynamics of SQ can capture the holographic RG flow up to marginal deformation and self-interaction with some constraints.

\section{Outlook}
In this paper, we propose a mathematical relationship between holographic Wilsonian renormalization group (HWRG) and stochastic quantization (SQ).
This relationship is illustrated with an example of holographic model in which the bulk action is the most general scalar field theory with arbitrary mass, coupling $\mathcal{L}_{\text{int}}\sim\lambda_{2n-2}\phi^{2n-2}$ and marginal deformation on AdS boundary(by employing asymptotic AdS as the bulk spacetime). In weakly coupled gravitational theory, we can consider single-particle and multi-particle states of the scalar field corresponding to single-trace and multi-trace operators, respectively, in the dual field theory via holographic dictionary. In general, their scale dependence has been widely discussed in terms of the holographic Wilsonian renormalization group flow. In our formalism, we can relate the scale dependence of operators described by a radial flow in AdS spacetime to the fictitious time flow of the stochastic correlation function in the relaxation process of the statistical system.

By identifying the stochastic time $t$ with the radial coordinate $r$ in AdS, we have established a dictionary between the radial flow of multi-trace deformations and the fictitious time flow of stochastic multi-point correlation functions in the presence of the marginal deformation on the boundary in momentum space. In the consequences of the matching process, we also obtain the relationship between the coupling constant of the Euclidean field theory in the stochastic frame and the multi-trace coupling constant in holographic data. It can be interpreted more intuitively by introducing the Feynman-like diagrammatic expressions of the stochastic quantization scheme. We define the stochastic 2-point propagator, noise field, and $n$-point vertex to express the solution of the Langevin equation. It turns out that these ingredients can constitute the connected diagrams of the stochastic correlation functions. 

We expect these results to provide a deep understanding of the holographic renormalization group flow from the bottom-up theories. The $n$-point correlation functions at finite momentum in the context of AdS/CFT have rich applications for studying strongly coupled phenomena in quantum field theories such as fluid/gravity duality \cite{Bredberg:2010ky, Basu:2011tt, Park:2016wch, Park:2022oek} in the context of holographic renormalization group flow \cite{Faulkner:2010jy,Heemskerk:2010hk, Sin:2011yh}. For example, in \cite{Sin:2011yh, Iqbal:2008by}, the shear viscosity does not run along the radial direction but if we consider anisotropic superfluid, it produces non-trivial RG flows of the shear viscosities \cite{Basu:2009vv, Oh:2012zu}. The shear viscosity is obtained from (retarded) Green's function of the fluid dynamics and so one can study such a fluid system by looking at stochastic correlation functions and their fictitious time evolution. Also, the usefulness of the momentum space in AdS/CFT is revealed in \cite{Albayrak:2018tam, Albayrak:2019yve, Albayrak:2020isk, Albayrak:2020fyp}, where the authors show that the scalar factor that appears in the tree-level Witten diagrams, which become much simpler in momentum space. We suggest that the transition amplitude in holographic models might be related to the stochastic correlation function. In fact, the non-equal time stochastic 2-point function corresponds to Green's function in AdS space with the appropriate initial boundary condition. One can develop the stochastic dynamics to describe the multi-point amplitude in the holographic models

To the present day, the stochastic quantization scheme has tended to be regarded as a fictitious phenomenon, just a tool for understanding (Euclidean) quantum mechanics.
In this study, however, we suggest that the non-equilibrium phenomenon of stochastic quantization might represent the process of RG flow in the radial direction. This also implies that it can generate entropy production along a stochastic trajectory \cite{Udo}, losing the energy of the field $\phi$ to its surroundings by a dissipative force term in the Langevin equation. Once one defines the total entropy of the system as 
\begin{equation}
s_{tot}=s+s_{\rm surr},
\end{equation}
the expectation value of the $\frac{\partial s_{tot}}{\partial t}$ is positive-semi definite implying the monotonic behavior of the total entropy, $s_{\rm tot}$.

Finally, for possible future directions, we want to mention the following topics.
\begin{itemize}
    \item To achieve a rigorous connection with the holographic renormalization group, it would be helpful to investigate the stochastic description of the holographic c-theorem to find a physical observable. The monotonicity of the stochastic Gibbs entropy\cite{Udo} might imply that it can become a probable candidate for the holographic c-function.
    \item We also suggest that it might be fascinating to find an actual physical system described in terms of the stochastic fictitious time. The connection to the dS/CFT holography would be interesting since the de Sitter time could play the same role as stochastic time via our formulation of the relationship between them. It would enable us to calculate the non-trivial RG flow of the power spectrum in standard cosmology in the language of stochastic quantization. Moreover, the behavior of inflaton might be explained by the stochastic dynamics of the scalar fields.
    \item Finally, we want to comment that the non-equal time stochastic correlator 
coincides with Green's function in AdS. Therefore, it might be related to the Witten diagram since the non-equal time corresponds to the different points in the bulk space from $t=r$ identification, which would be useful to understand bulk-to-bulk propagators and bulk-to-boundary propagators.
\end{itemize}

\section*{Acknowledgement}
J.H.O thanks God and he also thanks his Woojin and Yun-Jae. J.H.O and J.S.C thank every organizer at the SGC2023 conference, 1st Han-Gang Gravity Workshop, and APCTP Winter School on Fundamental Physics 2024 for their hospitality. J.H.O and J.S.C also thank for much of the useful discussion with the participants in these workshops. J.H.O and J.S.C are especially grateful to Prof. Ki-seok Kim, Prof. Jeong-Hyuck Park, Prof. Chanyong Park, and Prof. Dileep P. Jatkar for useful discussion. This work was supported by the National Research Foundation of Korea(NRF) grant funded by the Korea government(MSIT).(No.2021R1F1A1047930).  Finally, the authors of this paper agree with that Ji-seong Chae is the first author and Jae-Hyuk Oh is the corresponding author of this paper.

\begin{appendices}
\section{Evaluation of the Stochastic $(2n-2)$-correlation Functions}
\label{2n-2-correlation-function}
\subsection{The first type diagram}
The first type of stochastic $2n-2$ correlation function is given by
\begin{align}
&\langle \overbrace{\Phi(t)\cdots\Phi(t)}^{2n-2}\rangle^{[1]}_\textrm{S}
=\frac{(2n-2)!}{2!(2n-4)!} \cdot \text{Perm}\left\{ \langle \overbrace{\Phi^{(2)}_{k_1} \cdots \Phi^{(2)}_{k_{2n-4}}(t)}^{2n-4} \Phi^{(n)}_{k_{2n-3}}(t)\Phi^{(n)}_{k_{2n-2}}(t)\rangle \right\}.
\end{align}
We first calculate the correlation function without permutation:
\begin{align}
&\frac{(2n-2)!}{2!(2n-4)!}\langle \overbrace{\Phi^{(2)}_{k_1} \cdots \Phi^{(2)}_{k_{2n-4}}(t)}^{2n-4} \Phi^{(n)}_{k_{2n-3}}(t)\Phi^{(n)}_{k_{2n-2}}(t)\rangle=\frac{(2n-2)!}{2!(2n-4)!}\Bigg\langle S_{2n-4}(k_1,\dots,k_{2n-4};t)
\\ \nonumber
&\times -\frac{n}{2}\int^t_{t^\prime_0}\left[ \mathcal{K}_\nu(k_{2n-3},t;\bar{t}_n) \tau_n\bar{\mathcal{V}}_n(k_{n-1},\dots,k_{2n-3},k_{n-1};\bar{t}_n)\int\prod^{n-1}_{i=1}d^d\bar{k}_i \ S_{n-1}(\bar{k}_1,\dots,\bar{k}_{n-1};\bar{t}_n)     \right]d\bar{t}_n
\\ \nonumber
&\times -\frac{n}{2}\int^t_{t^\prime_0}\left[\mathcal{K}_\nu(k_{2n-2},t;\bar{t}'_n) \tau_n \bar{\mathcal{V}}_n(k_{1},\dots,k_{n-1},k_{2n-2};\bar{t}'_n)\int \prod^{n-1}_{l=1} d^d\bar{k}^\prime_l \ S_{n-1}(\bar{k}^\prime_1,\dots,\bar{k}^\prime_{n-1}; \bar{t}^\prime_{n}) \right]d\bar{t}^\prime_n
\\ \nonumber
&\times \delta^{(d)}\left(\sum^{n-1}_{m=1} \bar{k}_m +k_{2n-3}\right) \delta^{(d)}\left(\sum^{n-1}_{m^\prime=1}\bar{k}^\prime_{m^\prime} +k_{2n-2}\right)\Bigg\rangle.
\end{align}
We separate terms related to the noise field $\eta$:
\begin{align}
&\Bigg\langle S_{2n-4}(k_1,\dots,k_{2n-4};t)
\int\prod^{n-1}_{i=1}d^d\bar{k}_i \ S_{n-1}(\bar{k}_1,\dots,\bar{k}_{n-1};\bar{t}_n)
\int \prod^{n-1}_{l=1} d^d\bar{k}^\prime_l \ S_{n-1}(\bar{k}^\prime_1,\dots,\bar{k}^\prime_{n-1};\bar{t}^\prime_n)
\Bigg\rangle_\textrm{S}^\textrm{Tree}
\\ \nonumber
&= \Bigg\langle S_{2n-4}(k_1,\dots,k_{2n-4};t)
\ \int\prod^{n-2}_{i=1}d^d\bar{k}_i \ S_{n-2}(\bar{k}_1,\dots,\bar{k}_{n-2};\bar{t}_n)
\\ \nonumber
&\times \int d^d\bar{k}_{n-1}\left\{ \int^{\bar{t}_n}_{t_0} \mathcal{K}_\nu(\bar{k}_{n-1},\bar{t}_n;\bar{t}_{n-1})\eta_{k_{n-1}}(\bar{t}_{n-1})d\bar{t}_{n-1}\right\}
\\ \nonumber
&\times \int \prod^{n-2}_{l=1} d^d\bar{k^\prime}_lS_{n-2}(\bar{k}^\prime_1,\dots,\bar{k}^\prime_{n-2};\bar{t}^\prime_n)\
 \int d^d\bar{k^\prime}_{n-1}\left\{\int^{\bar{t^\prime_n}}_{t_0} \mathcal{K}_\nu(\bar{k}_{n-1}',\bar{t}_n';\bar{t}_{n-1}')\eta_{\bar{k^\prime}_{n-1}}(\bar{t^\prime}_{n-1}) d\bar{t^\prime}_{n-1}\right\}
\Bigg\rangle_\textrm{S}^\textrm{Tree}.
\end{align}
Note that the Markov property of the noise field $\eta$ results in the correlation function of the noise fields as a delta function given by
\begin{align}
&=(n-1)^2 \int d^d\bar{k}_{n-1}\left\{ \int^{\bar{t}_n}_{t_0} \mathcal{K}_\nu(\bar{k}_{n-1},\bar{t}_n;\bar{t}_{n-1}) d\bar{t}_{n-1}\right\}
 \int d^d\bar{k}^\prime_{n-1}\left\{\int^{\bar{t}^\prime_n}_{t_0} \mathcal{K}_\nu(\bar{k}_{n-1}',\bar{t}_n';\bar{t}_{n-1}') d\bar{t}^\prime_{n-1}\right\}
 \\ \nonumber
&\times
 \delta(\bar{t}_{n-1}-\bar{t^\prime}_{n-1})\delta^{(d)}(\bar{k}_{n-1}-\bar{k^\prime}_{n-1})
\\ \nonumber
&\times (2n-4)!\prod^{2n-4}_{j=n-1}\int^t_{t_0}\mathcal{K}_\nu(k_j,t;t'_j) dt^\prime_j
\prod^{n-2}_{i=1}\int d^d\bar{k}_i \int^{\bar{t}_n}_{t_0} \mathcal{K}_\nu(\bar{k}_i,\bar{t}_n;\bar{t}_i) d\bar{t}_i \prod^{n-2}_{r=1} \delta(t^\prime_{r+n-2}-\bar{t}_r)\delta^{(d)}(k_{r+n-2}-\bar{k}_r)
\\ \nonumber
&\times \prod^{n-2}_{l=1}\int^t_{t_0} \mathcal{K}_\nu(k_l,t;\bar{t}_l'') d{\bar{t}_l''}
\prod^{n-2}_{l=1} \int d^d \bar{k}_l' \int^{\bar{t^\prime}_n}_{t_0} \mathcal{K}_\nu(\bar{k}_l',\bar{t}_n';\bar{t}_l') d\bar{t}_l' \prod^{n-2}_{r^\prime=1} \delta({\bar{t''}_{r^\prime}}-\bar{t}^\prime_{r^\prime})\delta^{(d)}(k_r-\bar{k}^\prime_{r^\prime}).
\end{align}
We consider the connected diagram up to tree level. The factor of a number of all possible contractions of noise fields comes out which means that all the possible numbers of connected diagrams give the same contribution. If we identify the integration variables as a consequence of the delta function, we obtain
\begin{align}
&=\frac{(2n-2)!}{2!}\frac{n^2(n-1)^2}{4}
\\ \nonumber &\times \int^t_{t^\prime_0}\left[ \mathcal{K}_\nu(k_{2n-3},t;\bar{t}_n) \tau_n\bar{\mathcal{V}}_n(k_{n-1},\dots,k_{2n-3},k_{n-1};\bar{t}_n)\prod^{2n-4}_{i=n-1} \mathcal{K}_\nu(k_i,t) \mathcal{K}_\nu(k_i,\bar{t}_n)\frac{\tilde{\mathcal{Q}}_\nu(k_i,\bar{t}_n)}{\mathcal{K}_\nu(k_i,\bar{t}_n)}\right]d\bar{t}_n
\\ \nonumber
&\times \int^t_{t^\prime_0}\left[\mathcal{K}_\nu(k_{2n-2},t;\bar{t}'_n) \tau_n \bar{\mathcal{V}}_n(k_{1},\dots,k_{n-1},k_{2n-2};\bar{t}'_n)\prod^{n-2}_{j=1} \mathcal{K}_\nu(k_j,t) \mathcal{K}_\nu(k_j,\bar{t}_n')\frac{\tilde{\mathcal{Q}}_\nu(k_j,\bar{t}'_n)}{\mathcal{K}_\nu(k_j,\bar{t}'_n)} \right]d\bar{t}^\prime_n
 \\ \nonumber 
 &\times \left[\mathcal{O}(\bar{t}_n-\bar{t^\prime}_n)\int^{\bar{t}_n'}_{t_0}\frac{\mathcal{K}_\nu(k_{n-1},\bar{t}_n) \mathcal{K}_\nu(k_{n-1},\bar{t}_n')}{\mathcal{K}^2_\nu(k_{n-1},\bar{t}_{n-1}')} d\bar{t}_{n-1}' +(1-\mathcal{O}(\bar{t}_n-\bar{t^\prime}_n))\int^{\bar{t}_n}_{t_0}\frac{\mathcal{K}_\nu(k_{n-1},\bar{t}_n) \mathcal{K}_\nu(k_{n-1},\bar{t}_n')}{\mathcal{K}^2_\nu(k_{n-1},\bar{t}_{n-1})}d\bar{t}_{n-1}\right],
\end{align} 
Note that we need to consider the order of stochastic time to determine the order of integration and integration range, identifying the integration $\int^{\bar{t}_n}_{t_0}d\bar{t}_{n-1}$ and $\int^{\bar{t}'_n}_{t_0} d\bar{t}'_{n-1}$ by the delta function. There are two different cases: (\romannumeral 1) $\bar{t}_n > \bar{t}'_n$ and (\romannumeral 2) $\bar{t}_n > \bar{t}'_n$, and each case gives different value of the step function. For the first case, we need to integrate $\int^{\bar{t}_n}_{t_0}d\bar{t}_{n-1}$ in advance, otherwise, for the second case, $\int^{\bar{t}'_n}_{t_0} d\bar{t}'_{n-1}$. We simplify the terms and evaluate the integration as
\begin{align}
&=\frac{(2n-2)!}{2!}\frac{n^2(n-1)^2}{4}\prod^{2n-2}_{i=1}\mathcal{K}_\nu(k_i,t)\Bigg[
\\ \nonumber &\times \int^t_{t^\prime_0}\left[ \frac{1}{\mathcal{K}^2_\nu(k_{2n-3},\bar{t}_n)} \tau_n\prod^{2n-4}_{i=n-1}  \frac{\tilde{\mathcal{Q}}_\nu(k_i,\bar{t}_n)}{\mathcal{K}_\nu(k_i,\bar{t}_n)}\right]d\bar{t}_n\int^{\bar{t}_n}_{t^\prime_0}\left[ \frac{1}{\mathcal{K}^2_\nu(k_{2n-2},\bar{t}'_n)} \tau_n  \prod^{n-1}_{j=1} \frac{\tilde{\mathcal{Q}}_\nu(k_j,\bar{t}'_n)}{\mathcal{K}_\nu(k_j,\bar{t}'_n)} \right]d\bar{t}^\prime_n
\\ \nonumber
&+ \int^{\bar{t}_n'}_{t^\prime_0}\left[ \frac{1}{\mathcal{K}^2_\nu(k_{2n-3},\bar{t}_n)} \tau_n\prod^{2n-4}_{i=n-1}  \frac{\tilde{\mathcal{Q}}_\nu(k_i,\bar{t}_n)}{\mathcal{K}_\nu(k_i,\bar{t}_n)}\frac{\tilde{\mathcal{Q}}_\nu(k_{n-1},\bar{t}_n)}{\mathcal{K}_\nu(k_{n-1},\bar{t}_n)}\right]d\bar{t}_n\int^{t}_{t^\prime_0}\left[ \frac{1}{\mathcal{K}^2_\nu(k_{2n-2},\bar{t}'_n)} \tau_n  \prod^{n-2}_{j=1} \frac{\tilde{\mathcal{Q}}_\nu(k_j,\bar{t}'_n)}{\mathcal{K}_\nu(k_j,\bar{t}'_n)} \right]d\bar{t}^\prime_n\Bigg].
\end{align} 
Evaluate the first integration: 
\begin{align}
&=\frac{(2n-2)!}{2!}\frac{n(n-1)^2}{4}\prod^{2n-2}_{i=1}\mathcal{K}_\nu(k_i,t)\Bigg[ \int^t_{t^\prime_0}\left[ \frac{1}{\mathcal{K}^2_\nu(k_{2n-3},\bar{t}_n)} \tau_n\prod^{2n-4}_{i=n-1}  \frac{\tilde{\mathcal{Q}}_\nu(k_i,\bar{t}_n)}{\mathcal{K}_\nu(k_i,\bar{t}_n)}\right] 
\\ \nonumber &\times
\tau_n \Bigg[ \prod^{n-1}_{j=1} \frac{\tilde{\mathcal{Q}}_\nu(k_j,\bar{t}_n)}{\mathcal{K}_\nu(k_j,\bar{t}_n)} \frac{\tilde{\mathcal{Q}}_\nu(k_{2n-2},\bar{t}_n)}{\mathcal{K}_\nu(k_{2n-2},\bar{t}_n)} - \prod^{n-1}_{j=1}\left[\alpha_{k_j}(t'_0)-\alpha_{k_j}(t_0)\right]\left[\alpha_{k_{2n-2}}(t'_0)-\alpha_{k_{2n-2}}(t_0)\right]\Bigg]d\bar{t}_n
\\ \nonumber &+
\int^{t}_{t^\prime_0}\left[ \frac{1}{\mathcal{K}^2_\nu(k_{2n-2},\bar{t}'_n)} \tau_n  \prod^{n-2}_{j=1} \frac{\tilde{\mathcal{Q}}_\nu(k_j,\bar{t}'_n)}{\mathcal{K}_\nu(k_j,\bar{t}'_n)} \right]
\\ \nonumber
&\times \left[\tau_n \prod^{2n-3}_{i=n-1}  \frac{\tilde{\mathcal{Q}}_\nu(k_i,\bar{t}'_n)}{\mathcal{K}_\nu(k_i,\bar{t}'_n)}\frac{\tilde{\mathcal{Q}}_\nu(k_{n-1},\bar{t}'_n)}{\mathcal{K}_\nu(k_{n-1},\bar{t}'_n)} - \prod^{2n-3}_{i=n-1}\left[\alpha_{k_i}(t'_0)-\alpha_{k_i}(t_0)\right]\left[\alpha_{k_{n-1}}(t'_0)-\alpha_{k_{n-1}}(t_0)\right]\right]d\bar{t}^\prime_n\Bigg].
\end{align} 
We decide the order of the integration by the subset relations of the integration range which are $(t'_0,t) \supset (t'_0, \bar{t}_n)$ and $(t'_0,t) \supset (t'_0, \bar{t}'_n)$. If we expand the second and fourth lines, integrations become solvable:
\begin{align}
&=\frac{(2n-2)!}{2!}\frac{n(n-1)^2}{4}\prod^{2n-2}_{i=1}\mathcal{K}_\nu(k_i,t)\Bigg[ \int^t_{t^\prime_0}\left[ \frac{1}{\mathcal{K}^2_\nu(k_{2n-3},\bar{t}_n)} \tau_n\prod^{2n-4}_{i=1}  \frac{\tilde{\mathcal{Q}}_\nu(k_i,\bar{t}_n)}{\mathcal{K}_\nu(k_i,\bar{t}_n)}\frac{\tilde{\mathcal{Q}}_\nu(k_{2n-2},\bar{t}_n)}{\mathcal{K}_\nu(k_{2n-2},\bar{t}_n)}\frac{\tilde{\mathcal{Q}}_\nu(k_{n-1},\bar{t}_n)}{\mathcal{K}_\nu(k_{n-1},\bar{t}_n)}\right] d\bar{t}_n
\\ \nonumber
&- \int^t_{t^\prime_0}\left[ \frac{1}{\mathcal{K}^2_\nu(k_{2n-3},\bar{t}_n)} \tau_n\prod^{2n-4}_{i=n-1}  \frac{\tilde{\mathcal{Q}}_\nu(k_i,\bar{t}_n)}{\mathcal{K}_\nu(k_i,\bar{t}_n)}\right]\prod^{n-1}_{j=1}\left[\alpha_{k_j}(t'_0)-\alpha_{k_j}(t_0)\right]\left[\alpha_{k_{2n-2}}(t'_0)-\alpha_{k_{2n-2}}(t_0)\right]d\bar{t}_n
\\ \nonumber &+
\int^{t}_{t^\prime_0}\left[ \frac{1}{\mathcal{K}^2_\nu(k_{2n-2},\bar{t}'_n)} \tau_n  \prod^{2n-3}_{j=1} \frac{\tilde{\mathcal{Q}}_\nu(k_j,\bar{t}'_n)}{\mathcal{K}_\nu(k_j,\bar{t}'_n)}\frac{\tilde{\mathcal{Q}}_\nu(k_{n-1},\bar{t}'_n)}{\mathcal{K}_\nu(k_{n-1},\bar{t}'_n)} \right]d\bar{t}_n'
\\ \nonumber
&-\int^{t}_{t^\prime_0}\left[ \frac{1}{\mathcal{K}^2_\nu(k_{2n-2},\bar{t}'_n)} \tau_n  \prod^{n-2}_{j=1} \frac{\tilde{\mathcal{Q}}_\nu(k_j,\bar{t}'_n)}{\mathcal{K}_\nu(k_j,\bar{t}'_n)} \right] \prod^{2n-3}_{i=n-1}\left[\alpha_{k_i}(t'_0)-\alpha_{k_i}(t_0)\right]\left[\alpha_{k_{n-1}}(t'_0)-\alpha_{k_{n-1}}(t_0)\right]d\bar{t}^\prime_n\Bigg],
\end{align} 
where $\bar{k}_{n-1}=-\sum^{2n-3}_{i=n-1}k_i=-\sum^{n-2}_{j=1}k_j-k_{2n-2}=k_{n-1}$. If we evaluate the final integration and restore the permutation, the first kind diagram of the $(2n-2)$-point function is given by
\begin{align}
\label{2n-2-first-correlation}
    & \langle \overbrace{\Phi_{k}(t) \cdots \Phi_{k}(t)}^{2n-2} \rangle^{[1]} 
    \\ \nonumber &=\frac{(2n-2)!(n-1)^2n}{2! \cdot 4} \text{Perm}\Bigg[\prod^{2n-2}_{i=1}\tau^2_n\left\{\mathcal{K}_\nu(k_i,t)\right\} 
    \\ \nonumber
    &\times\Bigg[ \frac{2}{2n-1}\prod^{2n-2}_{i=1}\frac{\tilde{\mathcal{Q}}_\nu(k_i,t)}{\mathcal{K}_\nu(k_i,t)}\frac{\tilde{\mathcal{Q}}_\nu(k_{n-1},t)}{\mathcal{K}_\nu(k_{n-1},t)} - \frac{2}{2n-1}\prod^{2n-2}_{i=1}\left(\alpha(t^\prime_0)-\alpha(t_0)\right)\left(\alpha_{k_{n-1}}(t^\prime_0)-\alpha_{k_{n-1}}(t_0)\right) 
    \\ \nonumber 
    &-\frac{1}{n-1}\prod^{2n-3}_{i=n-1}\frac{\tilde{\mathcal{Q}}_\nu(k_i,t)}{\mathcal{K}_\nu(k_i,t)}\prod^{n-2}_{i=1}\left(\alpha(t^\prime_0)-\alpha(t_0)\right) \left(\alpha_{k_{n-1}}(t^\prime_0)-\alpha_{k_{n-1}}(t_0)\right) \left(\alpha_{k_{2n-2}}(t^\prime_0)-\alpha_{k_{2n-2}}(t_0)\right) 
    \\ \nonumber
    &-\frac{1}{n-1}\prod^{n-2}_{i=1}\frac{\tilde{\mathcal{Q}}_\nu(k_i,t)}{\mathcal{K}_\nu(k_i,t)}\frac{\tilde{\mathcal{Q}}_\nu(k_{2n-2},t)}{\mathcal{K}_\nu(k_{2n-2},t)}\prod^{2n-3}_{i=n-1}\left(\alpha(t^\prime_0)-\alpha(t_0)\right) \left(\alpha_{k_{n-1}}(t^\prime_0)-\alpha_{k_{n-1}}(t_0)\right) 
    \\ \nonumber
    &+ \frac{2}{n-1}\prod^{2n-2}_{i=1}\left(\alpha(t^\prime_0)-\alpha(t_0)\right) \left(\alpha_{k_{n-1}}(t^\prime_0)-\alpha_{k_{n-1}}(t_0)\right) \Bigg].
\end{align}
Furthermore, we unify the same terms under all possible permutations. This gives a simple result of the first type of the $(2n-2)$- stochastic correlation function given by
\begin{align}
    & \langle \overbrace{\Phi_{k}(t) \cdots \Phi_{k}(t)}^{2n-2} \rangle^{[1]} 
    \\ \nonumber &=\frac{(2n-2)!(n-1)^2n}{4} \text{Perm}\Bigg[\prod^{2n-2}_{i=1}\tau^2_n\left\{\mathcal{K}_\nu(k_i,t)\right\} 
    \\ \nonumber
    &\times\Bigg[ \frac{1}{2n-1}\prod^{2n-2}_{i=1}\frac{\tilde{\mathcal{Q}}_\nu(k_i,t)}{\mathcal{K}_\nu(k_i,t)}\frac{\tilde{\mathcal{Q}}_\nu(k_{n},t)}{\mathcal{K}_\nu(k_{n},t)} - \frac{1}{2n-1}\prod^{2n-2}_{i=1}\left(\alpha(t^\prime_0)-\alpha(t_0)\right)\left(\alpha_{k_{n}}(t^\prime_0)-\alpha_{k_{n}}(t_0)\right) 
    \\ \nonumber
    &-\frac{1}{n-1}\prod^{2n-2}_{i=n}\frac{\tilde{\mathcal{Q}}_\nu(k_i,t)}{\mathcal{K}_\nu(k_i,t)}\prod^{n}_{i=1}\left(\alpha(t^\prime_0)-\alpha(t_0)\right)
    + \frac{1}{n-1}\prod^{2n-2}_{i=1}\left(\alpha(t^\prime_0)-\alpha(t_0)\right) \left(\alpha_{k_{n}}(t^\prime_0)-\alpha_{k_{n}}(t_0)\right) \Bigg].
\end{align}


\subsection{The second type diagram}
The integral expression of the second type of the $(2n-2)$-stochastic correlation function is given by
\begin{align}
&\langle \overbrace{\Phi(t)\cdots\Phi(t)}^{2n-2}\rangle^{[2]}_\textrm{S}
=(2n-2)\cdot \text{Perm}\left\{ \left\langle \overbrace{\Phi^{(2)}_{k_1} \cdots \Phi^{(2)}_{k_{2n-3}}(t)}^{2n-3} \Phi^{(2n-2,1)}_{k_{2n-2}}(t)\right\rangle \right\}.
\end{align}
One can calculate the second type diagram with a similar progression to the first type diagram. We consider the correlation function without permutation as
\begin{align}
    &=(2n-2) \Bigg\langle S_{2n-3}(k_1,\dots,k_{2n-3};t) \frac{n^2(n-1)}{4}\int^t_{t_0''} \Bigg[\mathcal{K}_\nu(k_{2n-2},t;t')\text{Perm}\Bigg\{\int\prod^{n-2}_{l=1}d^d\tilde{k}_l \ S_{n-2}(\tilde{k}_1,\dots,\tilde{k}_{n-2};t^\prime)
    \\ \nonumber
    &\times d^d\tilde{k}_{n-1}\left[\int^{t^\prime}_{t^\prime_0}\mathcal{K}_\nu(\tilde{k}_{n-1},t';t'')\tau_n\bar{\mathcal{V}}_n(k_1,\dots,k_n;t'')\int \prod^{n-1}_{m=1} d^d\bar{k}_m \ S_{n-1}(\bar{k}_1,\dots,\bar{k}_{n-1};t'')\right] dt'' \Bigg\}
    \\ \nonumber
    &\times \tau_n\bar{\mathcal{V}}_n(k_n,\dots,k_{2n-2},k_n;t')\Bigg]dt^\prime \delta^{(d)}\left(\sum^{n-1}_{p=1}\bar{k}_p-\tilde{k}_{n-1}\right)\delta^{(d)}\left(\sum^{n-1}_{q=1}k_q-k_{2n-2}\right).
\end{align}
We separate the terms related to the noise field given by
\begin{align}
    &\left\langle  S_{2n-3}(k_1,\dots,k_{2n-3};t)\int\prod^{n-2}_{l=1}d^d\tilde{k}_l \ S_{n-2}(\tilde{k}_1,\dots,\tilde{k}_{n-2};t^\prime)\int \prod^{n-1}_{m=1} d^d\bar{k}_m \ S_{n-1}(\bar{k}_1,\dots,\bar{k}_{n-1};t'')\right\rangle
    \\ \nonumber
    &= (2n-3)!\prod^{n-1}_{j=1}\prod^{2n-3}_{s=n}\int^t_{t_0} \mathcal{K}_\nu(k_j,t;t'_j)\mathcal{K}_\nu(k_s,t;\bar{t}_s') dt^\prime_j d\bar{t}^\prime_s\int \prod^{n-2}_{l=1} d^d\tilde{k}_l 
    \left[\int^{t^\prime}_{t_0}\mathcal{K}_\nu(\tilde{k}_l,t';\tilde{t}_l)d\tilde{t}_l \right]
    \\ \nonumber
        &\times \int\prod^{n-1}_{m=1} d^d\bar{k}_m \left[\int^{t''}_{t_0}\mathcal{K}_\nu(\bar{k}_m,t'';\bar{t}_m)d\bar{t}_m\right]\prod^{n-1}_{r=1}\delta(t^\prime_r-\bar{t}_{r})\delta^{(d)}(k_{r}-\bar{k}_r)
        \prod^{n-2}_{r^\prime=1}\delta(\bar{t}^\prime_{r^\prime+n-1}-\tilde{t}_{r^\prime})\delta^{(d)}(k_{r^\prime+n-1}-\tilde{k}_{r^\prime})
        \\ \nonumber
        &=(2n-3)!\prod^{n-1}_{i=1}\int^{t''}_{t_0}\frac{\mathcal{K}_\nu(k_i,t)\mathcal{K}_\nu(k_i,t'')}{\mathcal{K}^2_\nu(k_i,\bar{t}_j)}d\bar{t}_j\prod^{2n-3}_{j=n}\int^{t'}_{t_0}\frac{\mathcal{K}_\nu(k_j,t)\mathcal{K}_\nu(k_j,t')}{\mathcal{K}^2_\nu(k_j,\tilde{t}_j)}d\tilde{t}_j.
\end{align}
Then, the correlation function of the second type diagram without permutation is given by
\begin{align}
    &\langle \overbrace{\Phi(t)\cdots\Phi(t)}^{2n-2}\rangle^{[2]}_\textrm{S, without Perm}
    = (2n-2)! \frac{n^2(n-1)}{4}\tau^2_n\prod^{2n-2}_{i=1}\left[\mathcal{K}_\nu(k_i, t)\right]\int^t_{{t''_0}}\left[\frac{1}{\mathcal{K}^2_\nu( k_{2n-2}, t^\prime)}\tau_n\prod^{2n-3}_{l=n}\frac{\tilde{\mathcal{Q}}_\nu(k_l,t')}{\mathcal{K}_\nu(k_l,t')}\right]dt'
    \\ \nonumber 
    &\times \text{Perm}\left\{\int^{t^\prime}_{t^\prime_0}\left[\frac{1}{\mathcal{K}^2_\nu( k_{n}, t'')}\tau_n\prod^{n-1}_{m=1}\frac{\tilde{\mathcal{Q}}_\nu(k_m,t'')}{\mathcal{K}_\nu(k_m,t'')}\right]dt''\right\}.
\end{align}
We restore the permutation and perform the $\int^{t'}_{t_0'} dt''$ integration first as
\begin{align}
    &\langle \overbrace{\Phi(t)\cdots\Phi(t)}^{2n-2}\rangle^{[2]}_\textrm{S}= (2n-2)! \frac{n(n-1)}{4} \text{Perm}\Bigg[\tau^2_n\prod^{2n-2}_{i=1}\left[\mathcal{K}_\nu(k_i, t)\right]\int^t_{{t''_0}}\left[\frac{1}{\mathcal{K}^2_\nu( k_{2n-2}, t^\prime)}\tau_n\prod^{2n-3}_{l=n}\frac{\tilde{\mathcal{Q}}_\nu(k_l,t')}{\mathcal{K}_\nu(k_l,t')}\right]
    \\ \nonumber 
    &\times \left[\prod^{n}_{m=1}\frac{\tilde{\mathcal{Q}}_\nu(k_m,t')}{\mathcal{K}_\nu(k_m,t')}-\prod^{n}_{m=1}\left[\alpha(t^\prime_0)-\alpha(t_0)\right]\right]dt^\prime\Bigg].
\end{align}
    Finally, the complete second type of the $(2n-2)$-stochastic correlation function is given by
    \begin{align}
    &\langle \overbrace{\Phi(t)\cdots\Phi(t)}^{2n-2}\rangle^{[2]}_\textrm{S}
    \\ \nonumber
        &= (2n-2)! \frac{n(n-1)}{4} \text{Perm}\Bigg\{\tau^2_n\prod^{2n-2}_{i=1}\left[\mathcal{K}_\nu( k_i, t)\right]\Bigg[\Bigg[\frac{1}{2n-1}\prod^{2n-2}_{l=1}\frac{\tilde{\mathcal{Q}}_\nu(k_l,t)}{\mathcal{K}_\nu(k_l,t)}\frac{\tilde{\mathcal{Q}}_\nu(k_n,t)}{\mathcal{K}_\nu(k_n,t)}
        \\ \nonumber
        &-\frac{1}{2n-1}\prod^{2n-2}_{l=1}\left[\alpha_{k_l}(t''_0)-\alpha_{k_l}(t_0)\right]\left[\alpha_{k_n}(t''_0)-\alpha_{k_n}(t_0)\right]
    -\frac{1}{n-1}\prod^{2n-2}_{l=n}\frac{\tilde{\mathcal{Q}}_\nu(k_l,t)}{\mathcal{K}_\nu(k_l,t)}\prod^{n}_{m=1}\left[\alpha_{k_m}(t^\prime_0)-\alpha_{k_m}(t_0)\right]
   \\ \nonumber
   &+\frac{1}{n-1}\prod^{2n-2}_{l=n}\left[\alpha_{k_l}(t''_0)-\alpha_{k_l}(t_0)\right]\prod^{n}_{m=1}\left[\alpha(t^\prime_0)-\alpha(t_0)\right]\Bigg]\Bigg\}.
\end{align}

\subsection{The third type diagram}
The integral expression of the third type of the (2n-2)-point correlation function is given by
\begin{align}
 &\langle \overbrace{\Phi(t)\cdots\Phi(t)}^{2n-2}\rangle^{[3]}_\textrm{S}=(2n-2)\text{Perm}\left\{\left\langle\overbrace{\Phi^{(2)}_{k_1}\cdots\Phi^{(2)}_{k_{2n-3}}}^{2n-3}\Phi^{(2n-2,2)}_{k_{2n-2}}(t)\right\rangle\right\}. 
\end{align}
We calculate the correlation function without permutation:
\begin{align}
&(2n-2)\left\langle\overbrace{\Phi^{(2)}_{k_1}\cdots\Phi^{(2)}_{k_{2n-3}}}^{2n-3}\Phi^{(2n-2,2)}_{k_{2n-2}}(t)\right\rangle
\\ \nonumber 
&=(2n-2)(n-1)\Bigg\langle S_{2n-3}(k_1,\dots,k_{2n-3};t)\int^t_{t_0''}dt^\prime\Bigg[\mathcal{K}_\nu(k_{2n-2},t;t')\int\prod^{2n-3}_{i=1}d^d\hat{k} \ S_{2n-3}(\hat{k}_1,\dots,\hat{k}_{2n-3};t^\prime)
\\ \nonumber
&\times \bar{\mathcal{V}}_{2n-2}(k_1,\dots,k_{2n-2};t')\left[-\frac{n^2}{4}\prod^{2n-2}_{j=1}\tau_{k_j}\text{Perm}\left\{\tau^2_{k_1+\cdots+k_{n-1}}\frac{I_\nu(\left\vert -\sum^{n-1}_{i=1}k_i\right\vert t^\prime)}{K_\nu(\left\vert -\sum^{n-1}_{i=1}k_i\right\vert t^\prime)}\right\}+\frac{\bar{\lambda}_{2n-2}}{n-1}\bar{K}^{(2n-2)}_\nu(k,t^\prime)-\tau_{2n-2}\right]\Bigg]
\\ \nonumber
&\times \delta^{(d)}\left(\sum^{2n-3}_{j=1}k_j-k_{2n-2}\right)\Bigg\rangle.
\end{align}
We calculate the correlation functions of the noise fields given by
\begin{align}
&\Bigg\langle S_{2n-3}(k_1,\dots,k_{2n-3};t)\int\prod^{2n-3}_{i=1}d^d\hat{k} \ S_{2n-3}(\hat{k}_1,\dots,\hat{k}_{2n-3};t^\prime)\Bigg\rangle
\\ \nonumber 
&= \prod^{2n-3}_{j=1}\int^t_{t_0} \mathcal{K}_\nu(k_j,t;t_j') dt^\prime_j \int \prod^{2n-3}_{i=1}d^d\hat{k}_i\left[\int^{t^\prime}_{t_0} \mathcal{K}_\nu(\hat{k}_i,t';\hat{t}_i)d\hat{t}_i \right](2n-3)!\prod^{2n-3}_{r=1}\delta(t^\prime_r-\hat{t}_r)\delta^{(d)}(k_r-\hat{t}_r).
\end{align}
Then, the third type of $2n-2$-correlation function is given by
\begin{align}
&\langle \overbrace{\Phi(t)\cdots\Phi(t)}^{2n-2}\rangle^{[3]}_\textrm{S}
=(n-1)(2n-2)!\text{Perm}\Bigg\{\left[\mathcal{K}_\nu( k_j, t')\right]\int^t_{t_0''}\Bigg[ \frac{1}{\mathcal{K}^2_\nu( k_{2n-2}, t)}\prod^{2n-3}_{j=1}\frac{\tilde{\mathcal{Q}}_\nu(k_j,t')}{\mathcal{K}_\nu(k_j,t')}
\\ \nonumber
&\times \left[\prod^{2n-2}_{j=1}\tau_{k_j} \cdot \tau_{k}^2\left\{-\frac{n^2}{4} \left(\frac{I_\nu(\left\vert k \right\vert t^\prime)}{K_\nu(\left\vert k\right\vert t^\prime)}-\alpha_k(t_0)\right)-\frac{n^2}{4}\alpha_k(t_0)\right\}+\frac{\bar{\lambda}_{2n-2}}{n-1}\bar{K}^{(2n-2)}_\nu(k,t^\prime)-\tau_{2n-2}\right]\Bigg]dt^\prime\Bigg\},
\end{align}
where we use the momentum conservation $k_1+k_2+\cdots+k_{n-1}=-k$ and $\tau_k=\tau_{-k}$.
In the last line, we expand the terms and perform integration by part.
Then, we can simply obtain the third type of $(2n-2)$-correlation function given by 
\begin{align}
& \langle \overbrace{\Phi_{k}(t) \cdots \Phi_{k}(t)}^{2n-2} \rangle^{[3]}=(2n-2
)!(n-1) \text{Perm}\Bigg[ \prod^{2n-2}_{j=1}\left[\mathcal{K}_\nu( k_j, t)\right]\Bigg[\prod^{2n-2}_{j=1}\tau_{k_j} \cdot \tau_{k}^2
\\ \nonumber
&\times \frac{1}{2n-1}\Bigg[-\frac{n^2}{4}\left\{\prod^{2n-2}_{i=1}\frac{\tilde{\mathcal{Q}}_\nu(k_i,t)}{\mathcal{K}_\nu(k_i,t)}\frac{\tilde{\mathcal{Q}}_\nu(k,t)}{\mathcal{K}_\nu(k,t)}\right.
-\left.\prod^{2n-2}_{i=1}[\alpha_{k_i}(t''_0)-\alpha_{k_i}(t_0)][\alpha_{k}(t''_0)-\alpha_k(t_0)]\right\}\Bigg]
\\ \nonumber
&+\frac{1}{2n-2}\left[-\frac{n^2}{4}\alpha_k(t_0)\left\{\prod^{2n-2}_{i=1}\frac{\tilde{\mathcal{Q}}_\nu(k_i,t)}{\mathcal{K}_\nu(k_i,t)}-\prod^{2n-2}_{i=1}\left[\alpha_{k_i}(t''_0)-\alpha_{k_i}(t_0)\right]\right\} \right]\Bigg]
    \\ \nonumber 
    &-\frac{1}{2n-2}\Bigg[\prod^{2n-2}_{i=1}\left\{\tilde{\mathcal{Q}}_\nu(k_i,t)\right\}\frac{\bar{\lambda}_{2n-2}}{n-1}\bar{K}^{(2n-2)}_\nu(k,t)+\prod^{2n-2}_{i=1}\left\{\mathcal{K}_\nu(k_i,t)\right\}\frac{\bar{\lambda}_{2n-2}}{n-1}\bar{Q}_\nu^{(2n-2)}(k,t)
    \\ \nonumber
    &-\tau_{2n-2}\left[\prod^{2n-2}_{i=1}\tilde{\mathcal{Q}}_\nu(k_i,t)-\prod^{2n-2}_{i=1}\mathcal{K}_\nu(k_i,t)\left(\alpha_{k_i}(t''_0)-\alpha_{k_i}(t_0)\right)\right]\Bigg],
\end{align}
where we define the unsolvable integration term as
\begin{align}
    \bar{Q}_\nu^{(2n-2)}(k,t) \equiv \int^t_0 {t^\prime}^{(n-2)d-2}\prod^{2n-2}_{i=1}Q_\nu(\vert k_i\vert t^\prime) dt^\prime.
\end{align}
\end{appendices}
\begin{appendices}
    \section{Relation of $(2n-2)$-multiple Trace Deformation and the Stochastic $(2n-2)$-point Function}
    \label{2n-2-relation-appendix}
The stochastic description of the $(2n-2)$-multiple trace deformation is given by
    \begin{align}
\label{relation-2n-2point-2n-2trace-appendix}
\left.\frac{\delta^{2n-2}S_B}{\prod^{2n-2}_{i=1}\delta\Phi(k_i,r)}\right\rvert^{r=t}&=-\left\langle \prod^{2n-2}_{i=1}\Phi(k_i,t)\right\rangle_\textrm{S}\prod^{2n-2}_{j=1}\left\langle\Phi(k_j,t)\Phi(-k_j,t)\right\rangle^{-1}_\textrm{S}
\\ \nonumber
&+\frac{(2n-2)!n^2}{2(n!)^2}\prod^{2n-2}_{i=1}\left\langle\Phi(k_i,t)\Phi(-k_i,t)\right\rangle^{-1}_\textrm{S}\times\textrm{Perm}\left[\left\langle\left\{\prod^{n-1}_{j=1}\Phi(k_j,t)\right\}\Phi(q,t)\right\rangle_\textrm{S}\right.
\\ \nonumber
&\left.\times\left\langle\Phi(q,t)\Phi(-q,t)\right\rangle^{-1}_\textrm{S}\left\langle\left\{\prod^{2n-2}_{l=n}\Phi(k_l,t)\right\}\Phi(-q,t)\right\rangle_\textrm{S}\right]
\\ \nonumber
&-\frac{1}{2}\frac{\delta^{2n-2}S_E}{\prod^{2n-2}_{i=1}\delta\Phi(k_i,t)}.
\end{align}
For the simple calculation, we firstly calculate the second and third lines of (\ref{relation-2n-2point-2n-2trace-appendix}):
\begin{align}
\label{2n-2-relation-first-final}
     &\frac{n^2}{2(n!)^2}\prod^{2n-2}_{i=1}\left\langle\Phi(k_i,t)\Phi(-k_i,t)\right\rangle^{-1}_\textrm{S}\textrm{Perm}\Bigg[\left\langle\left\{\prod^{n-1}_{j=1}\Phi(k_j,t)\right\}\Phi(q,t)\right\rangle_\textrm{S}
\left\langle\Phi(q,t)\Phi(-q,t)\right\rangle^{-1}_\textrm{S}
\\ \nonumber &\times \left\langle\left\{\prod^{2n-2}_{l=n}\Phi(k_l,t)\right\}\Phi(-q,t)\right\rangle_\textrm{S}\Bigg]
\\ \nonumber &= \frac{n^2}{8} \tau_n^2 \tilde{\mathcal{V}}_{2n-2}(k_1,\dots,k_{2n-2};t) 
\\ \nonumber &\times \Bigg[\prod^{2n-2}_{i=1}[\frac{\tilde{\mathcal{Q}}_\nu(k_i,t)}{\mathcal{K}_\nu(k_i,t)}\frac{\tilde{\mathcal{Q}}_\nu(k,t)}{\mathcal{K}_\nu(k,t)}
-\prod^{n-1}_{i=1}[\alpha(t^\prime_0)-\alpha(t_0)][\alpha_k(t^\prime_0)-\alpha_k(t_0)]\prod^{2n-2}_{i=n}\frac{\tilde{\mathcal{Q}}_\nu(k_i,t)}{\mathcal{K}_\nu(k_i,t)}
\\ \nonumber 
&-\prod^{2n-2}_{i=n}[\alpha(t^\prime_0)-\alpha(t_0)][\alpha_k(t^\prime_0)-\alpha_k(t_0)]\prod^{n-1}_{i=1}\frac{\tilde{\mathcal{Q}}_\nu(k_i,t)}{\mathcal{K}_\nu(k_i,t)}
+ \prod^{2n-2}_{i=1}[\alpha(t^\prime_0)-\alpha(t_0)][\alpha_k(t^\prime_0)-\alpha_k(t_0)]^2\frac{\mathcal{K}_\nu(k,t)}{\tilde{\mathcal{Q}}_\nu(k,t)} \Bigg].
\end{align}
Next, we calculate the summation of the first and second types of $2n-2$-correlation function as
\begin{align}
& -\frac{1}{(2n-2)!}\sum_{i=1}^2\langle \overbrace{\Phi_{k}(t) \cdots \Phi_{k}(t)}^{2n-2} \rangle^{[i]}
    \\ \nonumber &=-\frac{n(n-1)}{4} \text{Perm}\Bigg[\prod^{2n-2}_{i=1}\tau^2_n\left\{\mathcal{K}_\nu(k_i,t)\right\} 
    \\ \nonumber
    &\times\Bigg[ \frac{n-1}{2n-1}\prod^{2n-2}_{i=1}\frac{\tilde{\mathcal{Q}}_\nu(k_i,t)}{\mathcal{K}_\nu(k_i,t)}\frac{\tilde{\mathcal{Q}}_\nu(k_{n},t)}{\mathcal{K}_\nu(k_{n},t)} - \frac{n-1}{2n-1}\prod^{2n-2}_{i=1}\left(\alpha(t^\prime_0)-\alpha(t_0)\right)\left(\alpha_{k_{n}}(t^\prime_0)-\alpha_{k_{n}}(t_0)\right) 
    \\ \nonumber
    &-\prod^{2n-2}_{i=n}\frac{\tilde{\mathcal{Q}}_\nu(k_i,t)}{\mathcal{K}_\nu(k_i,t)}\prod^{n}_{i=1}\left(\alpha(t^\prime_0)-\alpha(t_0)\right) 
    + \prod^{2n-2}_{i=1}\left(\alpha(t^\prime_0)-\alpha(t_0)\right) \left(\alpha_{k_{n}}(t^\prime_0)-\alpha_{k_{n}}(t_0)\right) \Bigg]
\\ \nonumber
        &-\frac{n(n-1)}{4} \text{Perm}\Bigg\{\tau^2_n\prod^{2n-2}_{i=1}\left[\mathcal{K}_\nu( k_i, t)\right]\Bigg[\Bigg[\frac{1}{2n-1}\prod^{2n-2}_{l=1}\frac{\tilde{\mathcal{Q}}_\nu(k_l,t)}{\mathcal{K}_\nu(k_l,t)}\frac{\tilde{\mathcal{Q}}_\nu(k_n,t)}{\mathcal{K}_\nu(k_n,t)}
        \\ \nonumber
        &-\frac{1}{2n-1}\prod^{2n-2}_{l=1}\left[\alpha_{k_l}(t''_0)-\alpha_{k_l}(t_0)\right]\left[\alpha_{k_n}(t''_0)-\alpha_{k_n}(t_0)\right]
    -\frac{1}{n-1}\prod^{2n-2}_{l=n}\frac{\tilde{\mathcal{Q}}_\nu(k_l,t)}{\mathcal{K}_\nu(k_l,t)}\prod^{n}_{m=1}\left[\alpha_{k_m}(t^\prime_0)-\alpha_{k_m}(t_0)\right]
   \\ \nonumber
   &+\frac{1}{n-1}\prod^{2n-2}_{l=n}\left[\alpha_{k_l}(t''_0)-\alpha_{k_l}(t_0)\right]\prod^{n}_{m=1}\left[\alpha(t^\prime_0)-\alpha(t_0)\right]\Bigg]\Bigg\}.
\end{align}
At this moment, we can not fix the quantities $\{\alpha_{k_i}(t_0')\}$ and $\{\alpha_{k_i}(t_0'')\}$ from the initial time condition of the integration, where $\{\cdot\}$ denotes the set of $\alpha_{k_i}$ for all moemntum labels. But, for a simple description, we bring the final results in advance, which is straightforward to derive: $\{\alpha_k(t_0')\}=\{\alpha_k(t_0'')\}=0$. Then, we obtain 
\begin{align}
\label{2n-2-first-and-second-correlation-function}
     &=-\frac{n^2}{8} \text{Perm}\Bigg\{\tau^2_n\prod^{2n-2}_{i=1}\left[\mathcal{K}_\nu( k_i, t)\right]\Bigg[\Bigg[\frac{2n-2}{2n-1}\prod^{2n-2}_{l=1}\frac{\tilde{\mathcal{Q}}_\nu(k_l,t)}{\mathcal{K}_\nu(k_l,t)}\frac{\tilde{\mathcal{Q}}_\nu(k_n,t)}{\mathcal{K}_\nu(k_n,t)}
        \\ \nonumber
        &-\frac{2n-2}{2n-1}\prod^{2n-2}_{l=1}\left[-\alpha_{k_l}(t_0)\right]\left[-\alpha_{k_n}(t_0)\right]
    -2\prod^{2n-2}_{l=n}\frac{\tilde{\mathcal{Q}}_\nu(k_l,t)}{\mathcal{K}_\nu(k_l,t)}\prod^{n}_{m=1}\left[-\alpha_{k_m}(t_0)\right]+2\prod^{2n-2}_{l=n}\left[-\alpha_{k_l}(t_0)\right]\prod^{n}_{m=1}\left[-\alpha_{k_m}(t_0)\right]\Bigg]\Bigg\}.
\end{align}
Using the results of (\ref{2n-2-relation-first-final}) and (\ref{2n-2-first-and-second-correlation-function}), we can get the following quantity:
\begin{align}
\label{2n-2-relation-first-and-second-part}
    &\Bigg[\frac{n^2}{2(n!)^2}\textrm{Perm}\Bigg[\left\langle\left\{\prod^{n-1}_{j=1}\Phi(k_j,t)\right\}\Phi(q,t)\right\rangle_\textrm{S}
\left\langle\Phi(q,t)\Phi(-q,t)\right\rangle^{-1}_\textrm{S}\left\langle\left\{\prod^{2n-2}_{l=n}\Phi(k_l,t)\right\}\Phi(-q,t)\right\rangle_\textrm{S}\Bigg]
\\ \nonumber 
&-\frac{1}{(2n-2)!}\sum_{i=1}^2\langle \overbrace{\Phi_{k}(t) \cdots \Phi_{k}(t)}^{2n-2} \rangle^{[i]}\Bigg]\prod^{2n-2}_{i=1}\left\langle\Phi(k_i,t)\Phi(-k_i,t)\right\rangle^{-1}
 \\ \nonumber &=\frac{n^2}{8} \text{Perm}\Bigg[\tau^2_n \tilde{\mathcal{V}}_{2n-2}(k_1,\dots,k_{2n-2};t) \Bigg[ \frac{1}{2n-1}\prod^{2n-2}_{i=1}\frac{\tilde{\mathcal{Q}}_\nu(k_i,t)}{\mathcal{K}_\nu(k_i,t)}\frac{\tilde{\mathcal{Q}}_\nu(k,t)}{\mathcal{K}_\nu(k,t)}
    -\frac{1}{2n-1}\prod^{2n-2}_{i=1}\left(-\alpha_{k_i}(t_0)\right)\left(-\alpha_{k}(t_0)\right) 
    \\ \nonumber 
    &+\prod^{2n-2}_{i=n}\left(-\alpha_{k_i}(t_0)\right)\left(-\alpha_{k}(t_0)\right)^2 \left(\frac{\tilde{\mathcal{Q}}_\nu(k,t)}{\mathcal{K}_\nu(k,t)}\right) ^{-1}-\prod^{2n-2}_{i=1}\left(-\alpha_{k_i}(t_0)\right) \left(-\alpha_{k}(t_0)\right) \Bigg]\Bigg],
\end{align}
where $\vert k_n\vert =\vert k_{2n-2}\vert =\vert k \vert$.

Afterwards, we calculate the terms related to the third type of the $(2n-2)$-correlation function. We calculate the following quantity as
\begin{align}
& -\frac{1}{(2n-2)!}\langle \overbrace{\Phi_{k}(t) \cdots \Phi_{k}(t)}^{2n-2} \rangle^{[3]}\prod^{2n-2}_{i=1}\left\langle\Phi(k_i,t)\Phi(-k_i,t)\right\rangle^{-1}
     \\ \nonumber &=\frac{n^2}{8}\text{Perm}\Bigg[ \prod^{2n-2}_{j=1}\tau_{k_j} \cdot \tau_{k}^2 \ \bar{\mathcal{V}}_{2n-2}(k_1,\dots,k_{2n-2};t)\Bigg[
\\ \nonumber
&\times \frac{2n-2}{2n-1}\left\{\frac{\tilde{\mathcal{Q}}_\nu(k,t)}{\mathcal{K}_\nu(k,t)}\right.
-\left.\prod^{2n-2}_{i=1}\frac{\mathcal{K}_\nu(k_i,t)}{\tilde{\mathcal{Q}}_\nu(k_i,t)}[-\alpha_{k_i}(t_0)][-\alpha_k(t_0)]\right\}
+\alpha_k(t_0)+\prod^{2n-2}_{i=1}\frac{\mathcal{K}_\nu(k_i,t)}{\tilde{\mathcal{Q}}_\nu(k_i,t)}\left[-\alpha_{k_i}(t_0)\right][-\alpha_k(t_0)] \Bigg]
    \\ \nonumber 
    &-\bar{\mathcal{V}}_{2n-2}(k_1,\dots,k_{2n-2};t)\frac{\bar{\lambda}_{2n-2}}{2n-2}\bar{K}^{(2n-2)}_\nu(k,t)+\tilde{\mathcal{V}}_{2n-2}(k_1,\dots,k_{2n-2};t)\frac{\bar{\lambda}_{2n-2}}{2n-2}\int^t_0 {t^\prime}^{(n-2)d-2}\prod^{2n-2}_{i=1}Q_\nu(\vert k_i\vert t^\prime) dt^\prime
    \\ \nonumber
    &-\frac{\tau_{2n-2}}{2}\left[\bar{\mathcal{V}}_{2n-2}(k_1,\dots,k_{2n-2};t)-\tilde{\mathcal{V}}_{2n-2}(k_1,\dots,k_{2n-2};t)\prod^{2n-2}_{i=1}\left(-\alpha_{k_i}(t_0)\right)\right].
\end{align}
Then, for the final part, all the remaining terms of (\ref{relation-2n-2point-2n-2trace-appendix}) are given by
\begin{align}
\label{2n-2-relation-second-part}
& -\frac{1}{(2n-2)!}\langle \overbrace{\Phi_{k}(t) \cdots \Phi_{k}(t)}^{2n-2} \rangle^{[3]}\prod^{2n-2}_{i=1}\left\langle\Phi(k_i,t)\Phi(-k_i,t)\right\rangle^{-1} -\frac{1}{2}\frac{1}{(2n-2)!}\frac{\delta^{2n-2}S_E}{\prod^{2n-2}_{i=1}\delta\Phi(k_i,t)}
 \\ \nonumber &=\frac{n^2}{8} \text{Perm}\Bigg[ \prod^{2n-2}_{j=1}\tau_{k_j} \cdot \tau_{k}^2 \ \tilde{\mathcal{V}}_{2n-2}(k_1,\dots,k_{2n-2};t)
\\ \nonumber
&\times \Bigg[-\frac{1}{2n-1}\prod^{2n-2}_{i=1}\frac{\tilde{\mathcal{Q}}_\nu(k_i,t)}{\mathcal{K}_\nu(k_i,t)}\frac{\tilde{\mathcal{Q}}_\nu(k,t)}{\mathcal{K}_\nu(k,t)}
+\frac{1}{2n-1}\prod^{2n-2}_{i=1}[-\alpha_{k_i}(t_0)][-\alpha_k(t_0)]
\Bigg]\Bigg]
    \\ \nonumber 
    &+\frac{\bar{\lambda}_{2n-2}}{2n-2}\tilde{\mathcal{V}}_{2n-2}(k_1,\dots,k_{2n-2};t)\left[\int^t_0 {t^\prime}^{(n-2)d-2}\prod^{2n-2}_{i=1}Q_\nu(\vert k_i\vert t^\prime) dt^\prime\right]
    \\ \nonumber
    &-\frac{\tau_{2n-2}}{2}\left[\prod^{2n-2}_{i=1}\left(-\alpha(t_0)\right)\tilde{\mathcal{V}}_{2n-2}(k_1,\dots,k_{2n-2};t)\right].
\end{align}
To complete the right-hand side of the relation (\ref{relation-2n-2point-2n-2trace-appendix}), we should add up (\ref{2n-2-relation-first-and-second-part}) and (\ref{2n-2-relation-second-part}). Finally, we can compare $(2n-2)$-multiple trace deformation and $(2n-2)$-stochastic correlation function by 
\begin{align}
    &-\left\langle \prod^{2n-2}_{i=1}\frac{1}{(2n-2)!}\Phi(k_i,t)\right\rangle_\textrm{S}\prod^{2n-2}_{j=1}\left\langle\Phi(k_j,t)\Phi(-k_j,t)\right\rangle^{-1}_\textrm{S}
\\ \nonumber
&+\frac{n^2}{2(n!)^2}\prod^{2n-2}_{i=1}\left\langle\Phi(k_i,t)\Phi(-k_i,t)\right\rangle^{-1}_\textrm{S}\times\textrm{Perm}\left[\left\langle\left\{\prod^{n-1}_{j=1}\Phi(k_j,t)\right\}\Phi(q,t)\right\rangle_\textrm{S}\right.
\\ \nonumber
&\left.\times\left\langle\Phi(q,t)\Phi(-q,t)\right\rangle^{-1}_\textrm{S}\left\langle\left\{\prod^{2n-2}_{l=n}\Phi(k_l,t)\right\}\Phi(-q,t)\right\rangle_\textrm{S}\right]
\\ \nonumber
&-\frac{1}{2}\frac{1}{(2n-2)!}\frac{\delta^{2n-2}S_E}{\prod^{2n-2}_{i=1}\delta\Phi(k_i,t)}
\\ \nonumber
&=\frac{n^2}{8} \prod^{2n-2}_{j=1}\tau_{k_j} \ \tilde{\mathcal{V}}_{2n-2}(k_1,\dots,k_{2n-2};t) \prod^{2n-2}_{i=n}\left(\alpha_{k_i}(t_0)\right)\text{Perm}\left[\tau_{k}^2 \left\{\left(\alpha_{k}(t_0)\right)^2   \frac{\mathcal{K}_\nu(k,t)}{\tilde{\mathcal{Q}}_\nu(k,t)}+ \alpha_{k}(t_0)\right\} \right]
    \\ \nonumber 
    &+\frac{\bar{\lambda}_{2n-2}}{2n-2}\tilde{\mathcal{V}}_{2n-2}(k_1,\dots,k_{2n-2};t)\left[\left[\int^t_0 {t^\prime}^{(n-2)d-2}\prod^{2n-2}_{i=1}\tilde{{Q}}_\nu(\vert k_i\vert t^\prime) dt^\prime\right]+\frac{\tau_{2n-2}}{2}\left[\prod^{2n-2}_{i=1}\left(-\alpha_{k_i}(t_0)\right)\right]\right]
    \\ \nonumber
    &= \prod^{2n-2}_{j=1}\tau_{k_j}  \  \tilde{\mathcal{V}}'_{2n-2}(k_1,\dots,k_{2n-2};t) \Bigg[-\frac{n^2}{8}\text{Perm}\left[ \tau_k^2\ddfrac{\alpha_{k}(t_0)I_\nu(\vert k\vert t) }{I_\nu(\vert k\vert t)-\alpha_k(t_0)K_\nu(\vert k\vert t)}\right]
    \\ \nonumber 
    &+\frac{\bar{\lambda}_{2n-2}}{2n-2}\left[\int^t_0 {t^\prime}^{(n-2)d-2}\prod^{2n-2}_{i=1}\tilde{{Q}}'_\nu(\vert k_i\vert t^\prime) dt^\prime\right]-\frac{\tau_{2n-2}}{2}\Bigg],
\end{align}
where we define
\begin{align}
    \tilde{\mathcal{Q}}'_\nu(k,t)\equiv t^{1/2}\left[K_\nu(\vert k\vert t)-\tilde{\alpha}_k(t_0)I_\nu(\vert k\vert t)\right] \text{,}\quad \tilde{\alpha}_k(t_0)\equiv 1/\alpha_k(t_0),
\end{align}
and 
\begin{align}
\tilde{\mathcal{V}}'_{2n-2}(k_1,\dots,k_{2n-2};t) \equiv \prod^{2n-2}_{i=1}\left[\tilde{\mathcal{Q}}'_\nu(k_i,t)\right]^{-1}.
\end{align}

\end{appendices}

\end{document}